%% file: ngc6302_overview_v1.tex
\documentclass[fleqn,usenatbib]{mnras}

\DeclareRobustCommand{\VAN}[3]{#2}
\let\VANthebibliography\thebibliography
\def\thebibliography{\DeclareRobustCommand{\VAN}[3]{##3}\VANthebibliography}

\usepackage{graphicx}	
\usepackage{amsmath}	
\usepackage{amssymb}	
\usepackage{pdflscape}	
\usepackage{caption}

\usepackage{pgfplotstable}
\usepackage{csvsimple}
\usepackage{booktabs} 

\usepackage{soul}

\usepackage{subcaption}

\newcommand{\Msun}{{$M_{\odot}$}}
\newcommand{\Lsun}{{$L_{\odot}$}}
\newcommand{\kms}{km\,s$^{-1}$}

\newcommand{\HST}{{\it HST}}

\newcommand\arcdeg{\mbox{$^\circ$}}

\newcommand{\HI}{\ion{H}{i}}

\newcommand{\HeII}{\ion{He}{ii}}

\newcommand{\NeII}{[\ion{Ne}{ii}]}

\newcommand{\NeVI}{[\ion{Ne}{vi}]}
\newcommand{\NeV}{[\ion{Ne}{v}]}
\newcommand{\NaVI}{[\ion{Na}{VI}]}

\newcommand{\MgV}{[\ion{Mg}{v}]}
\newcommand{\MgVII}{[\ion{Mg}{vii}]}
\newcommand{\AlVI}{[\ion{Al}{vi}]}
\newcommand{\SiVII}{[\ion{Si}{VII}]}

\newcommand{\ArII}{[\ion{Ar}{ii}]}
\newcommand{\ArIII}{[\ion{Ar}{iii}]}
\newcommand{\ArIV}{[\ion{Ar}{iv}]}
\newcommand{\ArV}{[\ion{Ar}{v}]}
\newcommand{\ClIII}{[\ion{Cl}{III}]}
\newcommand{\ClIV}{[\ion{Cl}{IV}]}
\newcommand{\CrIX}{[\ion{Cr}{ix}]}

\newcommand{\FeII}{[\ion{Fe}{ii}]}

\newcommand{\OII}{[\ion{O}{ii}]}

\newcommand{\NiII}{[\ion{Ni}{ii}]}
\usepackage{amssymb}

\makeatletter
\newcommand{\undersim}[1]{\mathrel{\mathpalette\@undersim{#1}}}
\newcommand{\@undersim}[2]{%
  \vcenter{%
    \ialign{%
      ##\cr
      $\m@th#1#2$\cr
      \noalign{\nointerlineskip\kern.2ex}
      $\m@th#1\sim$\cr
      \noalign{\kern-.4ex}
    }%
  }%
}

\makeatother

\pgfplotsset{compat=1.18}

\title[JWST/MIRI view of PN NGC 6302 I.]{
The \textit{JWST}/MIRI view of the planetary nebula NGC 6302 I.:
a UV irradiated torus and a hot bubble triggering PAH
formation
}
\author[Matsuura et al.]{%
Mikako Matsuura$^1$,
Kevin Volk$^2$, 
Patrick Kavanagh$^{3}$,
Bruce Balick$^{4}$,
Roger Wesson$^{1,5}$,\newauthor
Albert A.\ Zijlstra$^{6,7}$, 
Harriet L. Dinerstein$^{8}$, 
Els Peeters$^{9,10,11}$,
N.\ C.\ Sterling$^{12}$,
Jan Cami$^{9,10,11}$,  \newauthor
Michael J. Barlow$^5$,
Joel Kastner$^{13, 14, 15}$,
Jeremy R.\ Walsh$^{16}$,
L.~B.~F.~M.\ Waters$^{17, 18}$,
Naomi Hirano$^{19}$, \newauthor
Isabel Aleman$^{20}$, 
Jeronimo Bernard-Salas$^{21,22}$,
Charmi Bhatt$^{9,10}$,
Joris Blommaert$^{23}$,
Nicholas Clark$^{9}$,
 \newauthor
Olivia Jones$^{24}$, 
Kay Justtanont$^{25}$, 
F.\ Kemper$^{26, 27, 28}$,
Kathleen E. Kraemer$^{29}$,
Eric Lagadec$^{30}$, \newauthor
J.\ Martin Laming$^{31}$, 
F.\ J.\ Molster$^{32}$,
Paula Moraga Baez$^{13}$,
H.\ Monteiro$^{1, 33**}$
Anita M S Richards$^{6}$, \newauthor
Raghvendra Sahai$^{34}$,
G.~C.\ Sloan$^{2, 35}$, 
Maryam Torki$^{26}$,
Peter A.\ M.\ van Hoof$^{36}$,
Nicholas J.\ Wright$^{37}$,\newauthor
Finnbar Wilson$^1$,
Alexander Csukai$^{6}$
\\
$^1$
Cardiff Hub for Astrophysics Research and Technology (CHART),
School of Physics and Astronomy, Cardiff University, \\
The Parade, Cardiff CF24 3AA, UK\\
$^2$ 
Space Telescope Science Institute, 3700 San Martin Drive, Baltimore, MD 21218, USA\\
$^{3}$
Department of Physics, Maynooth University, Maynooth, Co Kildare, Ireland \\
$^{4}$
Department of Astronomy, University of Washington, Seattle, WA 98195-1580, USA \\
$^{5}$
Department of Physics and Astronomy, University College London, Gower Street, London WC1E 6BT, United Kingdom\\
$^{6}$ 
Jodrell Bank Centre for Astrophysics, Department of Physics \&\ Astronomy, The University of Manchester, Oxford Road, \\
Manchester M13 9PL, UK \\
$^{7}$
School of Mathematical and Physical Sciences, Macquarie University, Sydney, NSW 2109, Australia \\
$^{8}$
Department of Astronomy, University of Texas at Austin, Austin, TX 78712, USA \\
$^{9}$
Department of Physics and Astronomy, University of Western Ontario, London, Ontario, Canada \\ 
$^{10}$
Institute for Earth and Space Exploration, University of Western Ontario, London, Ontario, Canada \\
$^{11}$
SETI Institute, Mountain View, CA, USA \\
$^{12}$
University of West Georgia, 1601 Maple Street, Carrollton, GA 30118, USA \\
$^{13}$ 
Center for Imaging Science, 
  Rochester Institute of Technology, Rochester NY 14623, USA \\
$^{14}$ 
School of Physics and Astronomy,  Rochester Institute of Technology, Rochester NY 14623, USA \\
$^{15}$ 
Laboratory for Multiwavelength Astrophysics,  Rochester Institute of Technology, Rochester NY 14623, USA \\
$^{16}$
European Southern Observatory, Karl-Schwarzschild Strasse 2, D-85748 Garching, Germany \\
$^{17}$
Department of Astrophysics/IMAPP, Radboud University, PO Box 19
9010, 6500 GL Nijmegen, The Netherlands \\
$^{18}$
SRON Netherlands Institute for Space Research, Niels Bohrweg 4,
2333 CA Leiden, The Netherlands \\
$^{19}$
Academia Sinica Institute of Astronomy and Astrophysics, 11F of Astronomy-Mathematics Building, \\
AS/NTU, No.1, Sec. 4, Roosevelt Road, Taipei 106319, Taiwan, R.O.C. \\
$^{20}$ 
Laborat\'{o}rio Nacional de Astrof\'{i}sica, Rua dos Estados Unidos, 154, Bairro das Na\c{c}\~{o}es, Itajub\'{a}, MG, 37504-365, Brazil\\
$^{21}$ 
ACRI-ST, Centred\'aEtudes et de Recherche de Grasse (CERGA), 10Av. Nicolas Copernic, 06130 Grasse, France \\
$^{22}$ 
INCLASS Common Laboratory, 10 Av. Nicolas Copernic, 06130 Grasse, France \\
$^{23}$
Astronomy and Astrophysics Research Group, Department of Physics and Astrophysics, Vrije Universiteit Brussel, Pleinlaan 2, \\
B-1050 Brussels, Belgium \\
$^{24}$
UK Astronomy Technology Centre, Royal Observatory, Blackford Hill, Edinburgh EH9 3HJ, UK \\
$^{25}$
Chalmers University of Technology, Onsala Space Observatory, 439 92 Onsala, Sweden \\
$^{26}$ 
Institut de Ci\`encies de l'Espai (ICE, CSIC), Can Magrans, s/n, E-08193 Cerdanyola del Vall\`es, Barcelona, Spain  \\
$^{27}$
ICREA, Pg. Llu\'is Companys 23, E-08010 Barcelona, Spain  \\
$^{28}$
Institut d'Estudis Espacials de Catalunya (IEEC), E-08860 Castelldefels, Barcelona, Spain  \\
$^{29}$
 Institute for Scientific Research, Boston College, 140 Commonwealth Avenue, Chestnut Hill, MA 02467, USA \\
$^{30}$
 Universit\'e C\^{o}te d’Azur, Observatoire de la C\^{o}te d’Azur, CNRS, Lagrange,
96 Bd de l’Observatoire, 06300 Nice, France \\
$^{31}$
Space Science Division, Code 7684, Naval Research Laboratory, Washington, DC 20375, USA \\
$^{32}$
Leidse instrumentmakers School, Einsteinweg 61, 2333 CC Leiden, The Netherlands  \\
$^{33}$
Instituto de F\'isica e Qu\'imica, Universidade Federal de Itajub\'a, Av. BPS 1303 Pinheirinho, 37500-903 Itajub\'a, MG, Brazil \\
$^{34}$
Jet Propulsion Laboratory, 4800 Oak Grove Drive, California Institute of Technology,
Pasadena, CA 91109, USA \\
$^{35}$
Department of Physics and Astronomy, University of North Carolina, Chapel Hill, NC 27599-3255, USA \\
$^{36}$
Royal Observatory of Belgium, Ringlaan 3, B-1180 Brussels, Belgium \\
$^{37}$
Astrophysics RC/Oesearch Centre, Keele University, Newcastle, ST5 5BG, UK
}
\date{Accepted XXX. Received YYY; in original form ZZZ}

\pubyear{2025}

\begin{document}
\label{firstpage}
\pagerange{\pageref{firstpage}--\pageref{lastpage}}
\maketitle

\begin{abstract}
NGC 6302 is a spectacular bipolar planetary nebula (PN) whose spectrum exhibits fast outflows and highly ionized emission lines, indicating the presence of a very hot central star ($\sim$220,000 K). Its infrared spectrum reveals a mixed oxygen and carbon dust chemistry, displaying both silicate and polycyclic aromatic hydrocarbon (PAH) features.
Using the JWST Mid-Infrared Instrument (MIRI) and Medium Resolution Spectrometer, a mosaic map was obtained over the core of NGC 6302, covering the wavelength range of 5–28\,$\mu$m and spanning an area of $\sim$18.5\arcsec$\times$15\arcsec.
The spatially resolved spectrum reveals $\sim$200 molecular and ionized lines from species requiring ionisation potentials of up to 205\,eV. The spatial distributions highlight a complex structure at the nebula’s centre. Highly ionized species such as \MgVII\ and \SiVII\ show compact structures, while lower-ionization species such as H$^+$ extend much farther outwards, forming filament-defined rims that delineate a bubble.
Within the bubble, the H$^+$ and H$_2$ emission coincide, while the PAH emission appears farther out, indicating an ionization structure distinct from typical photodissociation regions, such as the Orion Bar. This may be the first identification of a PAH formation site in a PN. This PN appears to be shaped not by a steady, continuous outflow, but by a series of dynamic, impulsive bubble ejections, creating local conditions conducive to PAH formation.
A dusty torus surrounds the core, primarily composed of large ($\mu$m-sized) silicate grains with crystalline components. The long-lived torus contains a substantial mass of material, which could support an equilibrium chemistry and a slow dust-formation process.
\end{abstract}




\section{Introduction}

Planetary nebulae (PNe) are excellent testbeds for studying the physics and chemistry of photoionized and photon-dissociated regions (PDRs), which are irradiated by intense UV radiation fields. Some central stars of PNe possess dense, dusty tori or disks. Since these stars emit strongly in the UV, 
PNe containing dusty tori or disks provide opportunities to examine the transition from ionized to neutral and molecular gas within a torus, under varying UV radiation field strength, fast stellar wind, and dust attenuation. These processes are also relevant to star-forming regions and protoplanetary disks. A key advantage of PNe is their large angular extent, allowing them to be spatially resolved with modern telescopes such as the {\it JWST}.

Low- and intermediate-mass (1--8\,\Msun) evolved stars -- asymptotic giant branch (AGB) stars -- are an important source of stardust in galaxies \citep[e.g.][]{1998ApJ...501..643D, 2009MNRAS.396..918M}. The processing of this stardust in the interstellar medium (ISM) can be traced by comparing the composition of stardust to that of interstellar dust \citep{2004ApJ...609..826K}.  It is unclear, however, in what form stardust from AGB stars and their descendants, such as PNe, enters the ISM. This is because (1) evidence is accumulating that their winds have complex, non-spherical structures, such as disks or tori, often associated with binary companions \citep[e.g.][]{Balick.2002, 2011AJ....141..134S}; (2) it is unclear how freshly-made stardust is affected by the harsh UV radiation field or dynamical environment that prevails in PNe \citep{1998Natur.391..868W, 2003A&A...402..189W}, whose central stars have energetic fast winds and extreme radiation fields \citep{Balick.2023iff}. 
 The presence of the disks and tori could modify the dust grain size and composition by frequent collisions of grains during long timescale exposures to the UV radiation from the central stars.

\begin{figure*}
   \includegraphics[trim={0cm 0cm 0cm 0cm},clip, width=1.0\textwidth]{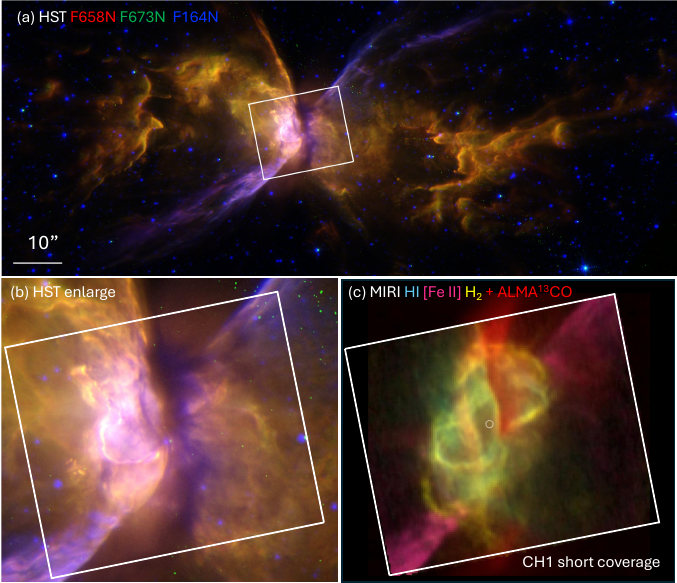}
\caption{
(a): The {\it JWST}/MIRI mapping area (the white box) of the CH1 short sub-band overlaid on the {\it HST} (\textit{Hubble Space Telescope}) three-colour composite image of NGC~6302. North is at the top, east is to the left.
The colour assignments of the {\it HST}/WFC3 image are F164N (blue), F673N (green) and F658N (red). The {\it HST} images are from  \citet{Kastner.2022} and \citet{Balick.2023iff}. Panel (b): A zoom-in of the core region of the {\it HST} image. Panel (c): The same zoom-in of the core region is now shown by features detected with {\it JWST}/MIRI and ALMA. 
The MIRI image is composed of \HI\, at 5.907\,$\mu$m,  \FeII\, at 5.34\,$\mu$m, H$_2$ at 6.91\,$\mu$m  and ALMA $^{13}$CO 2--1. The white circle marks the location of the central source.
}
  \label{hst-images} 
\end{figure*}

The MIRI/IFU (Integral Field Unit) spectrometer on-board {\it JWST}  provides an excellent opportunity to probe the composition of gas and dust in heavily UV-irradiated, but also dust-obscured regions, like the tori in bipolar PNe. In order to study extreme chemical and physical conditions, 
we chose the PN NGC 6302 for our target. This object is one of the most recognizable PNe due to its spectacular bipolar shape \citep{Balick.2002}, displayed in Fig.\,\ref{hst-images}. 
Based on the first deep optical images obtained of NGC 6302, 
\citet{DEvans59} described its optical appearance as ``two lobes of luminous gas, shaped like the wings of a butterfly, separated by a relatively dark lane'', where the dark lane is the dusty torus \citep{LesDin84, Matsuura:2009p23817}.  It has one of the highest estimated initial masses (5--6 \Msun) and the hottest central star (220,000\,K) of any Galactic PN, with high nitrogen abundance, making it a rare PN descended from an intermediate-mass star  \citep{Wright.2011}.

One of the important characteristics of NGC\,6302 is the presence of both oxygen-rich and carbon-rich dust within a single object.
Once formed and ejected, CO molecules lock up C and O atoms. Depending on the C/O abundance ratio, an excess of C atoms forms carbon-bearing molecules and dust, such as polycyclic aromatic hydrocarbons (PAHs), while an excess of O atoms forms oxygen-rich molecules and dust, such as silicates.
Both crystalline silicates and PAHs are detected in NGC\,6302, amid an O-rich gas chemistry \citep[C/O$\sim$0.4; ][]{Wright.2011}.
It has been hypothesized that the crystalline silicates formed in the dusty torus \citep{1998Natur.391..868W, Molster:1999fr}, where the high density and strong UV irradiation enabled slow annealing of dust grains or dust processing over long time scales \cite[$\sim10^5$ years; ][]{Molster:1999fr},
but this hypothesis has yet to be confirmed. 

We present {\it JWST} MIRI IFU mapping of the core of NGC\,6302,  allowing a detailed morphological snapshot of atomic lines, H$_2$, PAHs and crystalline silicate components.
The MIRI IFU map reveals the stratification of ionized gas in a bubble: atomic lines with higher ionization potential are emitted in a compact region, while lines with lower ionization potential are more extended.
H$_2$ is found in arc-like filaments at slightly larger radii. 
These new {\it JWST}  spatio-kinematic 3d maps capture for the first time how PAHs and crystalline silicates form, in a spatially resolved manner.
The detection of the central source is reported in a separate paper (Wesson et al. in preparation).

\section{Target: NGC 6302}

NGC\,6302 is among the best-studied PNe. It exhibits extreme bipolar, complex morphology,  the presence of very high excitation gas, high molecular mass and crystalline silicate dust 
 \citep{Balick.2002, Kastner.2022, 2007A&A...473..207P, Molster.2001}.
In the optical and near-infrared \HST\, images (Fig.\,\ref{hst-images}), the central region displays a torus, most of which is highly obscured by dust  \citep{Kastner.2022}. 
It has been suggested that this torus has confined more recent gas and dust ejections from the original AGB central star, now the PN central star, and shaped the bipolar nebula  \citep{Balick.2002}.

This torus contains a substantial mass of gas and dust, and  is slowly rotating ($<$0.1~km\,s$^{-1}$) as it expands ($\sim$10~km\,s$^{-1}$).
\citep{Rod82, 2008ApJ...673..934D, Santander-Garcia.2016}.
The total torus mass is estimated to be $\sim$2\,\Msun\, from CO lines \citep{2007A&A...473..207P}, with 0.03\,\Msun\, of dust \citep{2005MNRAS.359..383M}.

NGC\,6302 is a luminous PN \citep[total luminosity of $\sim$14,000 \Lsun; ][]{Wright.2011}.  Analysis of the nebular atomic lines shows that the central star is one of the hottest among PNe \citep{Ashley.1988aml}, with an approximate effective temperature of 220,000~K \citep{Wright.2011}. The central star that provides such a large energy has so far not been directly detected. \citet{2005MNRAS.359..383M} suggested that an infrared point source was the central star, but \citet{2009ApJ...707L..32S} argued that a faint optical star, located at the very centre but different from the infrared point source, to be the central star. Later, this optical star was found to be a foreground star with high proper motion, therefore not the central star \citep{Kastner.2022}. The true central star appears to be hidden by dust extinction or blinded by strong emission from ionized gas at optical wavelengths.

The nebula itself shows a complex morphology, with several pairs of long bipolar lobes with slightly different axes and ages \citep{Balick.2002,Balick.2023iff}.
Fig.\,\ref{hst-images}  shows a {\it Hubble Space Telescope (HST)} optical and near-infrared composite image.
The bright butterfly wings of the bipolar nebula extend over 1\,\arcmin\, \citep{Kastner.2022, Balick.2023iff}.  However, fainter nebulosity stretches  (end-to-end) as far as  7\,\arcmin\, \citep{2005AJ....130.2303M, 2024MNRAS.528..459P}, corresponding to a linear diameter of 2.1\,pc at a distance of 1.03$\pm$0.27\,kpc \citep{2020MNRAS.492.4097G}. Expansion measurements have indicated that these lobes were ejected 2250 years ago, post-dating the ejection of the torus \citep{2011MNRAS.416..715S}. 
\citet{Balick.2023iff} identified at least five other younger mass ejections within these wings

 The origin of such a complex nebula is controversial. \citet{Wright.2011} model it as a single star evolving on a post-AGB track. In contrast, \citet{2012ApJ...746..100S} include it as an Intermediate Luminosity Optical Transient (ILOT, such as a Red Nova), that evolved through an explosive event. 
 \citet{2014MNRAS.442.3162U} model the structure with interacting winds but found additional acceleration is required.  \citet{Balick.2023iff} point out that no ejection scenario accounts for the recurrent outbursts at diverse angles. Resolving this will require identifying the central star and  possible companion stars.

\subsection{Distance and foreground extinction} \label{distance_extinction}

As the central star of NGC\,6302 is not detected in the optical, its distance measurements rely on methods other than optical parallax. From expansion measurements, \citet{2005AJ....130.2303M} derive a distance of $1.04\pm0.16$\,kpc, improved to $1.17 \pm 0.14$\,kpc by \citet{2008MNRAS.385..269M}. A similar method was applied by \citet{2020MNRAS.492.4097G} resulting in a distance estimate of 1.03$\pm$0.27\,kpc. We adopt this distance.

NGC\,6302 lies near the Galactic plane ($b$=1.0557), hence may suffer from significant foreground ISM extinction.
The foreground extinction is estimated to be $c(H\beta) = 0.78 \pm 0.10$ \citep{2014A&A...563A..42R}, corresponding to $E(B-V)=0.53\pm0.07$\,mag.
 
\section{Observations and data reduction}

\subsection{Observations}

We observed NGC\,6302 with {\it JWST} \citep{Gardner2023} in Cycle 1 General Observers (GO) program 1742 (PI: Matsuura). 
The data were acquired on 2023 September 9, using the MIRI Medium Resolution Spectrometer  \citep[MRS;][]{Wells.2015won, Argyriou.2023dug} with IFU \citep{Wright.2023}.
The full spectral coverage of 4.9--27.9\,$\mu$m was obtained with four different channels (CH 1--4) (Table\,\ref{observing_log}).

A mosaic map of 5$\times$5 tiles was used to cover the heart of NGC\,6302 (Fig.\,\ref{hst-images}). 
The mapping covers approximately 18.5\arcsec$\times$15.5\arcsec\ (CH1 short) to 22.9\arcsec$\times$19.3\arcsec\ (CH4 long) with increasing areas at longer wavelengths. 
The mapping centre was defined as RA 17:13:44.3938 and Dec $-$37:06:12.36 (J2000), though the actual centres of the maps have slight (a few tenths of arcsec) differences among channels, due to differences in the dithered positions.
The observations were carried out using a four-point dither pattern, which is optimized for an extended object.
The exposure time per sub-band and per tile was 111--255\,s (Table\,\ref{observing_log}), and the readout pattern of FASTR1 was used.

Because the MIRI MRS observations can contain substantial emission from the zodiacal light and telescope thermal emission, the `background' was measured at the offset sky position at RA 17:13:30.66 and Dec $-$37:04:23.4 (J2000), and subtracted from the target data.
The numbers of groups were the same as the target, with the readout pattern of FASTR1. The dither 1 was chosen with a total of 2 dithers for each, optimised for a point source mode. The total integration was 2\,s for the background, making the total exposure times of 55.5\,s, 88.8\,s and 127.7\,s for short, medium and long sub-bands.

The pointing accuracy was evaluated based on two field stars within the MIRI mapping area: one on the east and the other on the west. 
The western star has a  \textit{Gaia} DR 3  identification as \textit{Gaia} DR3 5973805626168712320, with a proper motion of $pm_{\rm{ RA}}$=$-$1.071\,mas\,yr$^{-1}$ and $pm_{\rm{ Dec}}$=$-$7.487\,mas\,yr$^{-1}$ \citep{2023A&A...674A...1G}. 
With about a 6.5 year difference between the \textit{Gaia} and MIRI observations, the proper motion of this star is about $0.05$\arcsec. Compared with the angular resolution of MIRI \citep[0.36--0.4" in FWHM in CH 1; ][]{Argyriou.2023dug}, this proper motion is negligible.

The eastern star does not have a \textit{Gaia}  DR 3 identification but is identified as the Vista source VVV $J$171344.99$-$370615.59 \citep{2012A&A...537A.107S}.
The coordinates of these two stars were compared with those from the {\it HST} F160W image \citep{Kastner.2022}, which was astrometrically calibrated using the coordinates of multiple field stars with GAIA DR3 entries. 
The differences (MIRI$-${\it HST}) of the coordinates are 0.005\arcsec\ in the RA direction and 0.05\arcsec\ in the Dec direction for the eastern star and $-$0.009\arcsec\	in the RA direction	and 0.08\arcsec\ in the Dec direction for the western star.  This is consistent with \textit{JWST}'s positional/pointing accuracy of $<$0.1\arcsec\ \citep{Rigby.2023}.

\subsection{Data reduction}
To reduce the MRS data we used a development version of v1.14.0 JWST Calibration Pipeline \citep{2023zndo...7692609B} with versions 11.17.16 and ``jwst\_1202.pmap'' of the Calibration Reference Data System (CRDS) and CRDS context, respectively. We processed all level 1b (ramp) files through the \texttt{Detector1Pipeline} to produce level 2a (rate) images. Using the dedicated background exposures, we generated master background images for each MIRI/MRS sub-band and subtracted these from the science exposures. The resulting background subtracted level 2a files were processed through \texttt{Spec2Pipeline}, with the \texttt{residual\_fringe} step switched on to produce flux-calibrated level 2b (cal) images. These were then processed through \texttt{Spec3Pipeline} to produce spectral cube mosaics of NGC~6302 in all 12 MIRI/MRS sub-bands by setting the \texttt{cube\_build} parameter `output\_type' to `band'. 

\begin{table*}
\caption{{\it JWST}/MIRI observations of NGC\,6302.  \label{observing_log} }
\begin{filecontents*}{observing_log.csv}
Channel,Slice width,Pixel size,Sub-band,Wavelength,,R,R,Groups,Exposure time,Mapping area,,PSF,
CH1,0.177,0.196,Short,4.9,5.74,3320,3710,10,111,15.5,18.2,0.27,0.30
,,,Medium,5.66,6.63,3190,3750,16,178,15.5,18.2,0.29,0.32
,,,Long,6.53,7.65,3100,3610,23,255,15.4,18.2,0.32,0.36
CH2,0.280,0.196,Short,7.51,8.77,2990,3110,10,111,16.1,19.1,0.35,0.40
,,,Medium,8.67,10.13,2750,3170,16,178,16.1,19.3,0.39,0.44
,,,Long,10.02,11.70,2860,3300,23,255,16.2,19.4,0.44,0.49
CH3,0.390,0.245,Short,11.55,13.47,2530,2880,10,111,17.7,20.4,0.49,0.55
,,,Medium,13.34,15.57,1790,2640,16,178,17.9,20.6,0.55,0.62
,,,Long,15.41,17.98,1980,2790,23,255,17.5,20.4,0.61,0.70
CH4,0.656,0.273,Short,17.70,20.95,1460,1930,10,111,19.2,22.2,0.69,0.80
,,,Medium,20.69,24.48,1680,1770,16,178,19.2,22.2,0.79,0.91
,,,Long,24.19,27.9,1630,1330,23,255,18.9,22.2,0.90,1.03
\end{filecontents*}
\csvreader[tabular= l l l c r{@}{--}l r{@}{--}l r{@}{--}l c  r{@}{$\times$}l ,
				table head=\hline Channel &  Slice width (\arcsec)  & Pixel size (\arcsec) & Sub-band & \multicolumn{2}{c}{Wavelength ($\mu$m)} &  \multicolumn{2}{c}{$R$} & \multicolumn{2}{c}{PSF (\arcsec)} & $t_{\rm exp}$ (s) &\multicolumn{2}{c}{Map size (\arcsec)}  
               \\ \hline\hline, 
				table foot=\hline ] 
				{observing_log.csv}
				{} 
				{\csvcoli & \csvcolii   & \csvcoliii  & \csvcoliv & \csvcolv    & \csvcolvi  & \csvcolvii  & \csvcolviii   & \csvcolxiii &
                \csvcolxiv & \csvcolx  & \csvcolxi &
                \csvcolxii } 
\\				
The MIRI instrumental design and capabilities, including the full width at half maximum (FWHM) of the point spread function (PSF) \citet{Argyriou.2023dug}. The exposure time ($t_{\rm exp}$) and the approximate map size are specific to this program after the final data reduction.
\end{table*}


\subsection{The integrated MIRI spectrum}
Spectra were extracted from each of the MIRI/MRS sub-band mosaic cubes using the \texttt{aperture} module in the Astropy Photutils package\footnote{\url{https://photutils.readthedocs.io/en/stable/}} \citep{Bradley2022}. We applied the additional post-pipeline residual fringe correction to all our spectra, which is included in the JWST Calibration Pipeline package under the \texttt{extract\_1d} step. 
The integrated spectra were extracted from a large (5\arcsec\ at 5.5\,$\mu$m) aperture enclosing the central bright emission region through the MRS spectral channels. The aperture was allowed to `grow’ slightly with increasing wavelength to account for the effect of the increasing MIRI PSF. This ensured a consistent extraction region across the full spectral range of the MRS and that scaling of the spectral sub-bands was not required. 
This integrated MIRI spectrum is shown in Fig.~\ref{integrated-spectra}. It shows many lines, superposed on  dust and PAH features.

The spectra are affected by many issues resulting from the very bright and/or saturated emission lines from NGC~6302. Saturated lines are flanked by faint emission features on either side, caused by scattered light. In the case of the 10.52 $\mu$m [S~{\sc iv}] line, these are seen across a range of 0.1 $\mu$m, while the 7.65 $\mu$m [Ne~{\sc vi}] line shows wings from scattered light over a 0.2  $\mu$m  width. Strongly saturated lines can also cause negative artefacts (seen at 5.5 $\mu$m, for instance), which vary across the field and result from the column/row pull-up/pull-down effect. Finally, strong lines can cause cross-talk with other spectral segments. This is seen for instance at 6.8--6.88 $\mu$m where a weak bump in the spectrum is caused by cross-talk from the saturated 10.52 $\mu$m  line. This results from light scattering into the neighbouring channel. These effects are described in more detail in \citet{Argyriou.2023dug}. We identified all potential detector artefacts through careful inspection of all level 2b images, data cubes, and spectra, which were taken into account in our analysis and interpretation. Due to the impact of the bright/saturated lines on the shape of the continuum, we did not perform any scaling of the individual sub-band spectra.

\subsection{Ancillary data}

We use $^{13}$CO $J$=2--1 and H30$\alpha$ data at 231.90\,GHz from the ALMA project 2012.1.00320.S (PI: Hirano). This project 
obtained 12-meter interferometric data of NGC 6302 with array configurations of C32-4 on 2014 March 9.
The central $\sim$50\arcsec\ region including the torus was covered with the 7-pointing mosaic mode. The bandpass, phase, and flux calibration sources were J1700$-$2610, J1720$-$3552, and Titan, respectively. This project further observed NGC 6302 with the Atacama Compact Array (ACA) 7m array and the Total power (TP) array. The 7m array observations were conducted during the period from 2013 October 6 to 2014 March 22, and the TP observations were taken in 2015 July between the 22nd and 24th.
The total integration time on source was 98.6~min for the 12-m array, 75~min for the 7-m array, and 276.5~min for the TP.

In post-processing, the 12-m, 7-m and TP data are combined using {\sc CASA} \citep{2022PASP..134k4501C}, with the beam size of $0.76\arcsec\times 0.66\arcsec $ (major and minor axes) with the position angle of $-$0.89\arcdeg.

\section{Analysis and results}

\subsection{Spectral features}

The integrated MIRI spectrum of the whole field shows a wealth of narrow gas emission lines, broad emission features from PAHs and a dust continuum. The identifications of representative atomic lines, PAHs and dust features are labelled in Fig.\,\ref{integrated-spectra}.

\subsubsection{Emission lines }

\input{linetable}
\input{ips2}

The line fitting code {\sc ALFA} \citep{Wesson.2016} was used to measure the central wavelength and flux of each emission line. A total of 142 lines were measured with {\sc ALFA}. Line fluxes were also manually measured using {\sc IDL} routines that integrate above a user-defined continuum, and perform simultaneous multi-component Gaussian fits to blended features. The direct measurements are particularly valuable for obtaining accurate fluxes of weak features and of those in regions of rapidly varying continuum, e.g.\ on broad PAH or dust features. An additional 36 lines were measured in this manner, for a total of 178 detected lines.

To identify features, initial line lists were compiled from the \citet{Beintema.1999} {\it ISO/SWS} 2.4--36\,$\mu$m spectrum of NGC~6302, and the {\it JWST/MIRI} spectra of the Ring Nebula (NGC\,6720) reported by \citet{Wesson.2023xe} and van Hoof et al. (in preparation). Because the central star of NGC\,6302 has a higher effective temperature (220,000\,K) than that of NGC\,6720 (130,000\,K), it is expected to exhibit emission from more highly-ionized species.

Line identifications were determined based on wavelength, ionization potential and morphology (as described in Sect.\,\ref{section-morphology}). Wavelengths and ionization potentials for atomic features were obtained from CHIANTI \citep{2023ApJS..268...52D}, NIST\footnote{\url{https://www.nist.gov/pml/atomic-spectra-database}}, the Atomic Line List \citep{Hoof.2018}\footnote{\url{https://linelist.pa.uky.edu/newpage/}}, and the ISO/SWS spectral list assembled by Peter van Hoof
\footnote{\url{https://www.mpe.mpg.de/ir/ISO/linelists/index.html}}. The wavelengths of H$_2$ lines were taken from \citet{Roueff.2019}.
Of the detected lines, 66 are identified as \HI, 54 as \HeII, 8 are H$_2$ lines and the remainder are forbidden lines (Table\,\ref{table-lines}). 


We detect numerous lines from very high-ionization species that are referred to as ``coronal lines'' \citep{1990ApJ...352..307G, Greenhouse.1994}\footnote{
These ions are seen in the Sun's corona, where they are understood to be produced under collisional ionization equilibrium at gas temperatures of $\sim10^5$--$10^6$~K.
In current astronomical usage, for example, as seen in classical novae and AGN,  ``coronal lines'' arise from ions that require ionization energies of $\geq$ 100 eV \citep{Greenhouse.1994}. 
 In other astronomical sources, such as PNe, novae, these ions may alternatively be produced by high-energy photons, as appears to be in the case in NGC 6302.
}. 
 For example, MIRI detected a \SiVII\ 6.49$\mu$m line.
 This ion is also seen in the near-infrared spectrum  \citep{Ashley.1988aml}, and requires photon energies of 205.3~eV to form.
Additionally, transitions from several ions that require 120--190~eV to produce were detected (Table\,\ref{table-ionization-potential}), including \MgVII, \AlVI, and \CrIX.
%
%
%
%
%
%
%

The measured line fluxes 
integrated over the full MRS mapping area are more or less consistent with those measured with {\it ISO/SWS} \citep{Beintema.1999}. The {\it ISO/SWS} had aperture sizes from 14\arcsec$\times$20\arcsec\ (at 5\,$\mu$m) to 20\arcsec$\times$27\arcsec\ (at 28\,$\mu$m). Despite the different SWS aperture sizes and MIRI mapping area, these consistent measurements show that the majority of the mid-infrared line fluxes originate from this central region of the PN.

Strong lines that saturated the MIRI detectors are  \MgVII\, at 5.493~$\mu$m,   \MgV\, at 5.610~$\mu$m, \ArII\ at 6.985~$\mu$m, \NeVI\ at 7.652~$\mu$m, \NeII\ at 12.814~$\mu$m, and \NeV\ at 14.322~$\mu$m.

\subsubsection{PAHs} \label{sec-PAHs}

The abundance analysis of NGC~6302's ionized lines shows that the nebula is oxygen-rich 
\citep[C/O$\sim$0.4; ][]{Wright.2011}. It is known that NGC~6302 exhibits bands from carbonaceous PAH molecules \citep{1986MNRAS.221...63R, Molster.2001, 2005MNRAS.362.1199C}, despite being oxygen-rich.

The integrated spectrum shows clear features from PAH molecules. These are visible at 5.85--6.5~$\mu$m, 11.2~$\mu$m  and a weak band at 12.0~$\mu$m is present. Additional features are expected at 7.7~$\mu$m and 8.6~$\mu$m but these spectral regions are affected by artefacts from the very strong atomic emission lines
at these wavelengths, which makes it difficult to confidently identify the features.  A potential 12.7~$\mu$m feature coincides with the saturated [Ne~{\sc ii}] 12.81~$\mu$m emission line and thus the presence or absence of the PAH feature cannot be determined.

\subsubsection{Dust features} \label{dust}

\begin{figure*}
      \includegraphics[trim={0cm 0cm 0cm 0cm},clip, width=1.05\textwidth]{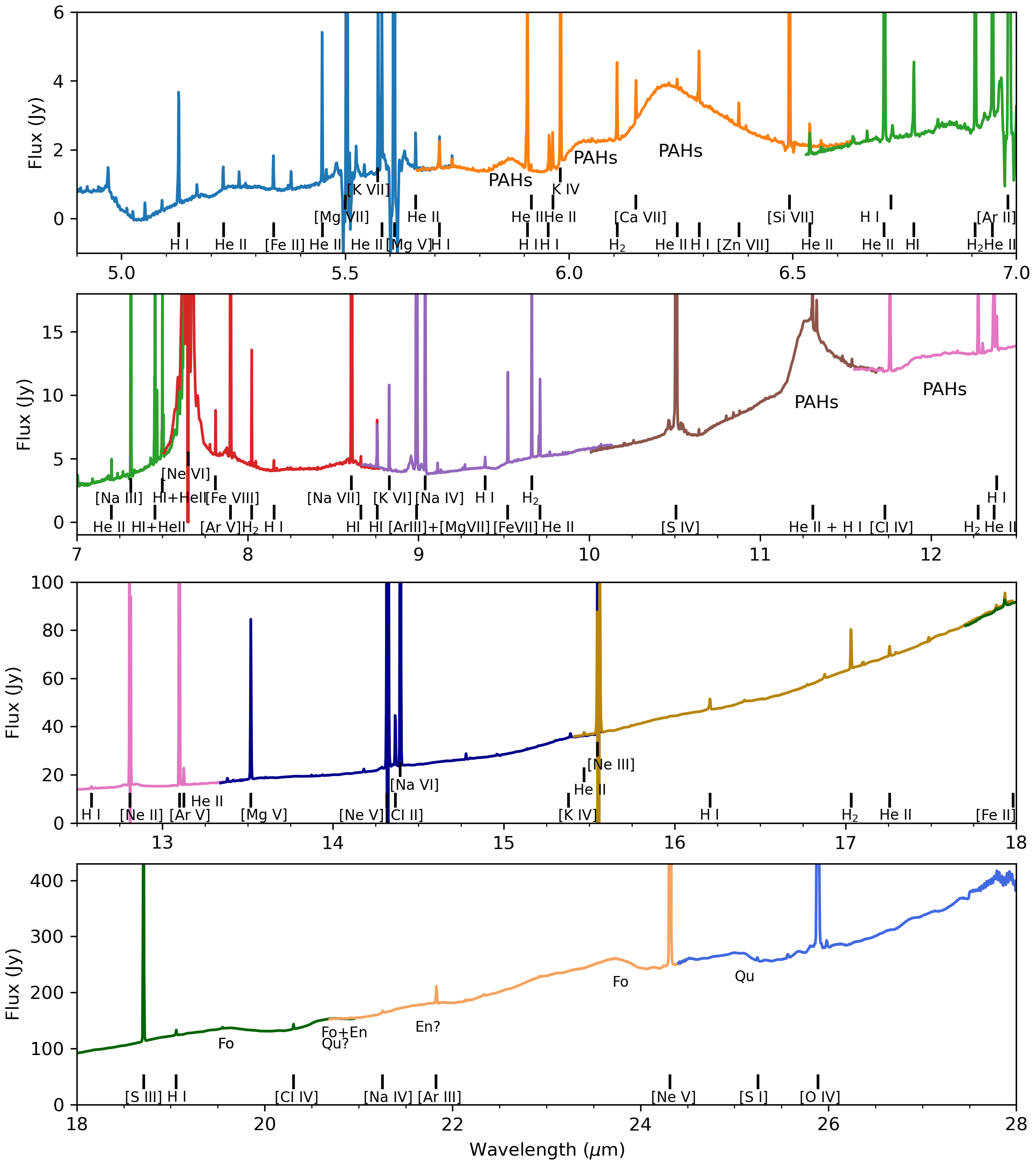}
\caption{The integrated spectra of NGC 6302 within the MIRI mapped area. Different colours are used to indicate different grating orders of MIRI. 
Representative identifications of lines and dust features are labelled. 
The labels of crystalline dust are: Fo (Forsterite), En (Enstatite) and Qu (Quartz).
Some bright atomic lines saturated the detectors and their line profiles can show negative excursions.
}
  \label{integrated-spectra} 
\end{figure*}

NGC\,6302 was already known to exhibit strong crystalline silicate features \citep[e.g.][]{Molster:1999fr, Molster.2001, Kemper.200273}. 
Most of these features are found at wavelengths longer than 18\,$\mu$m (Fig.~\ref{integrated-spectra}). Forsterite dominates most of the features \citep{Molster:2002dh, Chihara.2007}, but the presence of some enstatite features at 17--25\,$\mu$m has been suggested \citep{Hofmeister.2006}.
These features are detected, and are labelled in Fig.~\ref{integrated-spectra}.

The dust features have been compared with dust mass absorption coefficients in order to make their identifications.
When dust optical constants are provided, the dust mass absorption coefficients are calculated by the continuous distribution of spheroids  
\citep[CDS;][]{Min.2003}.
Optical constants and dust mass absorption coefficients are taken from  \citet{Suto.2006}, \cite{Koike.2010} and  \citet{Zeidler.2011} (forsterite),
\citet{Murata.2009cl} and \citet{Zeidler.2015} (enstatite), and \citet{Zeidler.2013e78} (quartz).
Approximately 5\,\% of the dust mass is composed of crystalline silicates with forsterite and quartz dominating (Appendix\,\ref{sect-appen-crystalline}, as opposed to amorphous silicates.

In addition to these known crystalline silicate features, the high-quality MIRI spectra reveal additional features. 
Using the optical constants taken from \citet{Zeidler.2013e78},
the feature at 25\,$\mu$m is likely due to quartz (SiO$_2$).
Quartz should also have broad features at 18 and 21\,$\mu$m, however, these features are blended with the forsterite and 
enstatite features.

We also compared with the dust mass absorption coefficients of  Ca-containing pyroxene \citep{k5m}, olivine \citep{Zeidler.2015} and fayalite \citep{Fabian.2001}, however, it seems that these constituents are not major contributors to the observed dust emission spectrum.

\subsection{Line maps}

 
 Once the lines are identified from the integrated spectra, the next stage is to build spectral maps.
%
%
%
%
Extraction of line images from the data cubes requires subtracting the continuum. For each line,
a number of wavelength pixels (four pixels by default) were selected to cover the emission line.  If adjacent continuum contained no other lines or features, the continuum level was taken as the average of segments consisting of the same number of wavelength pixels on the short and long wavelength side of the line.
The pixel-by-pixel mean of these two images was 
subtracted from the channel images.
The line images  were converted from  units of surface brightness (MJy\,sr$^{-1}$) to line flux  (W\,m$^{-2}$\,pixel$^{-1}$) using the wavelength spacing between the channels, and these individual line flux images were summed to make an output line flux image.  
A set of representative line maps are presented in Fig.\,\ref{five-color}.

In some cases, the continuum regions could not be selected immediately adjacent to the spectral line of interest.  In these instances, either continuum regions slightly further from the line of interest were used or the continuum level was taken from only one side of the line.  At the shorter wavelengths, where more lines are present, the continuum slope is usually small over the wavelength range around a line where the continuum regions are defined, and using a one-sided continuum estimation does not appear to
introduce much of a residual continuum in the line image.

In some instances for very weak lines, the attempt to make a line image failed because the line signal was of the same magnitude as the uncertainties or noise in the continuum image, but for many lines the output line images clearly trace a different morphology than the continuum at
that wavelength.  For some of the saturated lines,  some limited line image information can be obtained from the short-wavelength or long-wavelength wing of the line.



\subsection{The central source}

The nature of the photo-ionizing source at the heart of NGC\,6302 has long been uncertain due to its concealment by abundant dust towards the centre. Previous mid-infrared observations lacked the spatial resolution to confirm its location and sensitivity, while an apparent optical detection in {\it HST} images turned out to be a serendipitous alignment of a foreground star with the approximate geometric centre of the nebula \citep{2009ApJ...707L..32S, Kastner.2022}. 

Now, on the other hand, our JWST observations clearly reveal a bright, compact infrared source at the heart of the nebula, at RA=17:13:44.488$\pm$0.004, dec=-37:06:11.76$\pm$0.03. This confirms the tentative identification of the central source at this location in L-band imaging by \citet{2005MNRAS.359..383M}. No corresponding source is detected at this position at optical wavelengths. 
The nature of the central source will be discussed by Wesson et al. (in preparation).



\subsection{Morphology} \label{section-morphology}

\begin{figure*}
        \includegraphics[trim={0.5cm 0cm 7.5cm 0cm},clip, width=1.0\textwidth]{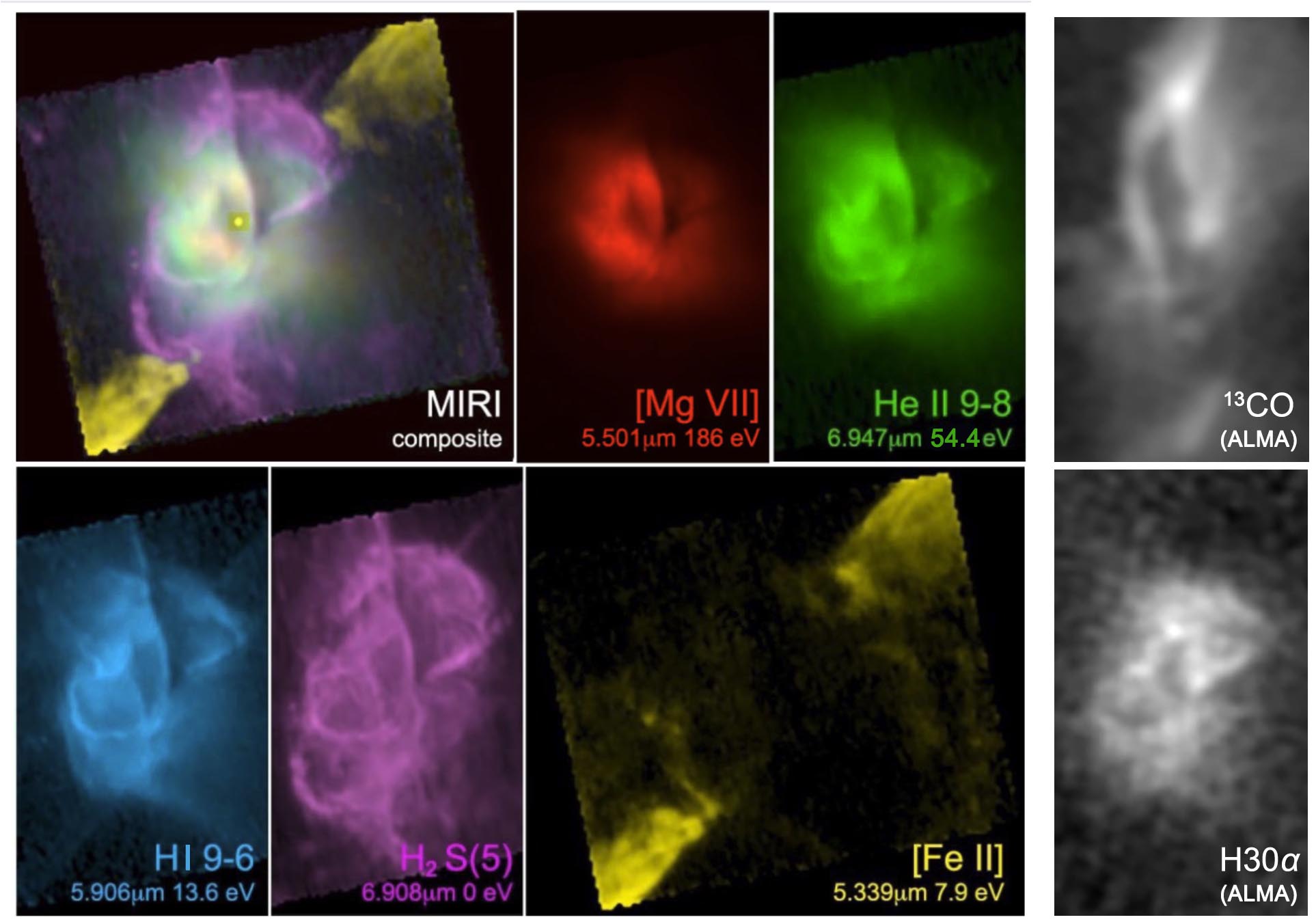}
\caption{Five-colour composite image of NGC~6302, for lines from a range of ionization potentials with 5.50\,$\mu$m \MgVII\, (186\,eV) the second highest ionization potential line, to molecular H$_2$. Each line indicates the ionization energy of the parent species.
  \label{five-color} }
\end{figure*}

Fig.~\ref{five-color} shows representative line maps, demonstrating the strength of MIRI’s IFU capability. 
In this figure, different colours are allocated to different lines. The MIRI composite image (top left) shows the trend of increasing spatial extent of the emission, when going from highly-ionized species such as \MgVII\, to lower ionization tracers, culminating in the H$_2$ line. This is well traced by different colours, with \MgVII\, being the most compact, and \HI\, and H$_2$ the most extended.

\begin{figure*}
    \includegraphics[trim={0cm 0cm 0cm 0cm},clip, width=0.98\textwidth]{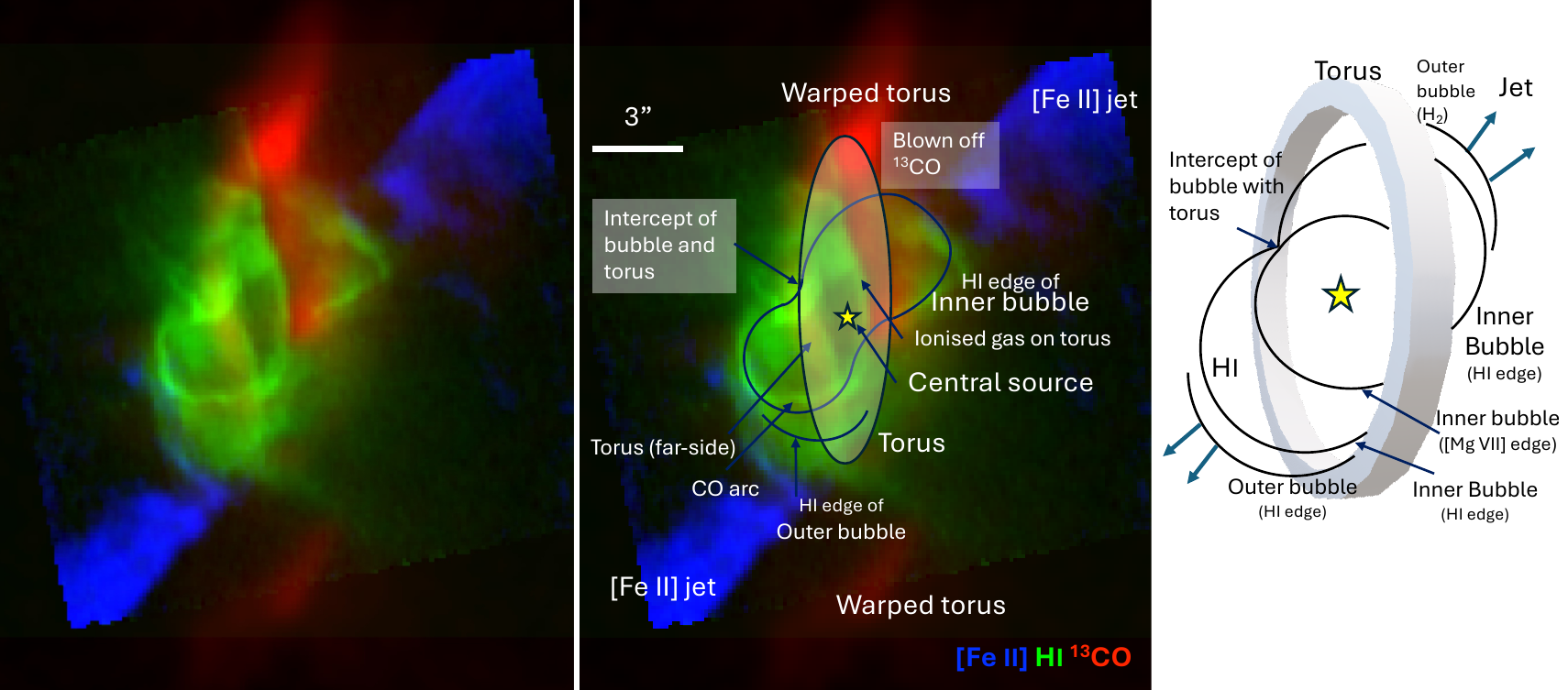}
\caption{
Three-colour composite image of NGC~6302: 5.91\,$\mu$m \HI\,9--6 (green), 5.34\,$\mu$m \FeII\, (blue) and ALMA $^{13}$CO $J$=2--1 (red). Several components, such as a warped torus, inner and outer bubbles and \FeII\, jets are labeled. The `CO arc' refers to a short segment of the arc which is visible in $^{13}$CO (left panel; also in Fig.\,\ref{H_H2_PAHs_13CO}), and is part of the inner bubble. The north part of the outer bubble is found only in H$_2$, which is shown in Figs.\,\ref{hst-images} and \ref{H_H2_PAHs_13CO}.
}
  \label{3color-images} 
\end{figure*}

The central region shows clear ionization stratification. \MgVII,  the second highest ionization emission (186.8 eV), forms a relatively smooth, compact ring. 
The ring is partly bisected by a vertical dark band, caused by extinction from the torus, as illustrated in Fig.\,\ref{3color-images}.  There is a hole inside the ring. The emission running from the north to south on the hole is due to the emission from the surface of the nearly edge-on torus. 

The \HeII\, image, which arises from recombining He$^{++}$  atoms, is centred similarly to \MgVII\, but lacks the clear circular ring structure, and instead appears elongated. The emitting region is larger and shows a well-defined edge along a position angle of approximately 20 degrees west of a North-South axis. 
Further out than \HeII, the \HI\ recombination lines, which trace H$^{+}$, show an elongated structure with a much clearer edge and arcs. Exterior to that, the H$_2$ image shows a second, larger arc at a position angle of approximately 20 degrees of the north-south line. 



A schematic picture of the central region of NGC~6302 is presented in Fig~\ref{3color-images}.
A torus runs north-south, which is detected in the $^{13}$CO 2--1 rotational line (see also Fig. \ref{hst-images}) and partly in the dust extinction (Sect.~\ref{dark-lane-fitting}). The torus appears distorted or warped \citep{Icke.2003} at the outermost regions. 

The nebula shows a series of H$^{+}$ arcs, seen in the \HI\ recombination line image. Some of these delineate two lobes with clear edges in Fig.\ref{3color-images}. 
We consider the two brighter arcs oriented south-east and north-west to be a single structure which we will call the `inner bubble.' It appears 
to be broken into two parts due to dust extinction across the torus. The inner bubble is peanut-shaped. The pinched waist of the peanut shape occurs where the bubble intercepts the torus. Elsewhere it interacts with less dense material and has expanded further (Sec.~\ref{inner-bubble}). 
There is a further H$^{+}$ rim exterior to the inner bubble, which we will call the `outer bubble.'

\subsubsection{[Fe II] and [Ni II] jets}

\begin{figure*}
      \includegraphics[trim={3.8cm 1.65cm 3.3cm 1.7cm},clip, width=0.3\textwidth]{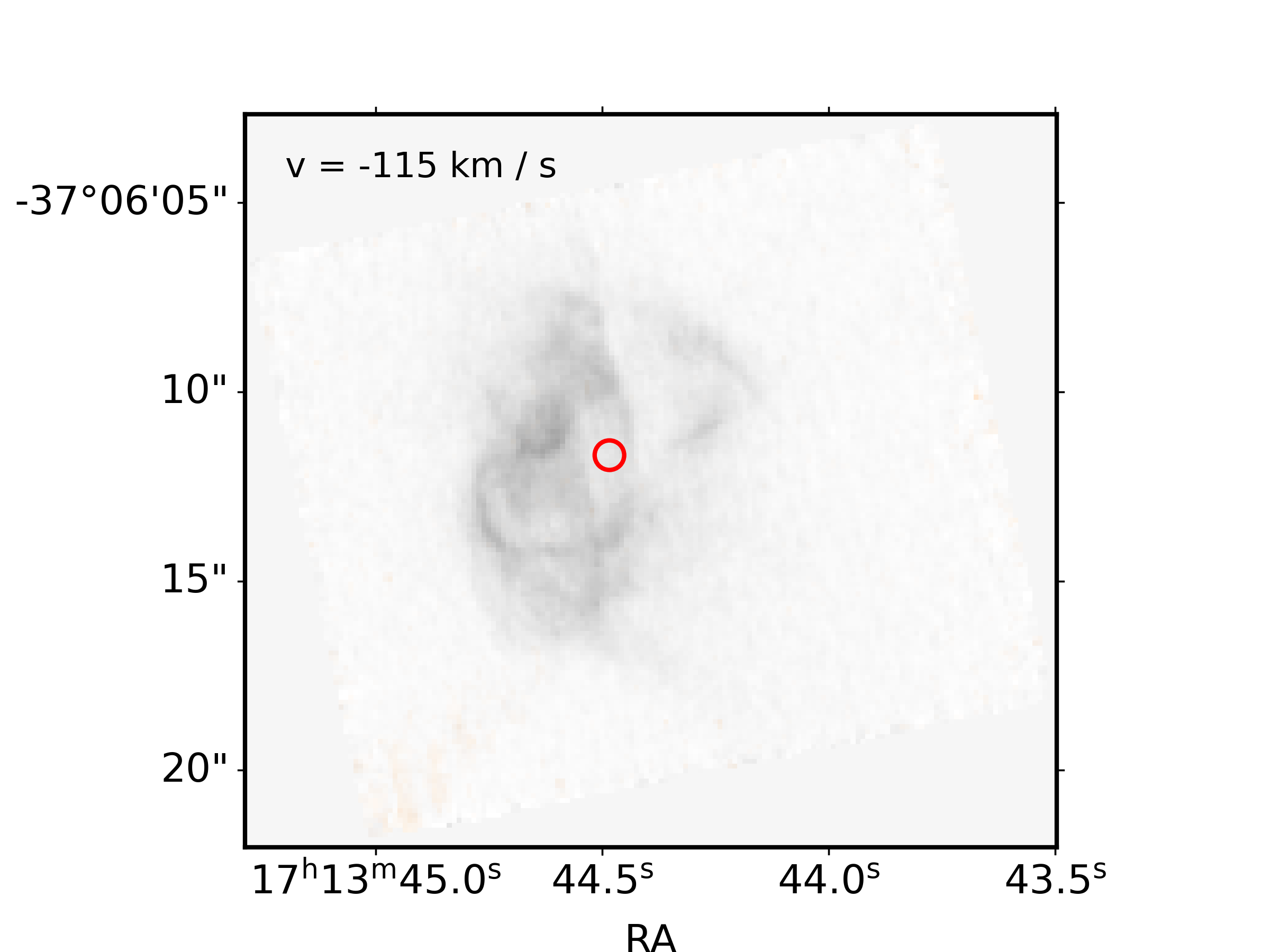}
    \includegraphics[trim={3.8cm 1.65cm 3.3cm 1.7cm},clip, width=0.3\textwidth]{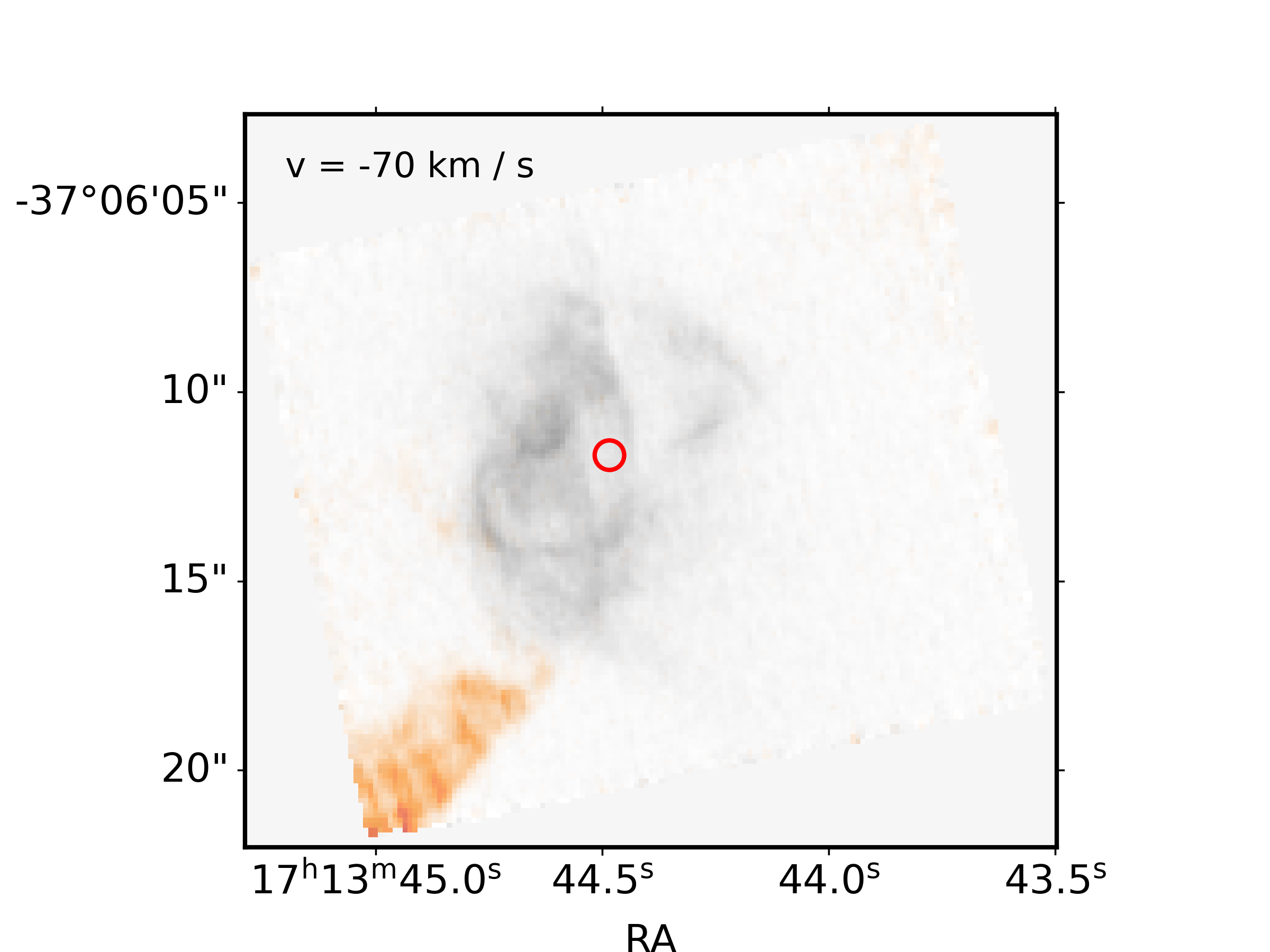}
    \includegraphics[trim={3.8cm 1.65cm 3.3cm 1.7cm},clip, width=0.3\textwidth]{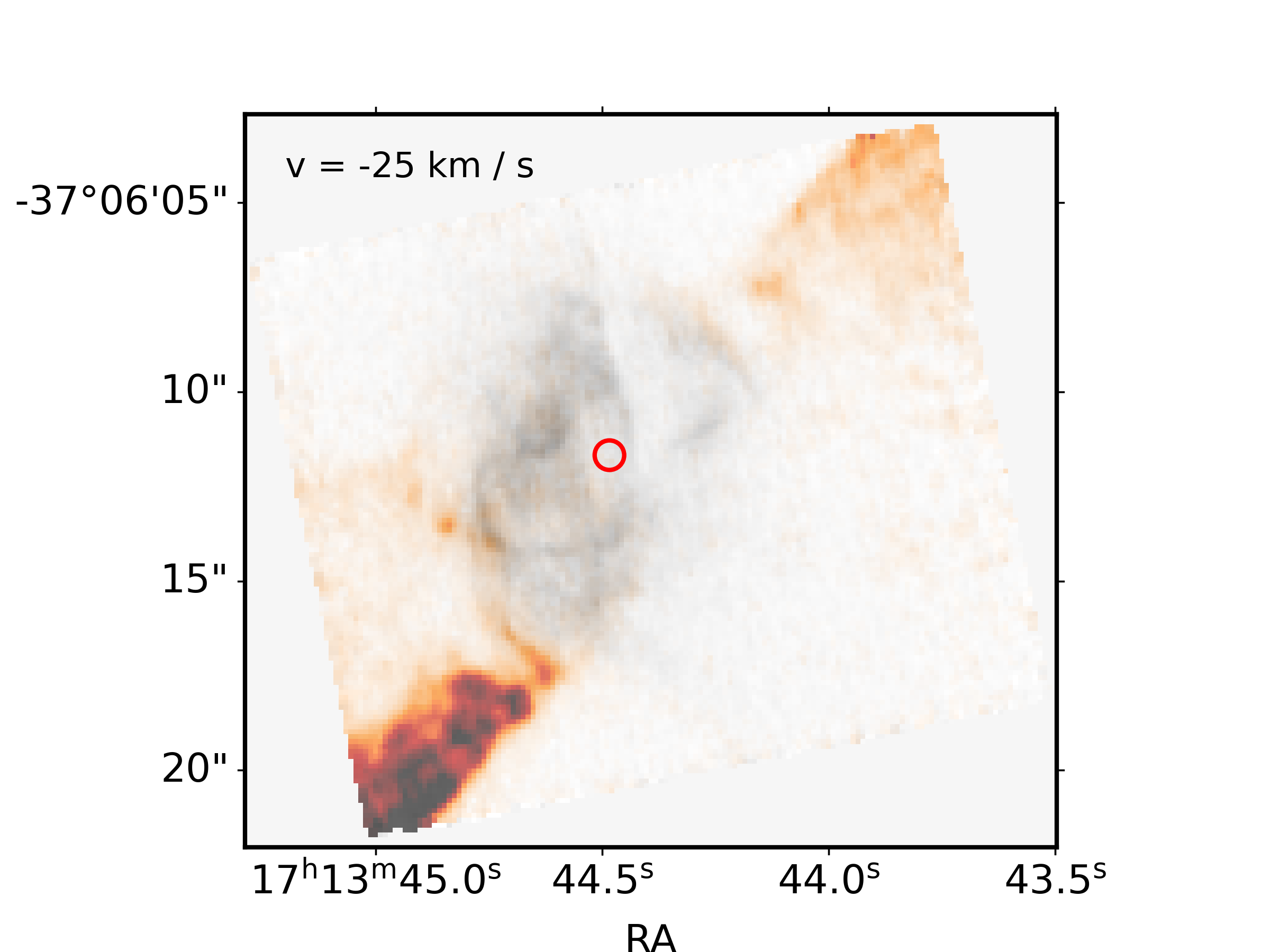}
    \includegraphics[trim={3.8cm 1.65cm 3.3cm 1.7cm},clip, width=0.3\textwidth]{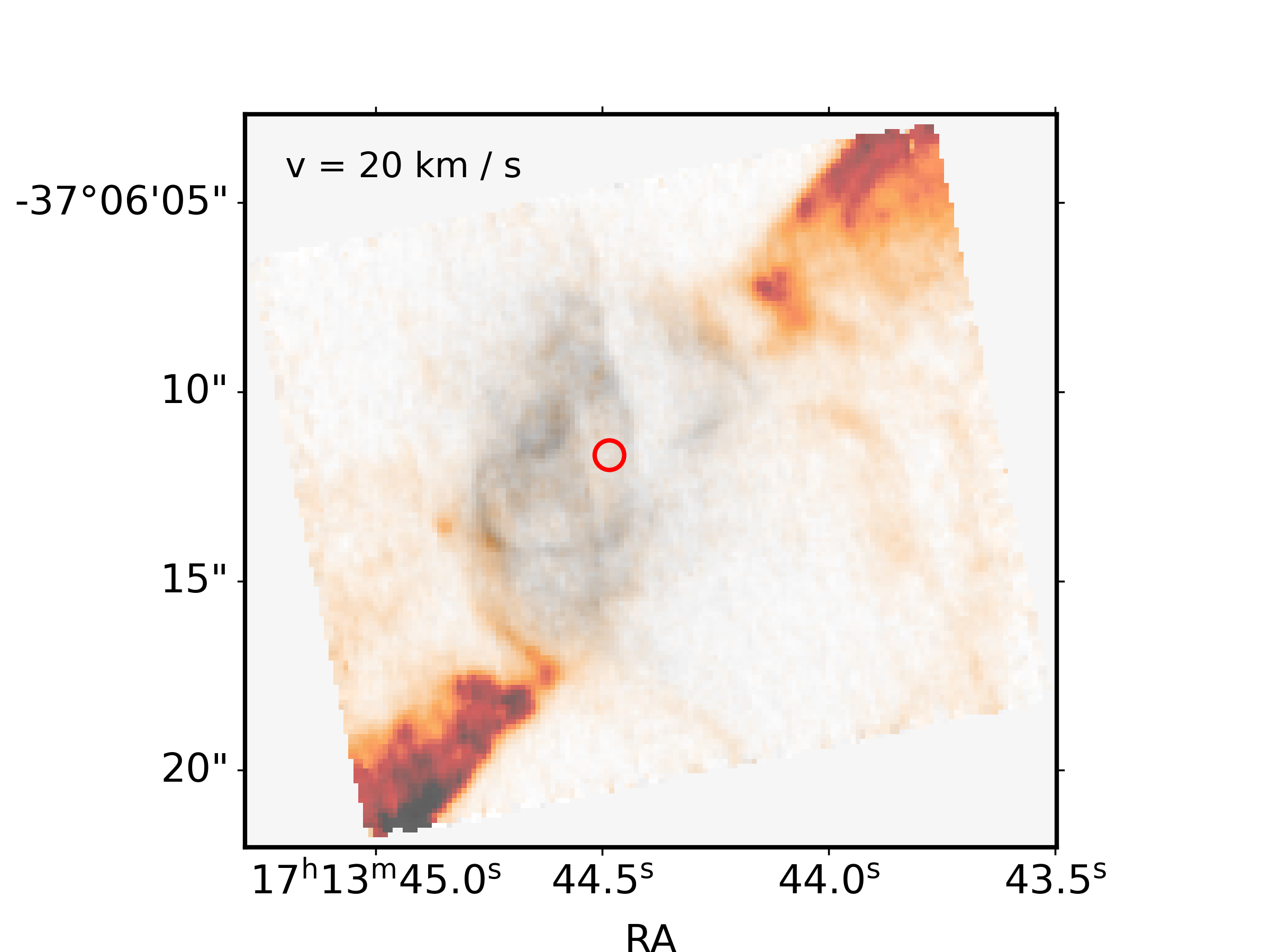} 
    \includegraphics[trim={3.8cm 1.65cm 3.3cm 1.7cm},clip, width=0.3\textwidth]{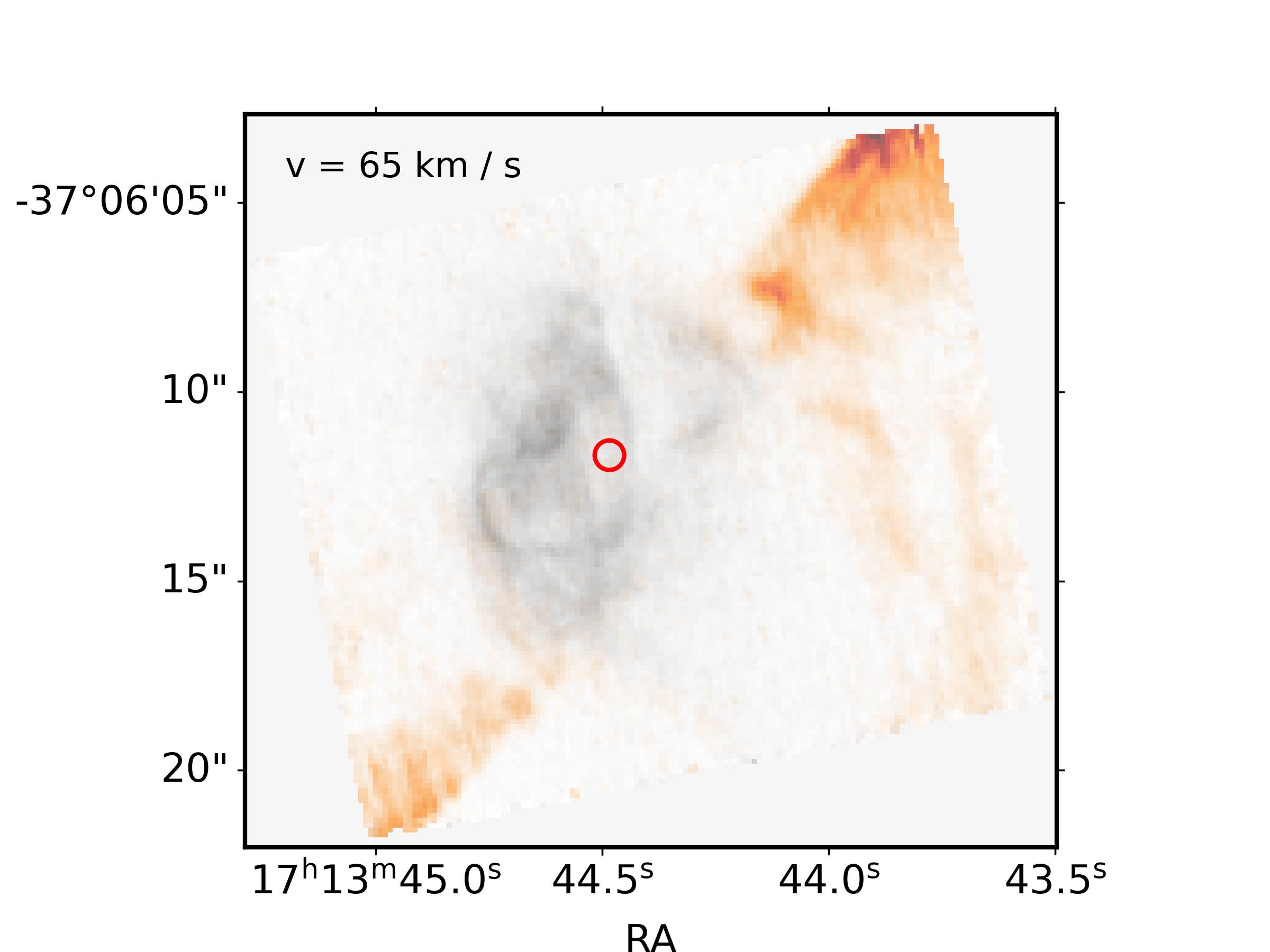}    
    \includegraphics[trim={3.8cm 1.65cm 3.3cm 1.7cm},clip, width=0.3\textwidth]{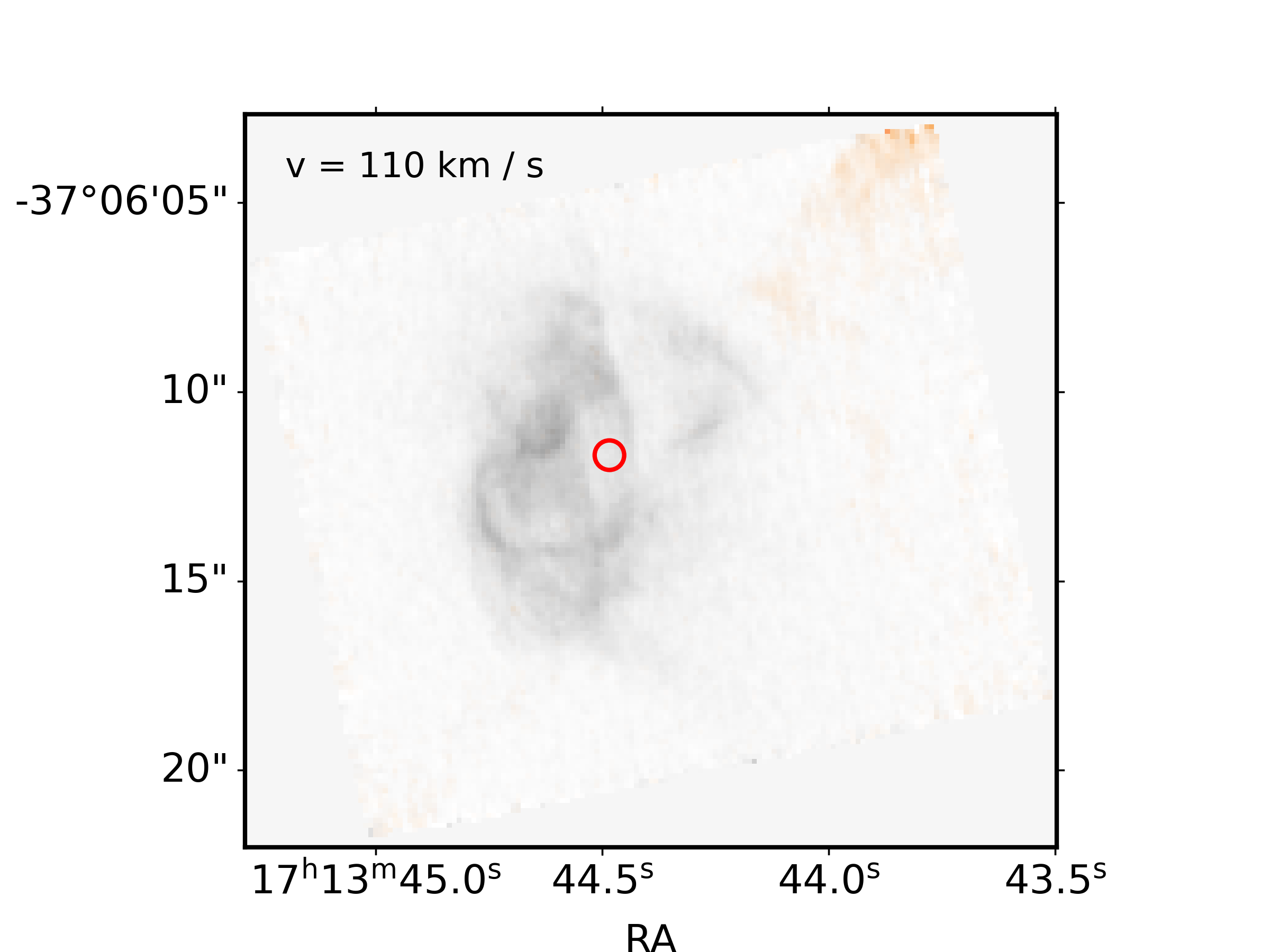}
\caption{
The MIRI channel map of the 5.3\,$\mu$m \FeII\ jet is shown. The jet, highlighted in red, extends from the southeast and the northwest edges of the outer bubble. The southeastern part of the jet is blue-shifted, while the northwestern part is red-shifted. 
The velocity labels represent rest-frame velocities in the Local Standard of Rest, obtained by subtracting the radial velocity of NGC\,6302 from the velocities measured with MIRI. 
For positional reference, the integrated image of the 5.13\,$\mu$m \HI\ 10--6 line is plotted in grey. The red circle shows the position of the central source. The orientation is north at the top, east to the left. The declination grid is marked at intervals of 5\arcsec.
Video version of the jet is available on line.
}
  \label{Fe-jet} 
\end{figure*}

The \FeII\ image reveals two jets extending in nearly opposite directions, with position angles of approximately $-35^\circ$ and $+145^\circ$ (Fig.\,\ref{five-color}). These jets were previously detected in {\it HST} images \citep{Kastner.2022, Balick.2023iff}.  
However, for the first time, JWST's MIRI MRS (Medium Resolution Spectrometer) has provided velocity-resolved observations of these jets, achieving a spectral resolution of $\Delta v \sim 45$ km\,s$^{-1}$ at 5\,$\mu$m, and potentially even better than $<40$ km\,s$^{-1}$ \citep{2023MNRAS.523.2519J}. The jets exhibit velocity spreads ranging from approximately $-115$ to $+110$ km\,s$^{-1}$ in the rest frame (Fig.\,\ref{Fe-jet}).

All six \FeII\ lines, along with a \NiII\ line (Fig.\,\ref{MIRI-images}) detected in the MIRI spectra, display jets with consistent morphology and velocity expansion. 

The \FeII\ jets originate outside the H$_2$ emission region, or outer bubble, but do not exhibit a jet structure interior to the outer bubble (Fig.\,\ref{3color-images}).  
At the outer bubble, there is an arc-like rim that is prominent in H$_2$. The bases of the pair of large \FeII\ “plumes” connect to this arc where it meets the edges of the butterfly wing (Fig.\,\ref{3color-images}).

Although the jets appear to be point-symmetric, their center of symmetry does not align with the central star but is instead offset to the south. 

In addition to the two point-symmetric jets, faint \FeII\ emission is detected throughout the outflow. This includes emission along the H$^{+}$ rim of the inner bubble, at the intercept, and within the interiors of both the eastern and western bipolar outflows.

The velocity map in Fig.~\ref{Fe-jet} represents the rest-frame velocity with respect to the Local Standard of Rest (LSR). The Doppler shift of the gas was measured using MIRI wavelengths in the Solar System Barycentric frame and then converted to $v_{\rm LSR}$. The systemic radial velocity of NGC 6302 \citep[$-30$\,\kms\ in $v_{\rm LSR}$;][]{Santander-Garcia.2016} was subsequently subtracted. The rest-frame wavelength of the \FeII\ line at 5.3401693 $\mu$m is adopted from the NIST database\footnote{\url{https://www.nist.gov/pml/atomic-spectra-database}}.

\subsection{Stratification: from ionized lines to H$_2$ and PAHs}

\begin{figure*}
   \includegraphics[trim={0cm 0cm 0cm 0cm},clip, width=1\textwidth]{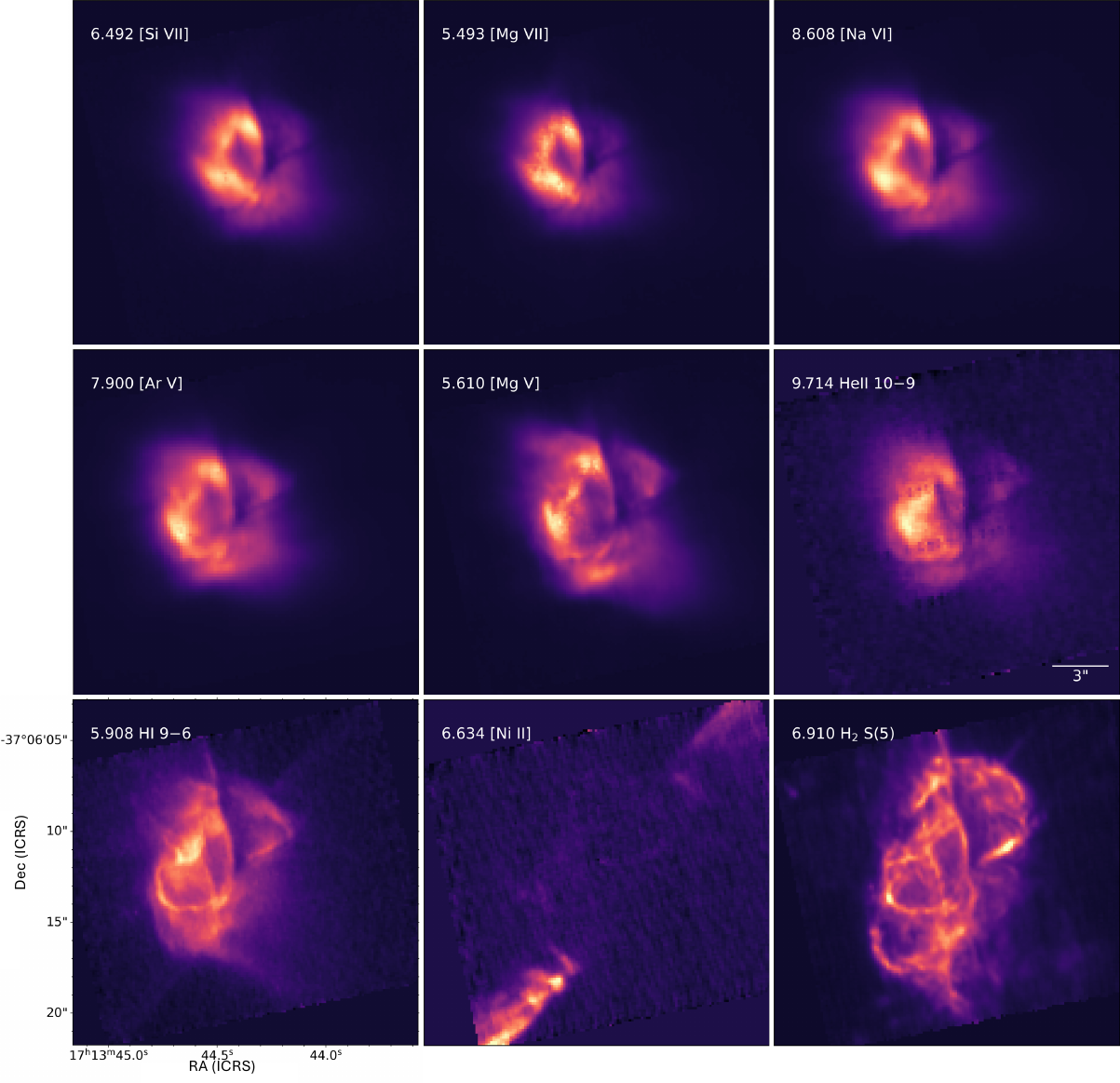}
\caption{NGC~6302 integrated line maps. 
From the left top  \SiVII, to the right column \NaVI\, and then to the middle bottom \NiII\, the atomic line maps are placed in order of a higher ionization line to a lower ionization line, followed by a H$_2$ molecular image. 
Higher ionization lines are more compact than lower ionization lines, and H$_2$ is much more extended, with many filaments within.  \NiII\, and \FeII\, (Fig.\,\ref{five-color}) show jet structures.
In H$_2$, there is a secondary arc outside of the inner bubble in the south-east tip. This corresponds to the outer bubble .
North is towards the top and east is to the left. Square dots in the 9.714\,$\mu$m \HeII\, image are artifacts usually resulting from an outlier detection on a
residual warm pixel.
}
  \label{MIRI-images} 
\end{figure*}

The MIRI IFU data enable us to detect and obtain maps of over 100 lines  from neutral atoms and ions with a large range of ionization potentials: from 7.6\,eV to above 200\,eV. 
Fig.\,\ref{five-color} demonstrates the stratification of several of these lines. Fig.\,\ref{MIRI-images} includes more atomic lines with slightly different ionization potential energies. In general, lines from high ionization-potential species tend to be compact, whereas lines from lower ionization-potential species trace more extended gas.
Several arc-like filaments are embedded in the more extended gas, as seen in Figs.\,\ref{MIRI-images} and \ref{H_H2_PAHs_13CO}. Fig.\,\ref{H_H2_PAHs_13CO} includes $^{13}$CO, which traces the torus, showing the relative positions of ionized gas to the torus. 
Fig.\,\ref{H_H2_PAHs_13CO} also demonstrates the distributions of H$_2$ and PAH emission.

\subsubsection{
The inner bubble and its stratification} \label{inner-bubble}

\begin{figure*}
    \includegraphics[trim={0cm 0cm 0cm 0cm},clip, width=1\textwidth]{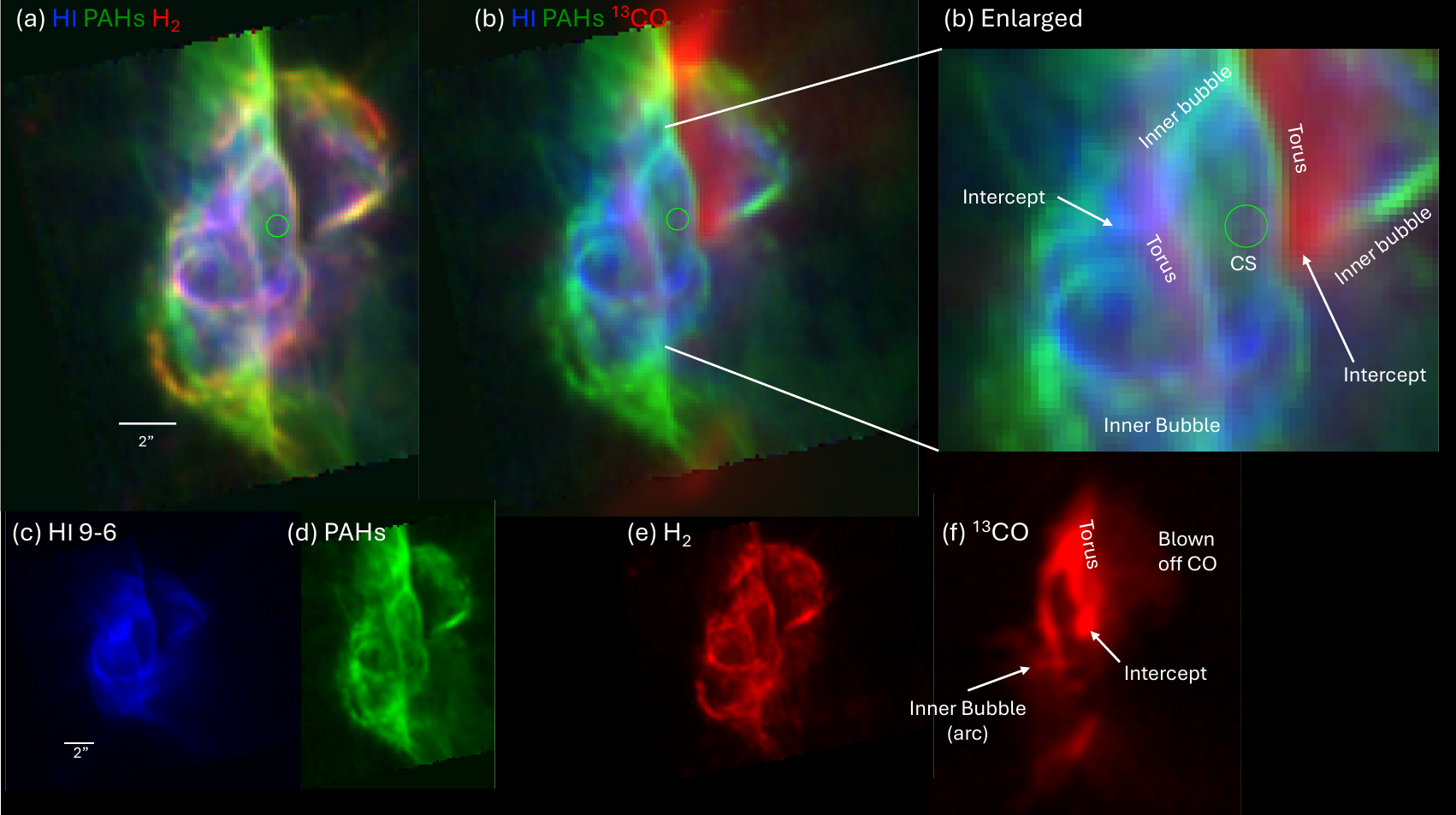}
\caption{Stratification of H$^{+}$, PAHs, and H$_2$, with a $^{13}$CO image displaying the torus. With MIRI's angular resolution, H$^{+}$, and H$_2$ mostly overlap at the bubble (a), as found in pink colour, while PAH emission is found outside of H$^{+}$ at the bubble (b enlarged). On the other side, on the outside of the bubble, PAHs are found inside of H$_2$.
$^{13}$CO traces the molecular torus: its front and rear sides, and the outer extended (warped) regions.
Additionally, $^{13}$CO emission is found at a part of the inner bubble (arc). In the enlarged image (b), stratification of  $^{13}$CO, H$^{+}$ and PAHs is found.
The green circle indicates the location of the central source (CS).
The H$^{+}$ gas is traced by the emission of 5.91\,$\mu$m \HI\ 9--6  line, while PAHs is from 6.3\,$\mu$m band, and H$_2$ is traced by 6.91\,$\mu$m H$_2$ 0--0 S(5) line.
}
  \label{H_H2_PAHs_13CO} 
\end{figure*}

Clear stratification is observed inside the  (labelled in Fig.\,\ref{H_H2_PAHs_13CO}): the physical implications of the bubble will be discussed in Sect.\,\ref{Sect-bubble}.
For example, lines from high 
ionization species, such as 
the \SiVII\, 6.49$\mu$m line (Fig.\,\ref{MIRI-images}),  
shows low intensity interior to this rim, 
and instead is concentrated in a more compact shell.

The line that traces ions of the second highest ionization potential (185~eV) 
is \MgVII.
The Mg$^6$ gas is also situated in a compact shell close to the centre (Fig.\,\ref{five-color}). 
On the other hand, the \HeII\, and \HI\, line emission is relatively weaker at Mg$^6$ gas emitting region, and is stronger further out. The \HI\, emission is strongest in filaments, broken in places. Such a stratification was previously suggested from long-slit spectra of four atomic lines at 2--4~$\mu$m \citep{Casassus.2000}, but is more clearly evident and spatially identified by MIRI.

An exception to the stratification is where the inner bubble `intercepts' the torus (Fig.\,\ref{H_H2_PAHs_13CO}). At the intercepts \MgVII, \HeII\, and \HI\ lines are all present as demonstrated in Fig.~\ref{five-color}. This intercept is the brightest point in the \HI\, line image. 

\subsubsection{Stratification of H$^{+}$, H$_2$ and PAHs at the inner bubble}

Stratification is also found among H$^{+}$ and H$_2$ lines and PAHs at the rim of the inner bubble and its exterior. Fig.~\ref{H_H2_PAHs_13CO} shows (a) the three-colour composite image of \HI, H$_2$ and PAH emission  and (b) the \HI, PAHs and $^{13}$CO image. The $^{13}$CO image is used to indicate the location of the molecular torus. Fig.~\ref{H_H2_PAHs_13CO} (b) is enlarged on the right top panel to see the detailed structures at the bubble and the torus. The H$^{+}$ gas is traced by the emission of 5.91\,$\mu$m \HI\ 9--6  line. The PAH image is a continuum subtracted image taken at 6.3$\mu$m, where the `continuum' (presumed to be mainly silicate emission) was estimated from 6.0 and 6.5\,$\mu$m, which accounts for about 30\% of the total intensity.   The H$_2$ image is from the S(5) line at 6.910\,$\mu$m.

Fig.~\ref{H_H2_PAHs_13CO} (a) shows that \HI\, and H$_2$ are emitted from the same location at the inner bubble (within the MIRI spatial resolution), as the bubble appears as pink in this  \HI\, (blue) and H$_2$ (red) overlay. In contrast, PAHs are found slightly  outwards from \HI, as shown in Fig.~\ref{H_H2_PAHs_13CO} (b) and (b) enlarged.


In the outer bubble, PAH emission overlaps that of H$_2$, or the H$_2$ emission is found outside the PAH emission. This indicates a difference between the stratification of the inner bubble and the larger outer bubble.  

The order of the stratification of H$^{+}$, H$_2$ and PAHs at the inner bubble of NGC~6302 looks different from a typical PDR stratification. 
In Orion Bar, the PAH emission appears immediately behind ($3\times10^{-3}$\,pc) the H$^{+}$, ionization front, and the H$_2$ ridge appears further away by 0.02\,pc \citep{Habart.2024ljv}. 

\subsubsection{Stratification in the torus} \label{Sect-stratification-torus}

Panel (b) enlargement in Fig.~\ref{H_H2_PAHs_13CO} illustrates the stratification of $^{13}$CO, H$^{+}$, and PAHs in the far-side torus. 
The $^{13}$CO emission appears as a long, near-vertical filament along the centre of the torus. PAH emission is present on both sides of the $^{13}$CO structure. The \HI\ and H$_2$ emissions are located alongside $^{13}$CO with a slight offset to the east (Fig.~\ref{H_H2_PAHs_13CO} (a) and enlarged panel (b)), whereas on the west side they overlap with $^{13}$CO.  
Additionally, both H$^{+}$ and H$_2$ exhibit more extended emission on the east side (Fig.~\ref{H_H2_PAHs_13CO} (c) and (e)).

\subsection{Dust extinction} \label{dark-lane-fitting}
\subsubsection{Extinction map}

The images show a prominent dark lane seen in extinction, running from the north to south,  over the full wavelength range covered by the MIRI IFU and {\it HST} (Figs.~\ref{hst-images} and \ref{five-color}). This dark lane is attributed to the torus, also seen in $^{13}$CO ALMA emission (Fig.~\ref{hst-images}). There is significant dust extinction from this torus of the nebula, out to wavelengths as long as 25~$\mu$m 
This extinction far exceeds the foreground
ISM extinction to NGC\,6302, $A(V)=1.6$\,mag from $E(B-V) = 0.53$\,mag (Sect.\,\ref{distance_extinction}).
To obtain the total optical depth at the reference wavelength of 5.91~$\mu$m at the \HI~9--6 line, the ALMA observation of NGC~6302 in the millimetre recombination line H30$\alpha$ was compared to the MIRI line flux.  The extinction at 1292.7609~$\mu$m (231.90093~GHz)  H30$\alpha$ line is assumed to be negligible.  The ratio image H30$\alpha$/\HI~9--6 has the largest value in the dark lane because of the reduction of the \HI~9--6 line intensity. The intrinsic line ratio of H30$\alpha$/\HI~9--6 is calculated from case B, where all the lines, apart from the Lyman lines, are optically thin \citep{Hummer.1987}.  A large grid of {\sc Cloudy} version 23.01 \citep{Chatzikos.2023} case B models were calculated covering the hydrogen density range from 10$^2$ to 10$^6$~cm$^{-3}$ and temperatures between 5,000 and 15,000~K.  The case B ratio shows the most variation for lower densities, which are not applicable for NGC~6302.  For densities at or above 80000~cm$^{-3}$, the typical density value from the {\sc MOCASSIN} model of the nebula from \citet{Wright.2011}, the range of the value of the line ratio is restricted to between about 0.00012 and 0.00016.  We used the minimum of this range to avoid having ratio values lower than the case B prediction within the bright part of the nebula in the MIRI images.

\begin{figure*}
    \includegraphics[trim={0cm 0cm 0cm 0cm},clip, width=0.98\textwidth]{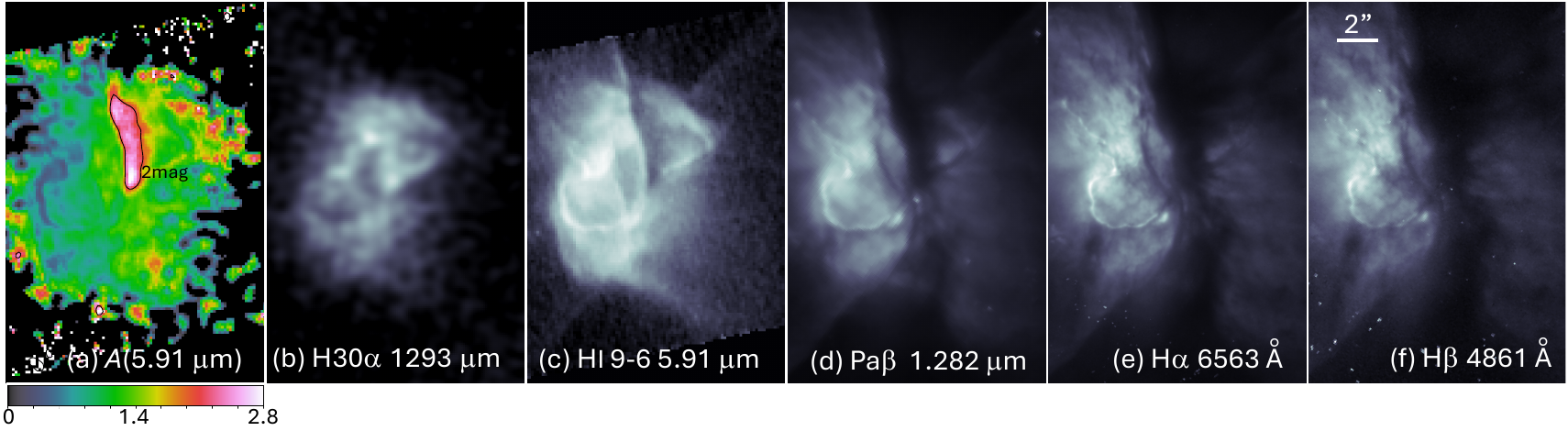}
\caption{(a) The relative extinction in magnitudes across the nebula at the \HI\,9--6 line (5.91~$\mu$m).  
The black contour line traces with $A (5.91~\mu {\rm m})=2$\, mag, the highly extinguished region by  the torus foreground to the bipolar outflow.
The relative extinction is derived from the comparison of the images of (b) the ALMA H30$\alpha$ line flux to (c) the \HI\, 9--6 line flux with respect to the minimum value of case B from {\sc Cloudy} models. The {\it HST} (d) Pa$\beta$, (e) H$\alpha$ and (f) H$\beta$ \citep{Kastner.2022} are plotted in comparison, 
showing the decrease of extinction with longer wavelength in the torus.
The northwest edge of the inner bubble, which serves as the background to the bipolar outflow, is visible only at infrared wavelengths (Fig.~\ref{3color-images}).
}
  \label{relative-extinction} 
\end{figure*}

To make the extinction map (Fig.~\ref{relative-extinction}), the case B minimum value of 0.00012 was used as the unextinguished reference ratio for calculating extinction values across the nebula.  
%
Consequently, the extinction values so derived represent the maximum possible extinction because the minimum case B ratio was used as a comparison.  If the larger ratio value of 0.000179 from the full {\sc CLOUDY} model is used instead, the total extinction at 5.91~$\mu$m in the middle of the dark dust lane 
at RA=17h13m44.40s and Dec=$-$37deg06m10.63s drops from 2.36 to 1.92 magnitudes.  So the extinction uncertainty ranges down 0.4\,mag from the values shown in Fig.~\ref{relative-extinction} for the high extinction areas of the map, although it seems likely that the actual unextinguished reference value to use for the calculation will be somewhere between the full model value and the case B minimum value given the range of ratio values seen from the image comparison.  


The extinction curve used here predicts an extinction of about 0.042 magnitudes at 5.907 $\mu$m from the foreground reddening of $E(B-V) = 0.53$\,mag.  This is much smaller than the extinction measured in Fig.~\ref{relative-extinction}. Hence, the large majority of the extinction originates inside NGC~6302.

\subsubsection{Wavelength dependence of the extinction}

Extinction can be measured by the reddening induced by the wavelength dependence of the dust absorption and scattering \citep[e.g.,][]{Rieke.1985}. One method is by using the observed  relative line intensities of hydrogen or helium recombination lines. There are several such lines detected in the MIRI data, such as the hydrogen recombination lines 
\HI~10--6 (5.13~$\mu$m), \HI~9--6 (5.91~$\mu$m), \HI~10--7 (8.76~$\mu$m), and \HI~8--7 (19.06~$\mu$m). However, dust extinction curves in the mid-infrared show a complex wavelength dependence caused by various silicate bands (Sect.~\ref{dust}), which can vary from typical ISM environments \citep[e.g.][]{2001ApJ...548..296W}. This makes this method difficult to apply. Indeed, the observed ratios of the mentioned lines, compared to predictions from a large grid of case B recombination line calculations from the {\sc CLOUDY} code \citep[version 23.01;][]{Chatzikos.2023} using standard extinction curves, did not give consistent results.

Instead, we determine the wavelength dependence of the extinction component from the torus directly from the spectral profiles of the MIRI IFU slices. 
A cut across the absorption band  was selected from the channel 1-short IFU slices, along Dec=$-$37deg06m10.63s. 
We assume that the emitted flux  varies smoothly across the absorption band, and can be approximated via a linear interpolation across the absorption feature. 
An interpolation was carried out from peak values near pixel $x=74$  (RA= 17h13m44.49s) to near pixel $x=103$ (RA=17h13m44.18s), and the line was interpolated to the position of the flux minimum near pixel $x=83$ (RA=17h13m44.40s).  The relative flux deficit to the interpolation was converted to an extinction value in magnitudes.  This process was carried out for the same cut in the channel-1 short, medium, and long data cubes.  For the other channels, the data cubes were resampled to the channel-1 pixel scale (0.13\arcsec\ per pixel) using the world coordinate system information for each cube, and then the same fitting process was carried out wavelength-by-wavelength for these nine resampled data cubes.  This process provides a measurement of the wavelength-dependent extinction across the absorption band over the full MIRI wavelength range.

\begin{figure*}
        \centering
            \includegraphics[trim={0cm 0cm 0cm 0cm},clip,width=0.80\textwidth]{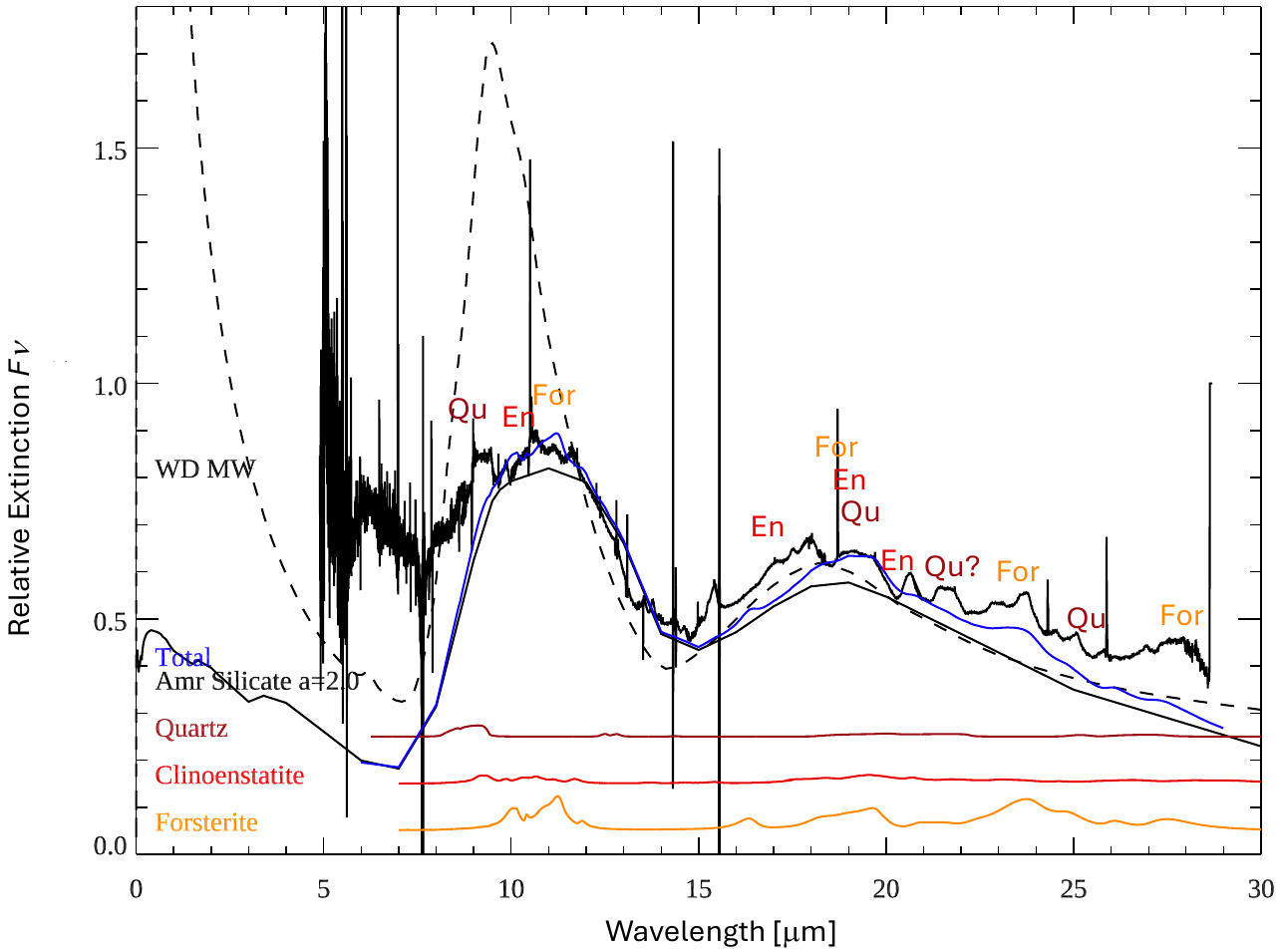}
        \caption{The derived  relative extinction values as a function of
	wavelength at the torus. The dust mass absorption coefficients $\kappa$ for the Milky Way ISM (WD MW) \citep{2001ApJ...548..296W}, amorphous silicate with 2~$\mu$m grain size \citep{1984ApJ...285...89D}, crystalline silicates, Quartz (Qu) \citep{Zeidler.2013e78}, Enstatite (En) \citep{Murata.2009cl} and Forsterite (For) \citep{Koike.2010} are also plotted. Major features are labeled. 
\label{magnitude-extinction}}
    \end{figure*}

Fig.~\ref{magnitude-extinction} shows the resulting wavelength variation of extinction. 
The figure also plots the dust mass absorption coefficient, $\kappa$, for model ISM grains \citep{2001ApJ...548..296W}.
Compared with standard ISM dust grains, the extinction curve for NGC 6302 is rather flat across the wavelength range, and the 10\,$\mu$m silicate feature is rather weak. This may be explained by the presence of larger grains, typically micron-sized grains. Such an extinction curve is represented by 2\,$\mu$m ISM grains, using the optical constants from  \citet{1984ApJ...285...89D}.
The small features around 10\,$\mu$m are mainly due to crystalline enstatite and quartz, on top of the amorphous silicate. The features longward of 15\,$\mu$m are due to crystalline enstatite, quartz and forsterite.

Combining this relative extinction map with the wavelength dependence shown in Fig.~\ref{magnitude-extinction} allows extinction estimates to be made across the central area of the nebula, under the assumption that the wavelength dependence is the same at all points in the central area.  Further work is needed to assess how valid an assumption this is, but together these give an initial estimate of the extinction effects for the MIRI data.

\section{Discussion}

\subsection{Dust in the torus}
\subsubsection{The extinction mass of the dusty torus}

The dense torus is notable, through its extinction, as a dark lane dissecting the visible nebula (Fig.~\ref{hst-images}). It is also seen in CO \citep[][this work]{Santander-Garcia.2016} and in sub-mm dust continuum emission \citep{2005MNRAS.359..383M}, both of which have the advantage of probing both the front and the back of the torus and  the nebula, while extinction only allows the front to be seen. The torus contains a large fraction of the mass of the overall dust mass in NGC\,6302. From JCMT data, \citet{2005MNRAS.359..383M} derive a torus dust mass of 0.03\,\Msun, or a total (gas + dust) mass of 3\,\Msun\, assuming a canonical gas-to-dust mass ratio of 100.


The extinction map can be used to estimate a torus mass. The extinction at 5.9\,$\mu$m can be converted to $A_V$ using the extinction curve of  \citet{2023ApJ...950...86G}, which gives $A(\lambda=5.9\,{\rm \mu m})/A(V) = 0.0261$\,mag, where we assume $R_V=3.1$.
The value for $A_V$ is then converted to a hydrogen column density using the relation of \citet{2009MNRAS.400.2050G}. The total mass is obtained by integrating the column density over the area.

With these assumptions, the area of the torus, defined as the region for which $A_{5.91 \mu \rm{m}}>2.0$\,mag (Fig. \ref{relative-extinction}), that corresponds to $A(V)>76.6$\,mag with a  standard ISM extinction curve \citep{2023ApJ...950...86G}, Fig.~\ref{relative-extinction} gives a total hydrogen mass of $M=1.6$\,\Msun. The method can only measure the torus mass in front of the nebula, and presumably the same amount is present behind the nebula. 
The total hydrogen mass becomes $M_{\rm torus}\sim 3$\,\Msun.

There are substantial uncertainties in this method. The conversion from extinction at 5.9 $\mu$m to $A_V$ and the conversion from $A_V$ to $N_{\rm H}$ both assume typical interstellar dust, whereas the dust in NGC\,6302 may be different. While the dust mass absorption coefficient $\kappa$ of astronomical silicates \citep{1984ApJ...285...89D} with a grain radius of 2\,$\mu$m, and the Milky Way ISM  dust with $R_V =3.1$ \citep{2001ApJ...548..296W} are more or less the same at 5.9~$\mu$m ($\sim$500 and 508 cm$^{-2}$~g, respectively), the $\kappa$  of silicates from \citet{1992A&A...261..567O} is twice as high at a radius of 2\,$\mu$m, hence the mass would be reduced by a factor of two.
The method also does not take into account asymmetric structure, which is expected from a warped torus: the dust extinction map captures the near-side of the north of the torus, which also has the highest $^{13}$CO intensity (Fig.\,\ref{H_H2_PAHs_13CO}).
The derived mass should be viewed as indicative, with a potential range of 0.8--3 \Msun.

Assuming that the line of sight depth of the dark lane is the same as its linear extent (5.5\arcsec, or $1.7\times 10^{17}$\,cm), the hydrogen density of the dark lane is $n_{\rm H} \sim 6 \times 10^6\,\rm cm^{-3}$.

The ionized gas has the highest density at the central region, $n_e=4\times 10^4\,\rm cm^{-3}$, with an electron temperature  of $T_e = 1.90\times10^4$\,K \citep{2014A&A...563A..42R}. These values, measured from optical lines, would indicate approximate pressure equilibrium between the dense lane and the ionized gas if the gas temperature in the lane is $T \sim 100\,$K.

\subsubsection{Dust formation in the torus}

The {\it ISO} detection of crystalline silicates, such as forsterite (Mg$_2$SiO$_4$) and enstatite (MgSiO$_3$), suggested that gas-phase condensation occurs close to chemical equilibrium in a disk or torus \citep{1998Natur.391..868W, Molster:1999fr, 2004ApJ...609..826K, 2006ApJ...644.1164N}. The {\it ISO} detection of other minerals, such as diopside (MgCaSi$_2$O$_6$) in NGC 6302 \citep{Kemper.200273}, supports this hypothesis and is reminiscent of the chemistry in planet-forming disks \citep[e.g.][]{1998A&A...332.1099G, 2009ApJS..182..477S, Olofsson.2010, Varga.2024}. 
For AGB stars in binary systems, a portion of the outflow may be trapped, so that it gains angular momentum and forms a disk or torus \citep{Mastrodemos.1999, 2012BaltA..21...88M}. In such a binary disk, high densities and temperatures can be sustained extended periods, persisting beyond the AGB phase and continuing through the post-AGB and PN stages.  
In contrast,
these conditions of crystalline silicate formation are difficult to reconcile with a rapidly cooling AGB wind, where changes in pressure, temperature, and density occur faster than the timescale required for chemical reactions. Consequently, most silicates in AGB outflows are amorphous, likely forming under non-equilibrium conditions \citep[e.g.,][]{Cherchneff.2006}, resulting in a predominantly glassy structure.

Previous observations of AGB stars, post-AGB stars and PNe with {\it ISO} and {\it Spitzer}  lacked the angular resolution necessary to distinguish between material in the stellar wind and that in the disk or torus \citep{2004ApJ...604..791M, Gielen:2009p25425}. Our spatially resolved observations with {\it JWST} MIRI definitely locate the crystalline silicate emission in the dense torus, providing strong confirmation of this scenario.  

Furthermore, MIRI’s detection of  the large sizes of the dust grains suggests that dust formation occurs over an extended timescale, allowing grains to grow to larger sizes.

\subsection{The effects of photoionization on the torus}

In the torus, a $^{13}$CO traces the cold molecular region and extends north to south, and PAH emission is present on both sides of the  $^{13}$CO strip (Sect.\,\ref{Sect-stratification-torus}; Fig.~\ref{H_H2_PAHs_13CO}). The ionized hydrogen \HI\ recombination line and H$_2$ emission appear adjacent to the $^{13}$CO with a slight offset to the east (Fig.\,~\ref{H_H2_PAHs_13CO}), whereas they overlap with the $^{13}$CO on the west side. 
We attribute this stratification within the torus to a hierarchical photodissociation region (PDR) or X-ray dominated region (XDR; Sect.\ref{Sect-bubble}) structure of different components within the torus, produced by the radiation from the central star. Projection of the torus structure towards our viewpoint causes such alignments progression of layers. 

\begin{figure}
    \includegraphics[trim={0cm 0cm 0cm 0cm},clip, width=0.5\textwidth]{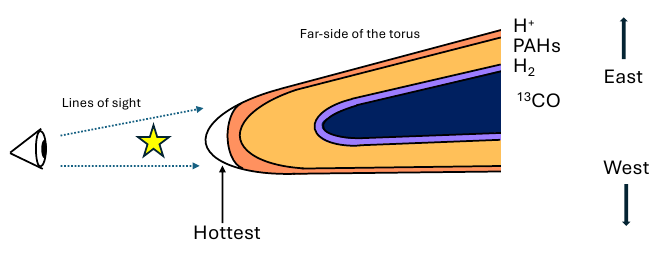}
\caption{A schematic side-view image of the torus relative to the central star. The innermost region of the torus is filled with molecular gas, represented by $^{13}$CO. As UV/X-ray radiation from the central star ionizes the gas, H$^{+}$, emission appears outside the $^{13}$CO region. PAHs are found in the intermediate region between $^{13}$CO and \HI, while H$_2$ is likely located close to the $^{13}$CO region.
}
  \label{schematic_torus} 
\end{figure}

Fig.~\ref{schematic_torus} shows a schematic picture of the far-side of the torus from the side view, illustrating these alignments. The torus is slightly flared, and the scale height increases with a greater distance from the central star. The $^{13}$CO is located at the innermost region of the torus in terms of the scale height. 
Due to the projection effect from the observers point of view, the $^{13}$CO lane runs in the middle of the torus in Fig.~\ref{H_H2_PAHs_13CO}. The H$^{+}$, H$_2$ and PAHs are found on both side of the torus. The torus is viewed almost vertically edge-on on the western side, but slightly tilted to the east, hence all H$^{+}$, H$_2$ and PAHs are aligned on the western side. They are offset in the schematic picture (Fig~\ref{schematic_torus}), but are not spatially resolved in the IFU reconstructed images, or the heavy dust extinction contained in the cold gas and dust region, traced by $^{13}$CO, is obscuring the far radial end of the torus.
The torus is slightly tilted on the eastern side from the observer's point of view, so that the flared part of the torus is spatially well-resolved on the eastern side of the torus. The $^{13}$CO emission is the midplane, and the inner edge of H$^{+}$ and H$_2$ follow, and PAHs are the outermost region. The H$^{+}$ and H$_2$ are extended from the inner edge to a higher scale height.
The stratification of the eastern side of the torus is similar to that of the inner bubble.

It seems likely that some of the torus component is blown off by the hot bubble, contributing to $^{13}$CO emission in the southern extent of the inner bubble.

When a torus is flared rather than flat, the central star can irradiate a higher scale height than a flat case; if the torus is flat, UV radiation can reach only the inner radius of the torus, due to its extinction.
This is well-modelled for proto-planetary disks \citep{Chiang.1997}, where the flare is caused by hydrostatic and radiative equilibrium. The radiation from the central star reaches farther in the case of a flared disk, so that the emission, in projected view, is more extended on the eastern side of the torus from the observer's point of view. \citet{Henning.2024w1b} showed that, due to chemical reactions in the irradiated disk, different molecules can be found in different parts of the proto-planetary disks. That may occur in the torus of NGC\,6302, too. In the case of H$_2$, this requires UV excitation from the central star, hence, it is only seen near the surface of the torus where there is little dust extinction. In contrast, the $^{13}$CO excitation temperature could range from 20--35\,K depending on the region \citep{2007A&A...473..207P}, so that it is in the inner, shielded regions  inside of the torus (Fig.\ref{schematic_torus}).

\subsection{The Expanding Bubbles} \label{Sect-bubble}

Although the stratification of the ionized gas can largely be explained by photoionization, the presence of multiple H$^{+}$ edges and rims, along with the peanut-shaped structures and their interaction with the torus, suggests characteristics of a hot bubble surrounding a high-temperature central star.

The original concept of a hot bubble involves a hot ($\sim 10^6$~K) stellar wind colliding with the surrounding ISM or an AGB wind \citep{Castor.1975, Kwok.1978, Balick.1987}. In the case of NGC~6302, the hot bubble is sweeping up material from the previously ejected nebula and the torus rather than from the ISM. The peanut-like shape of the H$^{+}$ rim (Sect.~\ref{section-morphology}) suggests that the hot bubble blown by the wind is interacting with the torus at the intersection. While the bubble can continue to expand into lower-density regions of the nebula, its expansion is stopped upon encountering the denser torus. This interaction causes the bubble to bend, resulting in its distinctive peanut-like morphology.

The structure of a hot bubble was modelled by radiative and hydrodynamic simulations.  \citet{Toala.2016} calculated the effects when the fast (initially over 1000~km\,s$^{-1}$) wind from PN central star encounters the slowly-expanding ($\sim$15~km\,s$^{-1}$) AGB material. 
This results in hot gas (over $10^6$~K) at the centre with a gradual temperature decrease outwards. After about 1000~years of the interaction, instabilities develop, causing formation of a filamentary structure of H$^+$ or rim of compressed ionized gas, while the hotter gas still fills the interior.
This morphological structure is well replicated in NGC~6302.
Our {\it JWST}/MIRI observations show that emission lines from higher ionization potential species predominantly fill the region near the central star, while the sharp edges of hydrogen recombination line emissions encircle the hotter gas.

\subsubsection{Photoionization, shocks and kinetic heating}

The stratification of ionized gas can be explained by two possible mechanisms: photoionization effects and kinetic effects. 
Since X-rays potentially contribute to both photoionization and the formation of a hot bubble, which generates shocks upon interacting with ambient material, we cannot determine which process primarily drives the stratification of ionized gas based on existing results alone.

First, we discuss the photoionization scenario.
The hot (220,000\,K) central star is capable of emitting from soft X-rays (100 eV $< h\nu<$ 1 keV) to far-UV  (6 eV $< h\nu<$ 13.6 eV)\citep{1985ApJ...291..722T, 2003ApJ...587..278W}  radiation \citep{Wright.2011}, leading to the expectation that the highest-ionization potential lines should be detected nearer the central
star. For the case of a 220,000 K blackbody, 0.38 of all photons emitted have energies $>$54.4\,eV capable of ionizing He$^+$ to He$^{2+}$, whereas only 0.00044 of all photons have energies $>$225\,eV capable of ionizing Mg$^{6+}$ to Mg$^{7+}$ \footnote{For their nebular photoionization models, the best-fitting 220,000 K, log g = 7.0 H-deficient model atmosphere used by Wright et al. emits about a hundred times more photons with energies greater than 225 eV than a 220,000 K blackbody - see their Fig. 5.}. Mg$^{7+}$ 3.028-$\mu$m emission has been detected, although not imaged, from the core of NGC 6302 by \citet{Casassus.2000}. The highest ion stages for which MRS imaging data are available are Si$^{6+}$, requiring 205~eV to create) and Mg$^{6+}$, requiring 187~eV to create. The 6.492-and 5.493-$\mu$m forbidden lines from these respective ions show comparable angular extents, which are both more compact than the extent of the \HeII\ 10--9 9.714-$\mu$m line (Fig.~\ref{MIRI-images}).

All five of the highly ionized species whose emission is imaged in Fig.~\ref{MIRI-images} (\SiVII, \MgVII, \NaVI, \ArV\ and \MgV) show a central hole. We attribute this lack of emission to the `hole' containing more highly ionized ion stages of the respective species, e.g. ion stages from Mg$^{7+}$ through to Mg$^{12+}$ in the case of Mg. Only two of the ions whose emission is shown in Fig.~\ref{MIRI-images} do not show the hole, namely \HI\ 9--6 and \HeII\ 10--9. These arise from
species that are already fully ionized, namely H$^+$ and He$^{2+}$, and so
cannot be ionized further.


In the case of a hot bubble driven by stellar winds, hotter gas could still be present in the central hole, but an alternative possibility is present. As the hot bubble expands, sweeping up the surrounding nebular material, it generates shocks that locally increase the gas temperature. The temperature at these shock fronts could exceed the temperature of gas near the central star \citep{Toala.2014zoom}. The high-temperature gas is enclosed by a thin rim of hydrogen recombination emission, as shock compression increases the local density while reducing the temperature \citep{Toala.2014zoom}. 
Future investigations into the gas within this central hole will help determine whether photoionization or shocks from the hot bubble play a more dominant role in shaping the observed stratification.

The characteristics of PNe with a hot bubble are that they are young (a few thousand years old) and show X-ray emission. The bipolar flows of NGC\,6302 have an age of around 2000 yr \citep{2011MNRAS.416..715S, Balick.2023iff}, so it fits this picture.  X-ray emission has not been detected in NGC~6302 \citep{Jr..2015}. This might be because NGC~6302 is extended to {\it Chandra}. {\it Chandra} X-ray survey PNe tends to favour detections from compact sources \citep[nebula radius smaller than 0.15\,pc; ][]{Freeman.2014}, and NGC\,6302 falls this borderline. 
A deep  X-ray exposure  could test the presence of hot gas inside the bubble.

Of course, NGC~6302 experienced outflows in different directions, with historical outflows older than 2000 years \citep{Balick.2023iff}. These ionized gases overlay in projection and shape the overall butterfly nebula.

\subsection{PAH formation  or excitation exterior to the inner bubble} \label{PAHformation}

In the Orion Bar, \textit{JWST} imaging and spectroscopy showed that the PAHs emission is located between the H$^{+}$ and H$_2$ layers.  \citep{Habart.2024ljv, 2024A&A...685A..74P}. The situation in NGC\,6302 is different, with PAH emission further from the H$^+$ rim and H$_2$ in the inner hot bubble.  The environment appears to affect the relation between H$_2$ and PAHs. This may be related to excitation, 
PAH formation, or PAH destruction.

We can rule out the feasibility of PAH destruction by the hot bubble to explain the IR PAH emitting region in NGC\,6302, based on the bonding energy argument. The ionization potential of 
\HI\, is 13.6~eV and the binding energy of H--H in an H$_2$ molecule is 4.52 eV. Two major bonds in PAHs are C=C and C-H. The C-H bonds are rather fragile and \citet{1985ApJ...290L..25A} 
mentioned an energy of 5~eV and above. The dissociation energy of a PAH is 4.6~eV 
\citep{2010A&A...510A..37M}. This is the dissociation energy by collisions not by photo-ionization, so that careful comparison is needed. 
This PAH dissociation energy is more or less similar to the H$_2$ binding energy, however,  the carbon cage of PAHs is more resilient to destruction, with a potential survival of up to 75~eV in collisional energy \citep{2010A&A...510A..37M}. The actual destruction is somewhere within this range \citep{2010A&A...510A..37M}. With this resilience of PAHs, it is easier to destroy H$_2$ than PAHs. 
Hence, PAHs could appear closer to the central star than H$_2$, if the physical conditions in the bubble would have destroyed PAHs and H$_2$. The observed arrangement is different, hence, the destruction of PAHs by the physical conditions in the bubble is an unlikely cause of the H$^+$, H$_2$ and PAHs order of stratification at the bubble. 
This excludes the scenario in which PAHs were formed during the AGB or post-AGB phase and subsequently destroyed inside the inner bubble during the PN phase.

Instead, the observed stratification can be explained if the expanding hot bubble causes a radiative shock, with a shock front somewhere within or interior to the \HI\, emitting region, where  UV photons from the radiative shock trigger an UV-induced PDR structure in the pre-shocked region, forming the H$^+$ and H$_2$ stratification, followed by the UV- and potentially X-ray induced chemistry 
and eventual PAH formation.

The sound speed of the bubble gas is approximately 16\,km\,s$^{-1}$, assuming a temperature of 10,000 K and the density of $10^6$\,cm$^{-3}$. The assumed density is from the estimated electron density (Sect.\,\ref{electron_density}), and the temperature is estimated by \citet{Wright.2011}. In the \HI\, emitting region, the temperature can be lower than that, hence, the sound speed can also be lower. With the expansion velocity of the gas of 13\,km\,s$^{-1}$, measured from Br$\gamma$ \citep{Casassus.2000}, the hot bubble can exceed the sound speed, and can cause mild shocks.
In the \HI\, emitting region, the temperature can be lower than that, hence, the sound speed can also be lower. With the expansion velocity of the gas of 13\,km\,s$^{-1}$, measured from Br$\gamma$ \citep{Casassus.2000}
\footnote{Following deconvolution in quadrature for their instrumental FWHM resolution of 15~km~s$^{-1}$, Table~3 of \citet{Casassus.2000} presents [Mg~{\sc viii}] 3.028-$\mu$m FWHM line widths of 18 and 24~km~s$^{-1}$ for two orthogonal slit positions on the core of NGC 6302, along with Br$\gamma$ FWHM values of 31 and 41~km~s$^{-1}$ at the same positions. Since each line width will have a contribution from thermal broadening, proportional to the square root of the ratio of the temperature divided by the mass of the species, as well as from macroscopic turbulent/expansion velocities, independent of the species mass, we can use these measured line widths for Mg and H to solve for the gas temperature and the macroscopic turbulent/expansion velocity. Using equations (1) and (2) of \citet{1995MNRAS.272..333B}, we obtain temperatures of 24,000~K and 42,000~K and corresponding turbulent/expansion velocities of 9.4 and 12.6~km~s$^{-1}$ for the two slit positions.}, the hot bubble can exceed the sound speed, and can cause mild shocks.

Energy generated by shocks can produce radiation, and UV radiation from the shocked region can affect the temperature structure of pre-shocked region, i.e. ahead of the shocks.
\citet{1979ApJS...39....1R}, \citet{1979ApJ...227..131S} and \citet{Hollenbach.1989f7}  examined the effect of the UV radiation from the shocked region on the pre-shocked (upstream) region.
\citet{1979ApJ...227..131S} showed that shock velocities higher than 110 km\,s$^{-1}$ generates sufficient UV radiation in the upstream (pre-shocked region) to ionise the gas. Lower shocks velocities  will not ionize the gas \citep{Allen.2008}, but still UV radiation from the shock front enters the neutral and the molecular H$_2$ region, if the pre-shocked region has a reasonably high density \citep{McKee.1980}.

The spectrum of the radiation field can also have an effect. The central star of NGC 6302 is much hotter ($2.2 \times 10^5$\,K)  and sufficient to generate soft X-ray emission. The exciting stars in the Orion Star forming region have temperatures of only 34,600\,K \citep{Abel.2019}. 
If the Orion Bar is a photo-dissociation region (PDR), irradiated by UV, the difference in NGC\,6302 could be due to X-ray radiation. 
Although X-rays have not been detected from NGC\,6302, the hot (220k\,K) central star can emit soft X-rays and far-UV \citep[Fig\,5 in][]{Wright.2011}. That can trigger an X-ray dominated region  \citep[XDR;][]{2022ARA&A..60..247W} in the ambient gas.
Soft X-rays are predominantly shielded by hydrogen \citep{Maloney.1996}, so that relatively high density at the H$^+$ rim of the bubble might shield X-ray, and subsequently forming H$_2$. 
An XDR can dissociate CO, leading to carbon chemistry \citep{Meijerink.2005r4}, and PAHs formation \citep{2003A&A...402..189W}. 
A PAH formation route through HCO$^+$ may benefit from X-ray irradiation \citep{2002ApJ...574L.167W}.
Indeed, HCO$^+$ has been detected in NGC\,6302 (Moraga Baez et al. in preparation) and also CH$_3^+$ (Bhatt et al. in preparation), which is also seeding carbon-rich chemistry.
The CO emission has been detected at the H$^+$ rim of the inner bubble \citep{Santander-Garcia.2016}, and if X-ray radiation is strong enough to dissociate CO into C or C$^+$, it can open the route to form PAHs via HCO$^+$. A characteristic of XDRs is that the transition between H$^+$, \HI\ and H$_2$ is rather gradual as a function of hydrogen number density, compared with PDR. This transition in the bubble of NGC 6302 is much sharper than the one observed in the Orion Bar. This could be because the decline of the density at this rim is steep. The remaining question is why PAHs are detected far behind H$_2$ or HCO$^+$.  PAHs can be excited by UV photons with energies larger than a few eV \citep{Draine.2021d6}, which should be present in this region,  hence, once PAHs are formed, they should be detected adjacent to the carbon chemistry region, i.e. HCO$^+$ and CH$_3^+$ \citep{Berne.20230dw}, but PAHs are displaced behind these molecules. It might be because the timescale of PAH formation is a slower process than the formation of H$_2$. Detailed chemical and PAH excitation models would be needed to explain PAH formation behind this inner bubble.

The displacement of IR PAHs emitting regions relative to H$_2$ emitting region in PDR regions can be influenced by differing shielding mechanisms affecting H$_2$ and CO. This concept is applied to the star-forming region N13 in the Small Magellanic Cloud (SMC). Recent JWST observations reveal that H$_2$ and PAHs in N13 are co-located or that PAHs are slightly behind H$_2$ \citep{2025arXiv250406247C}, contrasting with the Orion Bar, where PAHs are observed closer to the exciting star than H$_2$ \citep{2024A&A...685A..73H, 2024A&A...685A..74P}. The stratification observed in N13 is attributed to the fact that H$_2$ is capable of self-shielding \citep{Maloney.1996, 2010ApJ...716.1191W, 2014ApJ...795...37G}, whereas CO depends on dust extinction for protection against UV radiation. In the low-metallicity environment of the SMC, dust extinction is reduced, allowing UV photons to penetrate deeper compared to regions in the Milky Way. However, H$_2$ self-shielding remains effective, even at low metallicity. As a result, the stratification order of H$_2$ and PAHs is reversed in the SMC star-forming region N13 compared to the Orion Bar.  In the case of NGC 6302, the high effective temperature of its central star has to be considered, as it emits more energetic UV photons than the exciting star in the Orion Bar. However, dust processing in NGC 6302 may have reduced dust extinction at UV wavelengths, allowing deeper UV penetration and resulting in a stratification pattern similar to that observed in N13.

The interpretation of the hot bubble and UV irradiation of the upstream material applies to only the latest outflow (inner bubble). That region is strongly affected by the presence of the warped torus, hence, UV radiation can be easily shielded in certain directions (North-northwest and south-southeast). 
Stellar UV penetrates throughout the core and far into the larger nebula of NGC\,6302, suggesting that there is no ionization front in the bubble (or near it) to absorb all of the stellar UV photons.  Once PAHs are formed, stellar UV radiation can excite PAHs in the nebula, such as the outer bubble.

\section{Conclusions}

We have explored the heart of the extreme bipolar PN NGC 6302 using \textit{JWST} MRS mapping of its infrared spectrum at unprecedentedly-high angular-resolution. 
These new spectral maps provide the spatial distributions of nearly
200 emission lines of neutral atomic, ionic, and molecular species; PAH features; and crystalline and amorphous silicates. Combined with ALMA $^{13}$CO and H30$\alpha$, these atomic and molecular lines and dust features have revealed structures deep inside the dust-obscured region at the centre of the PN, providing fresh insight into the chemistry and shaping of this luminous, massive, prototypical bipolar PN.

NGC 6302 has a central expanding torus which contains dust grains within. The expanding torus, with approximate radius at infrared wavelengths of 5.5", or 5700 au, is the primary dust reservoir with a total dust plus gas mass of 0.8--3\,\Msun. Its mid-infrared spectra show a high fraction of crystalline silicates and inferred $\mu$m-sized dust grains, in contrast to typical ISM grain size distributions \citep[a few \AA\, to 1\,$\mu$m grain size; ][]{2001ApJ...548..296W}.
Our observations support the hypothesis of an equilibrium chemistry in a disk to form crystalline silicates and large dust grains.

The wind from the PN central star appears to have proceeded as a sequence of bursts rather than via a continuous outflow, resulting in bubbles and arc filaments. This may be due to the infall of material 
from a disk formed by mass transfer from a close companion, which may supply the central star with accretion energy capable of producing intermittent outflows.
The central region of the PN is not represented by steady state mass loss, but by more dynamic and impulsive energetic outflows.

Inside the inner bubble, a stratification of ionized species is found. Species with higher ionization potentials are found to have a more compact distribution, while the lower ionization potential species are found to be more extended. This shows that both the UV/X-ray radiation energy from the central star and perhaps the kinetic enregy of the hot bubble cause a higher degree of ionization in the innermost regions, which decreases towards the outside.
When the bubble breaks, 
the outflowing material shocks and heats previously ejected material, producing \FeII\, and \NiII\ line emission seen as the NW and SE jets.

{\it JWST} and ALMA capture the most recent stage of the wind, which forms the inner bubble, and blows some material off the torus.
The inner bubble should also trigger shocks when it interacts with the ambient circumstellar material, which presumably consists of  past outflows.  The shock emits UV radiation,
which penetrates the external regions and excites the H$_2$ and PAH emission.
This emitting region of PAHs is found outside of H$_2$, and that might suggest that PAHs is formed at the outer region of the bubble. This could be the first indication of capturing the PAH formation site in a PN, important for understanding PAH formation chemistry.

\section*{Acknowledgements}
This work is based on observations made with the NASA/ESA/CSA James Webb Space Telescope. The data were obtained from the Mikulski Archive for Space Telescopes at the Space Telescope Science Institute, which is operated by the Association of Universities for Research in Astronomy, Inc., under NASA contract NAS 5-03127 for JWST. These observations are associated with program \#1742.

This work has made use of data from the European Space Agency (ESA) mission {\it Gaia} (\url{https://www.cosmos.esa.int/gaia}), processed by the {\it Gaia} Data Processing and Analysis Consortium (DPAC, \url{https://www.cosmos.esa.int/web/gaia/dpac/consortium}). Funding for the DPAC has been provided by national institutions, in particular, the institutions participating in the {\it Gaia} Multilateral Agreement.

This study is based on the international consortium of ESSENcE (Evolved Stars and their Nebulae in the JWST era).

M.M. and R.W. acknowledge support from the STFC Consolidated grant (ST/W000830/1).
M.J.B. and R.W. acknowledge support from the European Research Council (ERC) Advanced Grant SNDUST 694520. 

A.A.Z. acknowledges funding through UKRI/STFC through grant ST/T000414/1.

H.L.D. acknowledges support from grant JWST-GO-01742.004 and NSF grants 1715332 and 2307117.

J.C. and E.P. acknowledge support from the University of Western Ontario, the Canadian Space Agency (CSA)[22JWGO1-14], and the Natural Sciences and Engineering Research Council of Canada. 

N.C.S.\ acknowledges support from NSF award AST-2307116.
 
G. G.-S. thanks Michael L. Norman and the Laboratory for Computational Astrophysics for the use of ZEUS-3D. The computations were performed at the Instituto de Astronom{\'i}a-UNAM at Ensenada.

P.J.K. acknowledges support from the Science Foundation Ireland/Irish Research Council Pathway programme under Grant Number 21/PATH-S/9360.

K.E.K. acknowledges support from grant JWST-GO-01742.010-A.

 C.K. and MT were partly supported by the Spanish program Unidad de Excelencia Mar\'{i}a de Maeztu CEX2020-001058-M, financed by MCIN/AEI/10.13039/501100011033.
 
 JML was supported by basic research funds of the Office of Naval Research.

R.S.'s contribution to the research described here was carried out at the Jet Propulsion
Laboratory, California Institute of Technology, under a contract with NASA, and partially
funded by grant JWST-GO-01742.005-A  from the STScI under NASA contract NAS5-03127

This research made use of {\sc photutils}, an {\sc astropy} package for
detection and photometry of astronomical sources (\citealt{larry_bradley_2023_7946442}).
The {\sc Montage} \footnote{\url{http://montage.ipac.caltech.edu/applications.html}} was used in python code to make false colour images, developed by \citet{2020MNRAS.493.2706C}.
The {\sc Montage} project is funded by the National Science Foundation under Grant Number ACI-1440620, and was previously funded by the National Aeronautics and Space Administration's Earth Science Technology Office, Computation Technologies Project, under Cooperative Agreement Number NCC5-626 between NASA and the California Institute of Technology.

This work makes use of the  ALMA data:  ADS/JAO.ALMA\#2012.1.00320.S. ALMA is a partnership of ESO (representing its member states), NSF (USA) and NINS (Japan), together with NRC (Canada), NSC and ASIAA (Taiwan), and KASI (Republic of Korea), in cooperation with the Republic of Chile. The Joint ALMA Observatory is operated by ESO, AUI/NRAO and NAOJ.

For the purpose of open access, the author has applied a CC BY public copyright licence (where permitted by UKRI, ‘Open Government Licence’ or ‘CC BY-ND public copyright licence’ may be stated instead) to any Author Accepted Manuscript version arising.

\section*{Data Availability}

\textit{JWST} data are available from the Barbara A. Mikulski Archive for Space Telescopes (MAST; \url{https://mast.stsci.edu}). 
The DOI of the data from this specific observing program is 10.17909/s1rn-1t84 \url{http://archive.stsci.edu/doi/resolve/resolve.html?doi=10.17909/s1rn-1t84}.
Reduced images will be available by 
request to the authors.



\bibliographystyle{mnras}
\bibliography{ngc6302_jwst} 

\appendix

\section{Electron density} \label{electron_density}


There are three good density diagnostics: \ArIII\, 8.99/21.83\,$\mu$m, \ArV\, 7.9/13.1,
and \ClIV\, 11.76/20.31 line ratios. From the integrated spectrum, these
give electron densities of 35000, 12800 and 12000\,cm$^{-3}$ respectively - very
similar to the values from optical diagnostics in the UVES
spectra of \citep{Pequignot.2011jr}: 12000, 16000 and 16000\,cm$^{-3}$ from \OII\, \ClIII\, and \ArIV\,
density diagnostic ratios respectively).

In the jet, the estimated electron densities are 6000, 10000, and 20000 cm$^{-3}$ from the \FeII\,17.94/6.72, 17.94/25.99, and 17.94/5.34 ratios, respectively  \citep{2024ApJ...968L..18T}. 

\section{Markov chain Monte Carlo fitting to silicate bands}
\label{sect-appen-crystalline}

\begin{figure}
    \includegraphics[width=8cm]{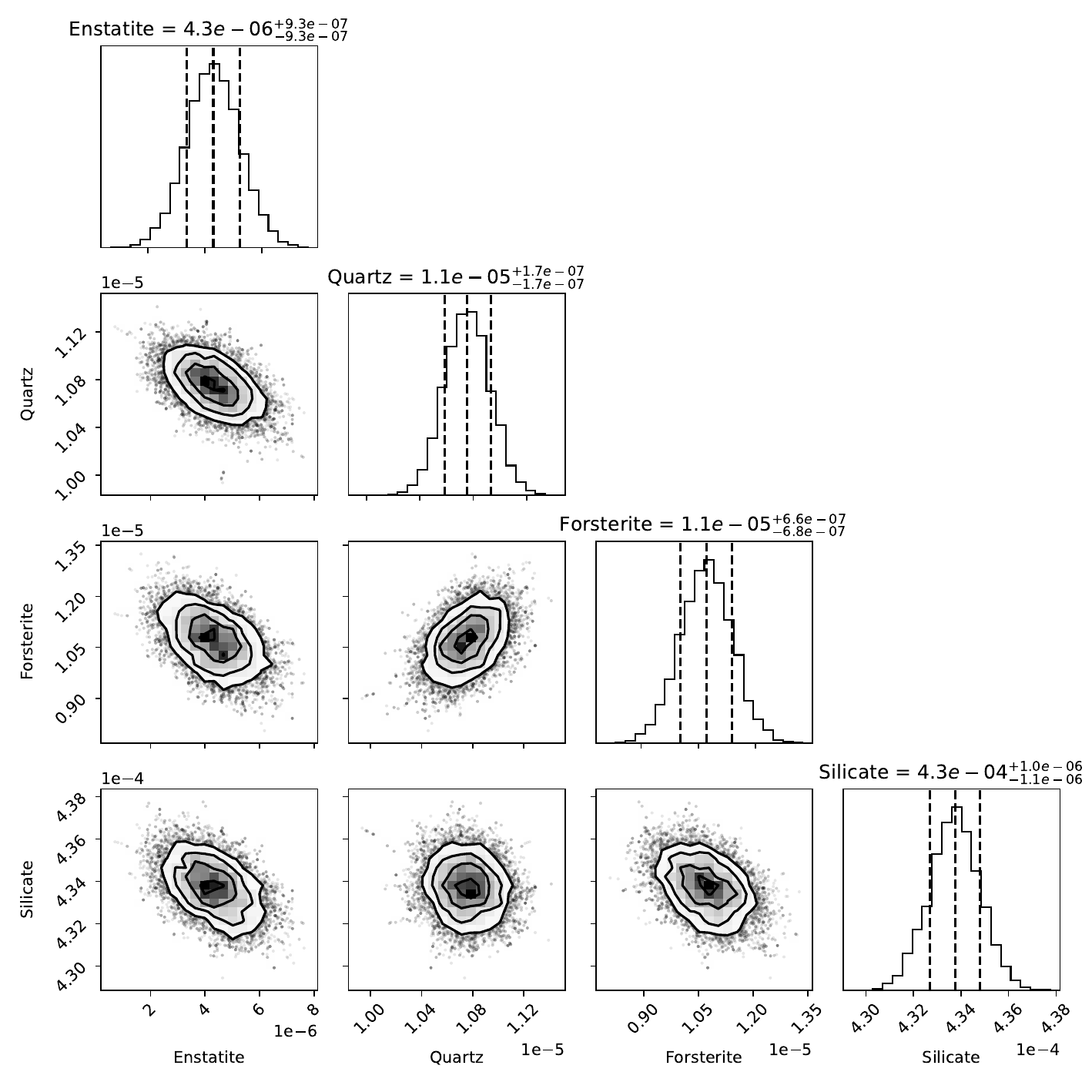}
\caption{\textsc{emcce} fit to the spectra with four different dust compositions. 
}
  \label{MCMC} 
\end{figure}

Fig.\,\ref{MCMC} demonstrates the Markov Chain Monte Carlo (MCMC) fitting used to estimate the fractions of amorphous and crystalline silicates.
The fitting is performed using \textsc{emcce}, developed by \citet{2013PASP..125..306F}, based on the MCMC method \citep{Goodman.2010}.
The figure shows the relative abundances, assuming amorphous silicates of 2\,$\mu$m size.

\bsp	
\label{lastpage}
\end{document}

%% file: linetable.tex
\begin{table}
\caption{Emission line fluxes ($10^{-12}$\,erg\,cm$^{-2}$\,s$^{-1}$) measured in the integrated MRS spectrum of NGC 6302. \label{table-lines}}
\begin{tabular}{lllll}
\hline
$\lambda_{obs}$ ($\mu$m) & $\lambda_{rest}$ ($\mu$m) & Flux & Species \\
\hline
4.9084 & 4.9088 & 0.020 $\pm$ 0.005 & He~{\sc ii} 28-13 \\
4.9236 & 4.9237 & 0.024 $\pm$ 0.003 & H~{\sc i} 23-7 \\
4.9532 & 4.9541 & 0.020 $\pm$ 0.004 & H$_2$ 1-1 S(9) \\
4.9699 & 4.9709 & 0.272 $\pm$ 0.046 & H~{\sc i} 22-7 + [Ni I] \\
5.0124 & 5.0126 & 0.032 $\pm$ 0.008 & He~{\sc ii} 27-13 \\
5.0252 & 5.0261 & 0.060 $\pm$ 0.008 & H~{\sc i} 21-7 \\
5.0524 & 5.0531 & 0.110 $\pm$ 0.008 & H$_2$ 0-0 S(8) \\
5.0908 & 5.0913 & 0.069 $\pm$ 0.009 & H~{\sc i} 20-7 \\
5.1257 & 5.1266 & 0.146 $\pm$ 0.015 & He~{\sc ii} 20-12 \\
5.1279 & 5.1287 & 0.793 $\pm$ 0.027 & H~{\sc i} 10-6 \\
5.1684 & 5.1693 & 0.058 $\pm$ 0.004 & H~{\sc i} 19-7\\
5.2276 & 5.2280 & 0.139 $\pm$ 0.016 & He~{\sc ii} 16-11 \\
5.2629 & 5.2637 & 0.083 $\pm$ 0.017 & H~{\sc i} 18-7 \\
5.2772 & 5.2777 & 0.018 $\pm$ 0.005 & He~{\sc ii} 25-13 \\
5.3396 & 5.3402 & 0.168 $\pm$ 0.014 & [Fe~{\sc ii}] \\
5.3725 & 5.3736 & 0.021 $\pm$ 0.003 & [Fe~{\sc ii}] \\
5.3791 & 5.3798 & 0.104 $\pm$ 0.010 & H~{\sc i} 17-7 \\
5.4484 & 5.4480 & 0.768 $\pm$ 0.031 & He~{\sc ii} 24-13 + [Fe~{\sc viii}] \\
5.4580 & 5.4583 & 0.023 $\pm$ 0.004 & He~{\sc ii} 19-12 \\
5.5036 & 5.4927 & Saturated & [Mg~{\sc VII}] \\
5.5244 & 5.5252 & 0.078 $\pm$ 0.027 & H~{\sc i} 16-7 \\
5.5732 & 5.5748 & 0.857 $\pm$ 0.065 & [K~{\sc vi}] \\
5.5799 & 5.5806 & 0.363 $\pm$ 0.067 & He~{\sc ii} 13-10 \\
5.5821 & 5.5828 & 0.763 $\pm$ 0.062 & He~{\sc ii} 11-9 \\
5.6092 & 5.6098 & Saturated & [Mg~{\sc v}] \\
5.6575 & 5.6582 & 0.166 $\pm$ 0.009 & He~{\sc ii} 23-13 \\
5.7080 & 5.7091 & 0.032 $\pm$ 0.004 & He~{\sc ii} 30-14 \\
5.7106 & 5.7115 & 0.148 $\pm$ 0.016 & H~{\sc i} 15-7 \\
5.7852 & 5.7870 & 0.009 $\pm$ 0.002 & [Cr~{\sc ix}] \\
5.8100 & 5.8109 & 0.005 $\pm$ 0.001 & H$_2$ 1-1 S(7) \\
5.8220 & 5.8227 & 0.005 $\pm$ 0.001 & He~{\sc ii} 29-14 \\
5.9046 & 5.9058 & 0.181 $\pm$ 0.041 & He~{\sc ii} 18-12 \\
5.9073 & 5.9082 & 1.019 $\pm$ 0.041 & H~{\sc i} 9-6 \\
5.9156 & 5.9165 & 0.019 $\pm$ 0.005 & He~{\sc ii} 22-13 \\
5.9533 & 5.9533 & 0.027 $\pm$ 0.003 & He~{\sc ii} 28-14 \\
5.9559 & 5.9568 & 0.163 $\pm$ 0.021 & H~{\sc i} 14-7 \\
5.9636 & 5.9646 & 0.181 $\pm$ 0.016 & He~{\sc ii} 15-11 \\
5.9812 & 5.9820 & 1.732 $\pm$ 0.020 & [K~{\sc iv}] \\
6.0540 & 6.0549 & 0.010 $\pm$ 0.002 & H~{\sc i} 42-8 \\
6.0652 & 6.0662 & 0.005 $\pm$ 0.001 & H~{\sc i} 41-8 \\
6.0780 & 6.0784 & 0.005 $\pm$ 0.001 & H~{\sc i} 40-8 \\
6.0908 & 6.0916 & 0.007 $\pm$ 0.002 & H~{\sc i} 39-8 \\
6.1076 & 6.1086 & 0.363 $\pm$ 0.007 & H$_2$ 0-0 S(6) \\
6.1500 & 6.154  & 0.206 $\pm$ 0.017 & [Ca~{\sc vii}] \\
6.2233 & 6.2243 & 0.016 $\pm$ 0.002 & H~{\sc i} 32-8 \\
6.2420 & 6.2431 & 0.044 $\pm$ 0.010 & He~{\sc ii} 21-13 \\
6.2506 & 6.2516 & 0.018 $\pm$ 0.002 & H~{\sc i} 31-8 \\
6.2882 & 6.2894 & 0.042 $\pm$ 0.005 & He~{\sc ii} 26-14 \\
6.2909 & 6.2919 & 0.241 $\pm$ 0.015 & H~{\sc i} 13-7 \\
6.3796 & 6.3816 & 0.097 $\pm$ 0.017 & [Zn~{\sc vii}] \\
6.3956 & 6.3969 & 0.017 $\pm$ 0.006 & H~{\sc i} 27-8 \\
6.4444 & 6.4455 & 0.011 $\pm$ 0.003 & H~{\sc i} 26-8 \\
6.4932 & 6.4923 & 3.610 $\pm$ 0.389 & [Si~{\sc vii}] + [Ni~{\sc vii}]?\\
6.5380 & 6.5394 & 0.095 $\pm$ 0.012 & He~{\sc ii} 17-12 \\
6.5636 & 6.5647 & 0.021 $\pm$ 0.005 & H~{\sc i} 24-8 \\
6.6347 & 6.6359 & 0.036 $\pm$ 0.004 & [Ni~{\sc ii}] \\
6.6373 & 6.6384 & 0.019 $\pm$ 0.002 & H~{\sc i} 23-8 \\
6.6664 & 6.6677 & 0.038 $\pm$ 0.004 & He~{\sc ii} 20-13 \\
\hline
\end{tabular}
Note: \MgVII\ at 5.50\,$\mu$m is saturated at the line center, but detected at side lobes.
\end{table}
\addtocounter{table}{-1}

\begin{table}
\caption{-- continued}
\begin{tabular}{lllll}
\hline
$\lambda_{obs}$ ($\mu$m) & $\lambda_{rest}$ ($\mu$m) & Flux & Species \\
\hline
6.7056 & 6.7067 & 3.107 $\pm$ 0.077 & [Cl~{\sc v}] \\
6.7212 & 6.7213 & 0.014 $\pm$ 0.002 & [Fe~{\sc ii}] \\
6.7232 & 6.7245 & 0.067 $\pm$ 0.007 & H~{\sc i} 22-8 + [Zn~{\sc vi}] \\
6.7682 & 6.7692 & 0.046 $\pm$ 0.014 & He~{\sc ii} 24-14 \\
6.7710 & 6.7720 & 0.326 $\pm$ 0.014 & H~{\sc i} 12-7 \\
6.8248 & 6.8259 & 0.031 $\pm$ 0.007 & H~{\sc i} 21-8 \\
6.9080 & 6.9095 & 2.108 $\pm$ 0.052 & H$_2$ 0-0 S(5) \\
6.9472 & 6.9480 & 1.956 $\pm$ 0.070 & He~{\sc ii} 9-8 \\
6.9844 & 6.9853 & Saturated & [Ar~{\sc ii}] \\
7.0920 & 7.0927 & 0.070 $\pm$ 0.005 & H~{\sc i} 19-8 \\
7.2040 & 7.2048 & 0.234 $\pm$ 0.006 & He~{\sc ii} 14-11 \\
7.2392 & 7.2400 & 0.045 $\pm$ 0.002 & He~{\sc ii} 19-13 \\
7.2704 & 7.2717 & 0.027 $\pm$ 0.002 & H~{\sc i} 18-8 \\
7.3168 & 7.3178 & 9.117 $\pm$ 0.410 & [Na~{\sc iii}] \\
7.4554 & 7.4568 & 0.909 $\pm$ 0.10 & He~{\sc ii} 12-10 \\
7.4588 & 7.4599 & 6.050 $\pm$ 0.405 & H~{\sc i} 6-5 \\
7.4696 & 7.4738 & 0.844 $\pm$ 0.025 & [Mn~{\sc vii}] \\
7.4941 & 7.4951 & 0.091 $\pm$ 0.010 & H~{\sc i} 17-8 \\
7.4981 & 7.4994 & 0.288 $\pm$ 0.030 & He~{\sc ii} 16-12 \\
7.5014 & 7.5025 & 1.712 $\pm$ 0.105 & H~{\sc i} 8-6 \\
7.5070 & 7.5081 & 0.436 $\pm$ 0.058 & H~{\sc i} 11-7 \\
7.6550 & 7.6524 & Saturated         & [Ne~{\sc vi}] \\
7.7797 & 7.7804 & 0.067 $\pm$ 0.013 & H~{\sc i} 16-8 \\
7.8123 & 7.8141 & 0.399 $\pm$ 0.048 & [Fe~{\sc vii}] \\
7.9007 & 7.9016 & 35.280 $\pm$ 0.395 & [Ar~{\sc v}] \\
8.0098 & 8.0103 & 0.024 $\pm$ 0.005 & He~{\sc ii} 25-15 \\
8.0241 & 8.0250 & 1.108 $\pm$ 0.012 & H$_2$ 0-0 S(4) \\
8.0372 & 8.0385 & 0.030 $\pm$ 0.009 & He~{\sc ii} 21-14 \\
8.0476 & 8.0490 & 0.072 $\pm$ 0.009 & He~{\sc ii} 18-13\\
8.1508 & 8.1549 & 0.016 $\pm$ 0.002 & He~{\sc ii} 30-16 \\
8.1540 & 8.1549 & 0.101 $\pm$ 0.003 & H~{\sc i} 15-8 \\
8.1711 & 8.1724 & 0.010 $\pm$ 0.002 & H~{\sc i} 29-9 \\
8.2352 & 8.2362 & 0.018 $\pm$ 0.002 & H~{\sc i} 28-9 \\
8.3076 & 8.3084 & 0.010 $\pm$ 0.003 & H~{\sc i} 27-9 \\
8.4118 & 8.4128 & 0.018 $\pm$ 0.006 & He~{\sc ii} 24-15 \\
8.4831 & 8.4849 & 0.012 $\pm$ 0.002 & H~{\sc i} 25-9 \\
8.6079 & 8.6106 & 20.001 $\pm$ 1.682 & [Na~{\sc vi}] \\
8.6595 & 9.6610 & 0.031 $\pm$ 0.004 & He~{\sc ii} 28-16 \\
8.6632 & 8.6645 & 0.085 $\pm$ 0.009 & H~{\sc i} 14-8 \\
8.7184 & 8.7206 & 0.029 $\pm$ 0.006 & H~{\sc i} 23-9 \\
8.7557 & 8.7565 & 0.082 $\pm$ 0.009 & He~{\sc ii} 20-14 \\
8.7591 & 8.7601 & 0.384 $\pm$ 0.062 & H~{\sc i} 10-7 \\
8.8289 & 8.8299 & 0.853 $\pm$ 0.111 & [K~{\sc vi}] \\
8.8684 & 8.8697 & 0.015 $\pm$ 0.002 & H~{\sc i} 22-9 \\
8.9190 & 8.9209 & 0.017 $\pm$ 0.003 & He~{\sc ii} 23-15 \\
8.9892 & 8.9914 & Saturated        & [Ar~{\sc iii}] + [Mg~{\sc vii}] \\
9.0399 & 9.0410 & 3.541 $\pm$ 0.160 & [Na~{\sc iv}] + H~{\sc i} 21-9\\
9.1126 & 9.1158 & 0.085 $\pm$ 0.004 & [Al~{\sc vi}] + He~{\sc ii} 15-12 \\
9.1321 & 9.1353 & 0.022 $\pm$ 0.003 & [Fe~{\sc ii}] \\
9.2596 & 9.2605 & 0.017 $\pm$ 0.002 & H~{\sc i} 20-9 \\
9.2725 & 9.2736 & 0.042 $\pm$ 0.004 & He~{\sc ii} 17-13 \\
9.3868 & 9.3882 & 0.017 $\pm$ 0.002 & He~{\sc ii} 26-16 \\
9.3907 & 9.3920 & 0.088 $\pm$ 0.011 & H~{\sc i} 13-8 \\
9.5234 & 9.5261 & 0.697 $\pm$ 0.075 & [Fe~{\sc vii}] \\
9.6638 & 9.6649 & 2.065 $\pm$ 0.050 & H$_2$ 0-0 S(3) \\
9.7055 & 9.7068 & 0.160 $\pm$ 0.012 & He~{\sc ii} 13-11 \\
9.7122 & 9.7135 & 0.599 $\pm$ 0.011 & He~{\sc ii} 10-9 \\
9.7691 & 9.7707 & 0.028 $\pm$ 0.001 & He~{\sc ii} 19-14 \\
9.8455 & 9.8470 & 0.007 $\pm$ 0.001 & H~{\sc i} 18-9 \\
9.8784 & 9.8796 & 0.008 $\pm$ 0.001 & He~{\sc ii} 25-16 \\
9.8784 & 9.88   & \*                 & [Ni~{\sc viii}] + [Zn~{\sc vii}] \\
\hline
\end{tabular}
Note: 9.88\,$\mu$m  [Ni~{\sc viii}] + [Zn~{\sc vii}] line is too weak to measure the intensity
\addtocounter{table}{-1}
\end{table}

\begin{table}
\caption{-- continued}
\begin{tabular}{lllll}
\hline
$\lambda_{obs}$ ($\mu$m) & $\lambda_{rest}$ ($\mu$m) & Flux & Species \\
\hline
10.2597 & 10.2613 & 0.023 $\pm$ 0.003 & H~{\sc i} 17-9 \\
10.5085 & 10.5105 & Saturated        & [S~{\sc iv}] \\
10.6425 & 10.6440 & 0.036 $\pm$ 0.002 & [Cr~{\sc vi}] \\
10.8019 & 10.8036 & 0.031 $\pm$ 0.004 & H~{\sc i} 16-9 \\
10.8443 & 10.8459 & 0.033 $\pm$ 0.004 & [Mn~{\sc vi}] \\
10.8524 & 10.8543 & 0.006 $\pm$ 0.001 & H~{\sc i} 25-10 \\
10.8804 & 10.88   & 0.019 $\pm$ 0.002 & [Co~{\sc vii}] \\
11.3019 & 11.3041 & 0.087 $\pm$ 0.012 & He~{\sc ii} 18-14 \\
11.3070 & 11.3087 & 0.548 $\pm$ 0.012 & H~{\sc i}~9-7 + [Ni~{\sc i}] ? \\
11.3315 & 11.3334 & 0.283 $\pm$ 0.020 & [Cl~{\sc i}] \\
11.4823 & 11.4824 & 0.042 $\pm$ 0.005 & [Ca~{\sc v}] \\
11.4896 & 11.4920 & 0.021 $\pm$ 0.003 & H~{\sc i} 22-10 \\
11.5381 & 11.5395 & 0.037 $\pm$ 0.004 & H~{\sc i} 15-9 \\
11.7612 & 11.7590 & 1.385 $\pm$ 0.026 & [Cl~{\sc iv}] \\
12.1562 & 12.1568 & 0.017 $\pm$ 0.004 & H~{\sc i} 20-10 \\
12.2470 & 12.2485 & 0.012 $\pm$ 0.004 & He~{\sc ii} 25-17 \\
12.2762 & 12.2786 & 0.939 $\pm$ 0.021 & H$_2$ 0-0 S(2) \\
12.3038 & 12.3107 & 0.039 $\pm$ 0.004 & [Fe~{\sc vi}] \\
12.3651 & 12.3668 & 0.345 $\pm$ 0.024 & He~{\sc ii} 14-12 \\
12.3703 & 12.3719 & 1.962 $\pm$ 0.024 & H~{\sc i} 7-6 \\
12.3856 & 12.3872 & 0.235 $\pm$ 0.016 & H~{\sc i} 11-8 \\
12.5863 & 12.5871 & 0.091 $\pm$ 0.008 & H~{\sc i} 14-9 \\
12.6088 & 12.6110 & 0.016 $\pm$ 0.003 & H~{\sc i} 19-10 \\
12.8125 & 12.8135 & Saturated         & [Ne~{\sc ii}] \\
13.1012 & 13.1022 & Saturated        & [Ar~{\sc v}] \\
13.1263 & 13.1283 & 0.543 $\pm$ 0.015 & He~{\sc ii} 11-10 \\
13.1862 & 13.1880 & 0.017 $\pm$ 0.004 & H~{\sc i} 18-10 \\
13.2137 & 13.2155 & 0.006 $\pm$ 0.001 & He~{\sc ii} 24-17 \\
13.3813 & 13.40   & 0.153 $\pm$ 0.020 & [F~{\sc v}]? \\
13.5188 & 13.5210 & 5.741 $\pm$ 0.118 & [Mg~{\sc v}] \\
13.6062 & 13.6080 & 0.028 $\pm$ 0.006 & He~{\sc ii} 19-15 \\
13.8737 & 13.8776 & 0.037 $\pm$ 0.002 & He~{\sc ii} 17-14 \\
13.9012 & 13.9044 & 0.012 $\pm$ 0.002 & He~{\sc ii} 21-16 \\
13.9388 & 13.9418 & 0.026 $\pm$ 0.007 & H~{\sc i} 17-10 \\
14.1750 & 14.1773 & 0.023 $\pm$ 0.003 & He~{\sc ii} 26-18 \\
14.1804 & 14.1831 & 0.112 $\pm$ 0.013 & H~{\sc i} 13-9 \\
14.3213 & 14.3217 & Saturated         & [Ne~{\sc v}] \\
14.3638 & 14.3678 & 1.631 $\pm$ 0.158 & [Cl~{\sc ii}] \\
14.3938 & 14.3964 & Saturated        & [Na~{\sc vi}] \\
14.7078 & 14.7098 & 0.023 $\pm$ 0.006 & H~{\sc i} 22-11 \\
14.7788 & 14.7710 & 0.119 $\pm$ 0.014 & [Fe~{\sc vi}] \\
14.9587 & 14.9623 & 0.022 $\pm$ 0.006 & H~{\sc i} 16-10 \\
15.3912 & 15.3960 & 0.105 $\pm$ 0.031 & [K~{\sc iv}] \\
15.4712 & 15.4713 & 0.097 $\pm$ 0.018 & He~{\sc ii} 15-13 \\
16.2011 & 16.2025 & 0.085 $\pm$ 0.018 & He~{\sc ii} 20-16 \\
16.2073 & 16.2091 & 0.325 $\pm$ 0.012 & H~{\sc i} 10-8 \\
16.4088 & 16.4117 & 0.033 $\pm$ 0.007 & H~{\sc i} 15-10 \\
16.8787 & 16.8806 & 0.145 $\pm$ 0.020 & H~{\sc i} 12-9 \\
17.0312 & 17.0348 & 1.201 $\pm$ 0.023 & H$_2$ 0-0 S(1) \\
17.2588 & 17.2609 & 0.272 $\pm$ 0.015 & He~{\sc ii} 12-11 \\
17.8838 & 17.8846 & 0.070 $\pm$ 0.020 & [P~{\sc iii}] \\
17.9338 & 17.9360 & 0.241 $\pm$ 0.016 & [Fe~{\sc ii}] \\
17.9825 & 17.9860 & 0.332 $\pm$ 0.092 & [Ca~{\sc vi}] \\
18.7110 & 18.7130 & 49.025 $\pm$ 0.366 & [S~{\sc iii}] \\
19.0590 & 19.0619 & 0.706 $\pm$ 0.037 & H~{\sc i} 8-7 \\
19.5510 & 19.5580 & 0.144 $\pm$ 0.031 & [Fe~{\sc vi}] \\
20.3070 & 20.3107 & 0.597 $\pm$ 0.078 & [Cl~{\sc iv}] \\
21.2570 & 21.29   & 0.352 $\pm$ 0.111 & [Na~{\sc iv}] \\
21.8270 & 21.8291 & 1.956 $\pm$ 0.074 & [Ar~{\sc iii}] \\
24.3170 & 24.3175 & 273.000 $\pm$ 3.424 & [Ne~{\sc v}] \\
25.2430 & 25.2490 & 0.327 $\pm$ 0.091 & [S~{\sc i}] \\
25.8850 & 25.8903 & 222.100 $\pm$ 3.697 & [O~{\sc iv}] \\
25.8970 & 25.8999 & 31.830 $\pm$ 3.697 & He~{\sc ii} 23-19 \\
25.9810 & 25.9884 & 0.896 $\pm$ 0.078 & [Fe~{\sc ii}] \\
\hline
\end{tabular}
\end{table}

%% file: ips2.tex
\begin{table}
\caption{Ionization potentials of species observed in the integrated
  MRS spectrum of NGC\,6302. The listed ionisation potentials are those of the next lowest ion, i.e.\ the photon energy needed to form the detected ion (collisionally-excited lines) or the recombining ion (permitted lines). \label{table-ionization-potential}}
\begin{tabular}{llllll}
\hline
Species & IP [eV]\quad & Species & IP [eV] & Species & IP [eV]  \\
\hline
H~{\sc i} & 13.6  & {}[P~{\sc iii}] & 19.8 &  {}[Cr~{\sc vi}] & 69.5\\
He~{\sc ii} & 54.4 &  {}[S~{\sc i}] & 0.0 &  {}[Cr~{\sc ix}] & 184.8 \\
{}[O~{\sc iv}] & 54.9 &  {}[S~{\sc iii}] & 23.3 & {}[Mn~{\sc vi}] & 72.4 \\
{}[F~{\sc v}] & 87.2 & {}[S~{\sc iv}] & 34.8 & {}[Mn~{\sc vii}] & 95.6 \\
{}[Ne~{\sc ii}] &  21.6&  {}[Cl~{\sc i}] &  0.0 & {}[Fe~{\sc ii}] &  7.9\\
{}[Ne~{\sc iii}] & 41.0 & {}[Cl~{\sc ii}] & 13.0 & {}[Fe~{\sc vi}] & 75.0 \\
{}[Ne~{\sc v}] & 97.1 &  {}[Cl~{\sc iv}] & 39.6 & {}[Fe~{\sc vii}] & 99.1 \\
{}[Ne~{\sc vi}] & 126.2 &  {}[Cl~{\sc v}] & 53.5 &  {}[Fe~{\sc viii}] & 125.0\\
{}[Na~{\sc iii}] & 47.3 &  {}[Ar~{\sc ii}] & 15.8 & {}[Co~{\sc vii}] & 102.0\\
{}[Na~{\sc iv}] & 71.6 &  {}[Ar~{\sc iii}] & 27.6 & {}[Ni~{\sc i}] &  0.0\\
{}[Na~{\sc vi}] & 138.4 &  {}[Ar~{\sc v}] & 59.8 & {}[Ni~{\sc ii}] &  7.6\\
{}[Mg~{\sc v}] & 109.3 & {}[K~{\sc v}] & 45.8 & {}[Ni~{\sc viii}] & 132.7\\
{}[Mg~{\sc vii}] & 186.8  & {}[K~{\sc vi}] & 82.7 & {}[Zn~{\sc vi}] & 82.6\\
{}[Al~{\sc vi}] & 153.8 & {}[Ca~{\sc v}] &  67.3 &  {}[Zn~{\sc vii}] & 108.0 \\
{}[Si~{\sc vii}] & 205.3 & {}[Ca~{\sc vii}] & 108.8  & & \\
\hline
\end{tabular}

\end{table}

%% file: ngc6302_overview_v1.bbl
\begin{thebibliography}{}
\makeatletter
\relax
\def\mn@urlcharsother{\let\do\@makeother \do\$\do\&\do\#\do\^\do\_\do\%\do\~}
\def\mn@doi{\begingroup\mn@urlcharsother \@ifnextchar [ {\mn@doi@}
  {\mn@doi@[]}}
\def\mn@doi@[#1]#2{\def\@tempa{#1}\ifx\@tempa\@empty \href
  {http://dx.doi.org/#2} {doi:#2}\else \href {http://dx.doi.org/#2} {#1}\fi
  \endgroup}
\def\mn@eprint#1#2{\mn@eprint@#1:#2::\@nil}
\def\mn@eprint@arXiv#1{\href {http://arxiv.org/abs/#1} {{\tt arXiv:#1}}}
\def\mn@eprint@dblp#1{\href {http://dblp.uni-trier.de/rec/bibtex/#1.xml}
  {dblp:#1}}
\def\mn@eprint@#1:#2:#3:#4\@nil{\def\@tempa {#1}\def\@tempb {#2}\def\@tempc
  {#3}\ifx \@tempc \@empty \let \@tempc \@tempb \let \@tempb \@tempa \fi \ifx
  \@tempb \@empty \def\@tempb {arXiv}\fi \@ifundefined
  {mn@eprint@\@tempb}{\@tempb:\@tempc}{\expandafter \expandafter \csname
  mn@eprint@\@tempb\endcsname \expandafter{\@tempc}}}

\bibitem[\protect\citeauthoryear{Abel, Ferland  \& O'Dell}{Abel
  et~al.}{2019}]{Abel.2019}
Abel N.~P.,  Ferland G.~J.,   O'Dell C.~R.,  2019, \mn@doi [The Astrophysical
  Journal] {10.3847/1538-4357/ab2a6e}, 881, 130

\bibitem[\protect\citeauthoryear{{Allamandola}, {Tielens}  \&
  {Barker}}{{Allamandola} et~al.}{1985}]{1985ApJ...290L..25A}
{Allamandola} L.~J.,  {Tielens} A.~G.~G.~M.,   {Barker} J.~R.,  1985, \mn@doi
  [\apjl] {10.1086/184435}, \href
  {https://ui.adsabs.harvard.edu/abs/1985ApJ...290L..25A} {290, L25}

\bibitem[\protect\citeauthoryear{Allen, Groves, Dopita, Sutherland  \&
  Kewley}{Allen et~al.}{2008}]{Allen.2008}
Allen M.~G.,  Groves B.~A.,  Dopita M.~A.,  Sutherland R.~S.,   Kewley L.~J.,
  2008, \mn@doi [ApJS] {10.1086/589652}, 178, 20

\bibitem[\protect\citeauthoryear{Argyriou et~al.,}{Argyriou
  et~al.}{2023}]{Argyriou.2023dug}
Argyriou I.,  et~al., 2023, \mn@doi [\aap] {10.1051/0004-6361/202346489}, 675,
  A111

\bibitem[\protect\citeauthoryear{Ashley \& Hyland}{Ashley \&
  Hyland}{1988}]{Ashley.1988aml}
Ashley M. C.~B.,  Hyland A.~R.,  1988, \mn@doi [ApJ] {10.1086/166578}, 331, 532

\bibitem[\protect\citeauthoryear{Balick}{Balick}{1987}]{Balick.1987}
Balick B.,  1987, \mn@doi [\aj] {10.1086/114504}, 94, 671

\bibitem[\protect\citeauthoryear{Balick \& Frank}{Balick \&
  Frank}{2002}]{Balick.2002}
Balick B.,  Frank A.,  2002, \mn@doi [\araa]
  {10.1146/annurev.astro.40.060401.093849}, 40, 439

\bibitem[\protect\citeauthoryear{Balick, Borchert, Kastner, Frank, Blackman,
  Nordhaus  \& Baez}{Balick et~al.}{2023}]{Balick.2023iff}
Balick B.,  Borchert L.,  Kastner J.~H.,  Frank A.,  Blackman E.,  Nordhaus J.,
    Baez P.~M.,  2023, \mn@doi [ApJ] {10.3847/1538-4357/acf5ea}, 957, 54

\bibitem[\protect\citeauthoryear{{Barlow}, {Crawford}, {Diego}, {Dryburgh},
  {Fish}, {Howarth}, {Spyromilio}  \& {Walker}}{{Barlow}
  et~al.}{1995}]{1995MNRAS.272..333B}
{Barlow} M.~J.,  {Crawford} I.~A.,  {Diego} F.,  {Dryburgh} M.,  {Fish} A.~C.,
  {Howarth} I.~D.,  {Spyromilio} J.,   {Walker} D.~D.,  1995, \mn@doi [\mnras]
  {10.1093/mnras/272.2.333}, \href
  {https://ui.adsabs.harvard.edu/abs/1995MNRAS.272..333B} {272, 333}

\bibitem[\protect\citeauthoryear{Beintema \& Pottasch}{Beintema \&
  Pottasch}{1999}]{Beintema.1999}
Beintema D.~A.,  Pottasch S.~R.,  1999, \aap, 347, 942

\bibitem[\protect\citeauthoryear{Bern\'e et~al.,}{Bern\'e
  et~al.}{2023}]{Berne.20230dw}
Bern\'e O.,  et~al., 2023, \mn@doi [Nature] {10.1038/s41586-023-06307-x},
  pp~1--4

\bibitem[\protect\citeauthoryear{Bradley}{Bradley}{2023}]{larry_bradley_2023_7946442}
Bradley L.,  2023, astropy/photutils: 1.8.0, \mn@doi{10.5281/zenodo.7946442}

\bibitem[\protect\citeauthoryear{Bradley et~al.,}{Bradley
  et~al.}{2022}]{Bradley2022}
Bradley L.,  et~al., 2022, astropy/photutils: 1.5.0,
  \mn@doi{10.5281/zenodo.6825092}, \url
  {https://doi.org/10.5281/zenodo.6825092}

\bibitem[\protect\citeauthoryear{{Bushouse} et~al.,}{{Bushouse}
  et~al.}{2023}]{2023zndo...7692609B}
{Bushouse} H.,  et~al., 2023, {JWST Calibration Pipeline}, Zenodo,
  \mn@doi{10.5281/zenodo.7692609}

\bibitem[\protect\citeauthoryear{{CASA Team} et~al.,}{{CASA Team}
  et~al.}{2022}]{2022PASP..134k4501C}
{CASA Team} et~al., 2022, \mn@doi [\pasp] {10.1088/1538-3873/ac9642}, \href
  {https://ui.adsabs.harvard.edu/abs/2022PASP..134k4501C} {134, 114501}

\bibitem[\protect\citeauthoryear{Casassus, Roche  \& Barlow}{Casassus
  et~al.}{2000}]{Casassus.2000}
Casassus S.,  Roche P.~F.,   Barlow M.~J.,  2000, \mn@doi [\mnras]
  {10.1046/j.1365-8711.2000.03208.x}, 314, 657

\bibitem[\protect\citeauthoryear{Castor, Weaver  \& McCray}{Castor
  et~al.}{1975}]{Castor.1975}
Castor J.,  Weaver R.,   McCray R.,  1975, \mn@doi [\apjl] {10.1086/181908},
  200, L107

\bibitem[\protect\citeauthoryear{{Chatzikos} et~al.,}{{Chatzikos}
  et~al.}{2023}]{Chatzikos.2023}
{Chatzikos} M.,  et~al., 2023, \mn@doi [\rmxaa]
  {10.22201/ia.01851101p.2023.59.02.12}, \href
  {https://ui.adsabs.harvard.edu/abs/2023RMxAA..59..327C} {59, 327}

\bibitem[\protect\citeauthoryear{{Chawner} et~al.,}{{Chawner}
  et~al.}{2020}]{2020MNRAS.493.2706C}
{Chawner} H.,  et~al., 2020, \mn@doi [\mnras] {10.1093/mnras/staa221}, \href
  {https://ui.adsabs.harvard.edu/abs/2020MNRAS.493.2706C} {493, 2706}

\bibitem[\protect\citeauthoryear{Cherchneff}{Cherchneff}{2006}]{Cherchneff.2006}
Cherchneff I.,  2006, \mn@doi [Astronomy and Astrophysics]
  {10.1051/0004-6361:20064827}, 456, 1001

\bibitem[\protect\citeauthoryear{Chiang \& Goldreich}{Chiang \&
  Goldreich}{1997}]{Chiang.1997}
Chiang E.~I.,  Goldreich P.,  1997, \mn@doi [\apj] {10.1086/304869}, 490, 368

\bibitem[\protect\citeauthoryear{Chihara, Koike  \& Tsuchiyama}{Chihara
  et~al.}{2007}]{Chihara.2007}
Chihara H.,  Koike C.,   Tsuchiyama A.,  2007, \mn@doi [\aap]
  {10.1051/0004-6361:20066009}, 464, 229

\bibitem[\protect\citeauthoryear{{Clark} et~al.,}{{Clark}
  et~al.}{2025}]{2025arXiv250406247C}
{Clark} I.~Y.,  et~al., 2025, \mn@doi [arXiv e-prints]
  {10.48550/arXiv.2504.06247}, \href
  {https://ui.adsabs.harvard.edu/abs/2025arXiv250406247C} {p. arXiv:2504.06247}

\bibitem[\protect\citeauthoryear{{Cohen} \& {Barlow}}{{Cohen} \&
  {Barlow}}{2005}]{2005MNRAS.362.1199C}
{Cohen} M.,  {Barlow} M.~J.,  2005, \mn@doi [\mnras]
  {10.1111/j.1365-2966.2005.09366.x}, \href
  {https://ui.adsabs.harvard.edu/abs/2005MNRAS.362.1199C} {362, 1199}

\bibitem[\protect\citeauthoryear{{Dere}, {Del Zanna}, {Young}  \&
  {Landi}}{{Dere} et~al.}{2023}]{2023ApJS..268...52D}
{Dere} K.~P.,  {Del Zanna} G.,  {Young} P.~R.,   {Landi} E.,  2023, \mn@doi
  [\apjs] {10.3847/1538-4365/acec79}, \href
  {https://ui.adsabs.harvard.edu/abs/2023ApJS..268...52D} {268, 52}

\bibitem[\protect\citeauthoryear{{Dinh-V-Trung}, {Bujarrabal},
  {Castro-Carrizo}, {Lim}  \& {Kwok}}{{Dinh-V-Trung}
  et~al.}{2008}]{2008ApJ...673..934D}
{Dinh-V-Trung} {Bujarrabal} V.,  {Castro-Carrizo} A.,  {Lim} J.,   {Kwok} S.,
  2008, \mn@doi [\apj] {10.1086/524373}, \href
  {https://ui.adsabs.harvard.edu/abs/2008ApJ...673..934D} {673, 934}

\bibitem[\protect\citeauthoryear{{Draine} \& {Lee}}{{Draine} \&
  {Lee}}{1984}]{1984ApJ...285...89D}
{Draine} B.~T.,  {Lee} H.~M.,  1984, \mn@doi [\apj] {10.1086/162480}, \href
  {https://ui.adsabs.harvard.edu/abs/1984ApJ...285...89D} {285, 89}

\bibitem[\protect\citeauthoryear{Draine, Li, Hensley, Hunt, Sandstrom  \&
  Smith}{Draine et~al.}{2021}]{Draine.2021d6}
Draine B.~T.,  Li A.,  Hensley B.~S.,  Hunt L.~K.,  Sandstrom K.,   Smith J.
  D.~T.,  2021, \mn@doi [\apj] {10.3847/1538-4357/abff51}, 917, 3

\bibitem[\protect\citeauthoryear{{Dwek}}{{Dwek}}{1998}]{1998ApJ...501..643D}
{Dwek} E.,  1998, \mn@doi [\apj] {10.1086/305829}, \href
  {https://ui.adsabs.harvard.edu/abs/1998ApJ...501..643D} {501, 643}

\bibitem[\protect\citeauthoryear{{Evans}}{{Evans}}{1959}]{DEvans59}
{Evans} D.~S.,  1959, \mn@doi [\mnras] {10.1093/mnras/119.2.150}, \href
  {https://ui.adsabs.harvard.edu/abs/1959MNRAS.119..150E} {119, 150}

\bibitem[\protect\citeauthoryear{Fabian, Henning, J{\"a}ger, Mutschke,
  Dorschner  \& Wehrhan}{Fabian et~al.}{2001}]{Fabian.2001}
Fabian D.,  Henning T.,  J{\"a}ger C.,  Mutschke H.,  Dorschner J.,   Wehrhan
  O.,  2001, \mn@doi [A\&A] {10.1051/0004-6361:20011196}, 378, 228

\bibitem[\protect\citeauthoryear{{Foreman-Mackey}, {Hogg}, {Lang}  \&
  {Goodman}}{{Foreman-Mackey} et~al.}{2013}]{2013PASP..125..306F}
{Foreman-Mackey} D.,  {Hogg} D.~W.,  {Lang} D.,   {Goodman} J.,  2013, \mn@doi
  [\pasp] {10.1086/670067}, \href
  {https://ui.adsabs.harvard.edu/abs/2013PASP..125..306F} {125, 306}

\bibitem[\protect\citeauthoryear{Freeman et~al.,}{Freeman
  et~al.}{2014}]{Freeman.2014}
Freeman M.,  et~al., 2014, \mn@doi [\apj] {10.1088/0004-637x/794/2/99}, 794, 99

\bibitem[\protect\citeauthoryear{{Gaia Collaboration} et~al.,}{{Gaia
  Collaboration} et~al.}{2023}]{2023A&A...674A...1G}
{Gaia Collaboration} et~al., 2023, \mn@doi [\aap]
  {10.1051/0004-6361/202243940}, \href
  {https://ui.adsabs.harvard.edu/abs/2023A&A...674A...1G} {674, A1}

\bibitem[\protect\citeauthoryear{{Gail}}{{Gail}}{1998}]{1998A&A...332.1099G}
{Gail} H.~P.,  1998, \aap, \href
  {https://ui.adsabs.harvard.edu/abs/1998A&A...332.1099G} {332, 1099}

\bibitem[\protect\citeauthoryear{{Gardner}, {Mather}, {Abbott}  \& et
  al.}{{Gardner} et~al.}{2023}]{Gardner2023}
{Gardner} J.~P.,  {Mather} J.~C.,  {Abbott} R.,   et al. 2023, \mn@doi [\pasp]
  {10.1088/1538-3873/acd1b5}, \href
  {https://ui.adsabs.harvard.edu/abs/2023arXiv230404869G} {135, 1048}

\bibitem[\protect\citeauthoryear{Gielen, Winckel, Matsuura, Min, Deroo, Waters
  \& Dominik}{Gielen et~al.}{2009}]{Gielen:2009p25425}
Gielen C.,  Winckel H.~V.,  Matsuura M.,  Min M.,  Deroo P.,  Waters L. B.
  F.~M.,   Dominik C.,  2009, \mn@doi [Astronomy and Astrophysics]
  {10.1051/0004-6361/200912060}, 503, 843

\bibitem[\protect\citeauthoryear{{Gnedin} \& {Draine}}{{Gnedin} \&
  {Draine}}{2014}]{2014ApJ...795...37G}
{Gnedin} N.~Y.,  {Draine} B.~T.,  2014, \mn@doi [\apj]
  {10.1088/0004-637X/795/1/37}, \href
  {https://ui.adsabs.harvard.edu/abs/2014ApJ...795...37G} {795, 37}

\bibitem[\protect\citeauthoryear{{G{\'o}mez-Gordillo}, {Akras},
  {Gon{\c{c}}alves}  \& {Steffen}}{{G{\'o}mez-Gordillo}
  et~al.}{2020}]{2020MNRAS.492.4097G}
{G{\'o}mez-Gordillo} S.,  {Akras} S.,  {Gon{\c{c}}alves} D.~R.,   {Steffen} W.,
   2020, \mn@doi [\mnras] {10.1093/mnras/staa060}, \href
  {https://ui.adsabs.harvard.edu/abs/2020MNRAS.492.4097G} {492, 4097}

\bibitem[\protect\citeauthoryear{Goodman \& Weare}{Goodman \&
  Weare}{2010}]{Goodman.2010}
Goodman J.,  Weare J.,  2010, \mn@doi [Communications in Applied Mathematics
  and Computational Science] {10.2140/camcos.2010.5.65}, 5, 65

\bibitem[\protect\citeauthoryear{{Gordon}, {Clayton}, {Decleir}, {Fitzpatrick},
  {Massa}, {Misselt}  \& {Tollerud}}{{Gordon}
  et~al.}{2023}]{2023ApJ...950...86G}
{Gordon} K.~D.,  {Clayton} G.~C.,  {Decleir} M.,  {Fitzpatrick} E.~L.,  {Massa}
  D.,  {Misselt} K.~A.,   {Tollerud} E.~J.,  2023, \mn@doi [\apj]
  {10.3847/1538-4357/accb59}, \href
  {https://ui.adsabs.harvard.edu/abs/2023ApJ...950...86G} {950, 86}

\bibitem[\protect\citeauthoryear{{Greenhouse}, {Grasdalen}, {Woodward},
  {Benson}, {Gehrz}, {Rosenthal}  \& {Skrutskie}}{{Greenhouse}
  et~al.}{1990}]{1990ApJ...352..307G}
{Greenhouse} M.~A.,  {Grasdalen} G.~L.,  {Woodward} C.~E.,  {Benson} J.,
  {Gehrz} R.~D.,  {Rosenthal} E.,   {Skrutskie} M.~F.,  1990, \mn@doi [\apj]
  {10.1086/168537}, \href
  {https://ui.adsabs.harvard.edu/abs/1990ApJ...352..307G} {352, 307}

\bibitem[\protect\citeauthoryear{Greenhouse, Feldman, Smith, Klapisch, Bhatia
  \& Bar-Shalom}{Greenhouse et~al.}{1994}]{Greenhouse.1994}
Greenhouse M.~A.,  Feldman U.,  Smith H.~A.,  Klapisch M.,  Bhatia A.~K.,
  Bar-Shalom A.,  1994, \mn@doi [IAU Symposium] {10.1017/s0074180900176284},
  159, 447

\bibitem[\protect\citeauthoryear{{G{\"u}ver} \& {{\"O}zel}}{{G{\"u}ver} \&
  {{\"O}zel}}{2009}]{2009MNRAS.400.2050G}
{G{\"u}ver} T.,  {{\"O}zel} F.,  2009, \mn@doi [\mnras]
  {10.1111/j.1365-2966.2009.15598.x}, \href
  {https://ui.adsabs.harvard.edu/abs/2009MNRAS.400.2050G} {400, 2050}

\bibitem[\protect\citeauthoryear{Habart, Peeters, Bern\'e, Trahin, Canin, Chown
   \& et al.}{Habart et~al.}{2024a}]{Habart.2024ljv}
Habart E.,  Peeters E.,  Bern\'e O.,  Trahin B.,  Canin A.,  Chown R.,   et al.
  2024a, \mn@doi [A\&A] {10.1051/0004-6361/202346747}, 685, A73

\bibitem[\protect\citeauthoryear{{Habart} et~al.,}{{Habart}
  et~al.}{2024b}]{2024A&A...685A..73H}
{Habart} E.,  et~al., 2024b, \mn@doi [\aap] {10.1051/0004-6361/202346747},
  \href {https://ui.adsabs.harvard.edu/abs/2024A&A...685A..73H} {685, A73}

\bibitem[\protect\citeauthoryear{Henning et~al.,}{Henning
  et~al.}{2024}]{Henning.2024w1b}
Henning T.,  et~al., 2024, \mn@doi [PASP] {10.1088/1538-3873/ad3455}, 136,
  054302

\bibitem[\protect\citeauthoryear{Hofmeister \& Bowey}{Hofmeister \&
  Bowey}{2006}]{Hofmeister.2006}
Hofmeister A.~M.,  Bowey J.~E.,  2006, \mn@doi [\mnras]
  {10.1111/j.1365-2966.2006.09894.x}, 367, 577

\bibitem[\protect\citeauthoryear{Hollenbach \& McKee}{Hollenbach \&
  McKee}{1989}]{Hollenbach.1989f7}
Hollenbach D.,  McKee C.~F.,  1989, \mn@doi [\apj] {10.1086/167595}, 342, 306

\bibitem[\protect\citeauthoryear{Hummer \& Storey}{Hummer \&
  Storey}{1987}]{Hummer.1987}
Hummer D.~G.,  Storey P.~J.,  1987, \mn@doi [\mnras] {10.1093/mnras/224.3.801},
  224, 801

\bibitem[\protect\citeauthoryear{Icke}{Icke}{2003}]{Icke.2003}
Icke V.,  2003, \mn@doi [Astronomy and Astrophysics]
  {10.1051/0004-6361:20030729}, 405, L11

\bibitem[\protect\citeauthoryear{{Jones} et~al.,}{{Jones}
  et~al.}{2023}]{2023MNRAS.523.2519J}
{Jones} O.~C.,  et~al., 2023, \mn@doi [\mnras] {10.1093/mnras/stad1609}, \href
  {https://ui.adsabs.harvard.edu/abs/2023MNRAS.523.2519J} {523, 2519}

\bibitem[\protect\citeauthoryear{{Kastner}, {Moraga Baez}, {Balick}, {Bublitz},
  {Montez}, {Frank}  \& {Blackman}}{{Kastner} et~al.}{2022}]{Kastner.2022}
{Kastner} J.~H.,  {Moraga Baez} P.,  {Balick} B.,  {Bublitz} J.,  {Montez} R.,
  {Frank} A.,   {Blackman} E.,  2022, \mn@doi [\apj]
  {10.3847/1538-4357/ac51cd}, \href
  {https://ui.adsabs.harvard.edu/abs/2022ApJ...927..100K} {927, 100}

\bibitem[\protect\citeauthoryear{Kemper, Molster, J{\"a}ger  \& Waters}{Kemper
  et~al.}{2002}]{Kemper.200273}
Kemper F.,  Molster F.~J.,  J{\"a}ger C.,   Waters L. B. F.~M.,  2002, \mn@doi
  [\aap] {10.1051/0004-6361:20021119}, 394, 679

\bibitem[\protect\citeauthoryear{{Kemper}, {Vriend}  \& {Tielens}}{{Kemper}
  et~al.}{2004}]{2004ApJ...609..826K}
{Kemper} F.,  {Vriend} W.~J.,   {Tielens} A.~G.~G.~M.,  2004, \mn@doi [\apj]
  {10.1086/421339}, \href
  {https://ui.adsabs.harvard.edu/abs/2004ApJ...609..826K} {609, 826}

\bibitem[\protect\citeauthoryear{Koike et~al.,}{Koike et~al.}{2000}]{k5m}
Koike C.,  et~al., 2000, A\&A

\bibitem[\protect\citeauthoryear{Koike, Imai, Chihara, Suto, Murata,
  Tsuchiyama, Tachibana  \& Ohara}{Koike et~al.}{2010}]{Koike.2010}
Koike C.,  Imai Y.,  Chihara H.,  Suto H.,  Murata K.,  Tsuchiyama A.,
  Tachibana S.,   Ohara S.,  2010, \mn@doi [ApJ] {10.1088/0004-637x/709/2/983},
  709, 983

\bibitem[\protect\citeauthoryear{Kwok, Purton  \& Fitzgerald}{Kwok
  et~al.}{1978}]{Kwok.1978}
Kwok S.,  Purton C.~R.,   Fitzgerald P.~M.,  1978, \mn@doi [\apj]
  {10.1086/182621}, 219, L125

\bibitem[\protect\citeauthoryear{{Lester} \& {Dinerstein}}{{Lester} \&
  {Dinerstein}}{1984}]{LesDin84}
{Lester} D.~F.,  {Dinerstein} H.~L.,  1984, \mn@doi [\apjl] {10.1086/184287},
  \href {https://ui.adsabs.harvard.edu/abs/1984ApJ...281L..67L} {281, L67}

\bibitem[\protect\citeauthoryear{Maloney, Hollenbach  \& Tielens}{Maloney
  et~al.}{1996}]{Maloney.1996}
Maloney P.~R.,  Hollenbach D.~J.,   Tielens A. G. G.~M.,  1996, \mn@doi [\apj]
  {10.1086/177532}, 466, 561

\bibitem[\protect\citeauthoryear{Mastrodemos \& Morris}{Mastrodemos \&
  Morris}{1999}]{Mastrodemos.1999}
Mastrodemos N.,  Morris M.,  1999, \mn@doi [Astrophysical Journal]
  {10.1086/307717}, 523, 357

\bibitem[\protect\citeauthoryear{{Matsuura} et~al.,}{{Matsuura}
  et~al.}{2004}]{2004ApJ...604..791M}
{Matsuura} M.,  et~al., 2004, \mn@doi [\apj] {10.1086/382064}, \href
  {https://ui.adsabs.harvard.edu/abs/2004ApJ...604..791M} {604, 791}

\bibitem[\protect\citeauthoryear{{Matsuura}, {Zijlstra}, {Molster}, {Waters},
  {Nomura}, {Sahai}  \& {Hoare}}{{Matsuura} et~al.}{2005}]{2005MNRAS.359..383M}
{Matsuura} M.,  {Zijlstra} A.~A.,  {Molster} F.~J.,  {Waters} L.~B.~F.~M.,
  {Nomura} H.,  {Sahai} R.,   {Hoare} M.~G.,  2005, \mn@doi [\mnras]
  {10.1111/j.1365-2966.2005.08903.x}, \href
  {https://ui.adsabs.harvard.edu/abs/2005MNRAS.359..383M} {359, 383}

\bibitem[\protect\citeauthoryear{{Matsuura} et~al.,}{{Matsuura}
  et~al.}{2009a}]{2009MNRAS.396..918M}
{Matsuura} M.,  et~al., 2009a, \mn@doi [\mnras]
  {10.1111/j.1365-2966.2009.14743.x}, \href
  {https://ui.adsabs.harvard.edu/abs/2009MNRAS.396..918M} {396, 918}

\bibitem[\protect\citeauthoryear{Matsuura et~al.,}{Matsuura
  et~al.}{2009b}]{Matsuura:2009p23817}
Matsuura M.,  et~al., 2009b, \mn@doi [\apj] {10.1088/0004-637x/700/2/1067},
  700, 1067

\bibitem[\protect\citeauthoryear{McKee \& Hollenbach}{McKee \&
  Hollenbach}{1980}]{McKee.1980}
McKee C.~P.,  Hollenbach D.~J.,  1980, \mn@doi [ARA\&A]
  {10.1146/annurev.aa.18.090180.001251}, 18, 219

\bibitem[\protect\citeauthoryear{{Meaburn}, {L{\'o}pez}, {Steffen}, {Graham}
  \& {Holloway}}{{Meaburn} et~al.}{2005}]{2005AJ....130.2303M}
{Meaburn} J.,  {L{\'o}pez} J.~A.,  {Steffen} W.,  {Graham} M.~F.,   {Holloway}
  A.~J.,  2005, \mn@doi [\aj] {10.1086/496978}, \href
  {https://ui.adsabs.harvard.edu/abs/2005AJ....130.2303M} {130, 2303}

\bibitem[\protect\citeauthoryear{{Meaburn}, {Lloyd}, {Vaytet}  \&
  {L{\'o}pez}}{{Meaburn} et~al.}{2008}]{2008MNRAS.385..269M}
{Meaburn} J.,  {Lloyd} M.,  {Vaytet} N.~M.~H.,   {L{\'o}pez} J.~A.,  2008,
  \mn@doi [\mnras] {10.1111/j.1365-2966.2007.12782.x}, \href
  {https://ui.adsabs.harvard.edu/abs/2008MNRAS.385..269M} {385, 269}

\bibitem[\protect\citeauthoryear{Meijerink \& Spaans}{Meijerink \&
  Spaans}{2005}]{Meijerink.2005r4}
Meijerink R.,  Spaans M.,  2005, \mn@doi [A\&A] {10.1051/0004-6361:20042398},
  436, 397

\bibitem[\protect\citeauthoryear{{Micelotta}, {Jones}  \&
  {Tielens}}{{Micelotta} et~al.}{2010}]{2010A&A...510A..37M}
{Micelotta} E.~R.,  {Jones} A.~P.,   {Tielens} A.~G.~G.~M.,  2010, \mn@doi
  [\aap] {10.1051/0004-6361/200911683}, \href
  {https://ui.adsabs.harvard.edu/abs/2010A&A...510A..37M} {510, A37}

\bibitem[\protect\citeauthoryear{Min, Hovenier  \& Koter}{Min
  et~al.}{2003}]{Min.2003}
Min M.,  Hovenier J.~W.,   Koter A.~D.,  2003, \mn@doi [\aap]
  {10.1051/0004-6361:20030456}, 404, 35

\bibitem[\protect\citeauthoryear{{Mohamed} \& {Podsiadlowski}}{{Mohamed} \&
  {Podsiadlowski}}{2012}]{2012BaltA..21...88M}
{Mohamed} S.,  {Podsiadlowski} P.,  2012, \mn@doi [Baltic Astronomy]
  {10.1515/astro-2017-0362}, \href
  {https://ui.adsabs.harvard.edu/abs/2012BaltA..21...88M} {21, 88}

\bibitem[\protect\citeauthoryear{Molster et~al.,}{Molster
  et~al.}{1999}]{Molster:1999fr}
Molster F.~J.,  et~al., 1999, \mn@doi [Nature] {10.1038/44085}, 401, 563

\bibitem[\protect\citeauthoryear{Molster et~al.,}{Molster
  et~al.}{2001}]{Molster.2001}
Molster F.~J.,  et~al., 2001, \mn@doi [\aap] {10.1051/0004-6361:20010465}, 372,
  165

\bibitem[\protect\citeauthoryear{Molster, Waters  \& Tielens}{Molster
  et~al.}{2002}]{Molster:2002dh}
Molster F.~J.,  Waters L. B. F.~M.,   Tielens A. G. G.~M.,  2002, \mn@doi
  [\aap] {10.1051/0004-6361:20011551}, 382, 222

\bibitem[\protect\citeauthoryear{Montez et~al.,}{Montez
  et~al.}{2015}]{Jr..2015}
Montez R.~J.,  et~al., 2015, \mn@doi [\apj] {10.1088/0004-637x/800/1/8}, 800, 8

\bibitem[\protect\citeauthoryear{Murata, Chihara, Koike, Noguchi, Takakura,
  Imai  \& Tsuchiyama}{Murata et~al.}{2009}]{Murata.2009cl}
Murata K.,  Chihara H.,  Koike C.,  Noguchi T.,  Takakura T.,  Imai Y.,
  Tsuchiyama A.,  2009, \mn@doi [ApJ] {10.1088/0004-637x/698/2/1903}, 698, 1903

\bibitem[\protect\citeauthoryear{{Nuth}, {Wilkinson}, {Johnson}  \&
  {Dwyer}}{{Nuth} et~al.}{2006}]{2006ApJ...644.1164N}
{Nuth} III J.~A.,  {Wilkinson} G.~M.,  {Johnson} N.~M.,   {Dwyer} M.,  2006,
  \mn@doi [\apj] {10.1086/503700}, \href
  {https://ui.adsabs.harvard.edu/abs/2006ApJ...644.1164N} {644, 1164}

\bibitem[\protect\citeauthoryear{Olofsson, Augereau, Dishoeck, MerÃ­n,
  Grosso, MÃ©nard, Blake  \& Monin}{Olofsson et~al.}{2010}]{Olofsson.2010}
Olofsson J.,  Augereau J.-C.,  Dishoeck E. F.~v.,  MerÃ­n B.,  Grosso N.,
  MÃ©nard F.,  Blake G.~A.,   Monin J.-L.,  2010, \mn@doi [Astronomy \&
  Astrophysics] {10.1051/0004-6361/200913909}, 520, A39

\bibitem[\protect\citeauthoryear{{Ossenkopf}, {Henning}  \&
  {Mathis}}{{Ossenkopf} et~al.}{1992}]{1992A&A...261..567O}
{Ossenkopf} V.,  {Henning} T.,   {Mathis} J.~S.,  1992, \aap, \href
  {https://ui.adsabs.harvard.edu/abs/1992A&A...261..567O} {261, 567}

\bibitem[\protect\citeauthoryear{{Peeters} et~al.,}{{Peeters}
  et~al.}{2024}]{2024A&A...685A..74P}
{Peeters} E.,  et~al., 2024, \mn@doi [\aap] {10.1051/0004-6361/202348244},
  \href {https://ui.adsabs.harvard.edu/abs/2024A&A...685A..74P} {685, A74}

\bibitem[\protect\citeauthoryear{P\'equignot, Morisset  \&
  Casassus}{P\'equignot et~al.}{2011}]{Pequignot.2011jr}
P\'equignot D.,  Morisset C.,   Casassus S.,  2011, \mn@doi [Proceedings of the
  International Astronomical Union] {10.1017/s1743921312011970}, 7, 470

\bibitem[\protect\citeauthoryear{{Peretto}, {Fuller}, {Zijlstra}  \&
  {Patel}}{{Peretto} et~al.}{2007}]{2007A&A...473..207P}
{Peretto} N.,  {Fuller} G.,  {Zijlstra} A.,   {Patel} N.,  2007, \mn@doi [\aap]
  {10.1051/0004-6361:20066973}, \href
  {https://ui.adsabs.harvard.edu/abs/2007A&A...473..207P} {473, 207}

\bibitem[\protect\citeauthoryear{{Pignata}, {Weidmann}, {Schmidt}, {Mudrik}  \&
  {Mast}}{{Pignata} et~al.}{2024}]{2024MNRAS.528..459P}
{Pignata} R.~A.,  {Weidmann} W.~A.,  {Schmidt} E.~O.,  {Mudrik} A.,   {Mast}
  D.,  2024, \mn@doi [\mnras] {10.1093/mnras/stad3568}, \href
  {https://ui.adsabs.harvard.edu/abs/2024MNRAS.528..459P} {528, 459}

\bibitem[\protect\citeauthoryear{{Rauber}, {Copetti}  \& {Krabbe}}{{Rauber}
  et~al.}{2014}]{2014A&A...563A..42R}
{Rauber} A.~B.,  {Copetti} M.~V.~F.,   {Krabbe} A.~C.,  2014, \mn@doi [\aap]
  {10.1051/0004-6361/201323363}, \href
  {https://ui.adsabs.harvard.edu/abs/2014A&A...563A..42R} {563, A42}

\bibitem[\protect\citeauthoryear{{Raymond}}{{Raymond}}{1979}]{1979ApJS...39....1R}
{Raymond} J.~C.,  1979, \mn@doi [\apjs] {10.1086/190562}, \href
  {https://ui.adsabs.harvard.edu/abs/1979ApJS...39....1R} {39, 1}

\bibitem[\protect\citeauthoryear{{Rieke} \& {Lebofsky}}{{Rieke} \&
  {Lebofsky}}{1985}]{Rieke.1985}
{Rieke} G.~H.,  {Lebofsky} M.~J.,  1985, \mn@doi [\apj] {10.1086/162827}, \href
  {https://ui.adsabs.harvard.edu/abs/1985ApJ...288..618R} {288, 618}

\bibitem[\protect\citeauthoryear{Rigby, Perrin, McElwain  \& et al.}{Rigby
  et~al.}{2023}]{Rigby.2023}
Rigby J.,  Perrin M.,  McElwain M.,   et al. 2023, \mn@doi [PASP]
  {10.1088/1538-3873/acb293}, 135, 048001

\bibitem[\protect\citeauthoryear{{Roche} \& {Aitken}}{{Roche} \&
  {Aitken}}{1986}]{1986MNRAS.221...63R}
{Roche} P.~F.,  {Aitken} D.~K.,  1986, \mn@doi [\mnras]
  {10.1093/mnras/221.1.63}, \href
  {https://ui.adsabs.harvard.edu/abs/1986MNRAS.221...63R} {221, 63}

\bibitem[\protect\citeauthoryear{{Rodriguez} \& {Moran}}{{Rodriguez} \&
  {Moran}}{1982}]{Rod82}
{Rodriguez} L.~F.,  {Moran} J.~M.,  1982, \mn@doi [\nat] {10.1038/299323a0},
  \href {https://ui.adsabs.harvard.edu/abs/1982Natur.299..323R} {299, 323}

\bibitem[\protect\citeauthoryear{Roueff, Abgrall, Czachorowski, Pachucki,
  Puchalski  \& Komasa}{Roueff et~al.}{2019}]{Roueff.2019}
Roueff E.,  Abgrall H.,  Czachorowski P.,  Pachucki K.,  Puchalski M.,   Komasa
  J.,  2019, \mn@doi [\aap] {10.1051/0004-6361/201936249}, 630, A58

\bibitem[\protect\citeauthoryear{{Sahai}, {Morris}  \& {Villar}}{{Sahai}
  et~al.}{2011}]{2011AJ....141..134S}
{Sahai} R.,  {Morris} M.~R.,   {Villar} G.~G.,  2011, \mn@doi [\aj]
  {10.1088/0004-6256/141/4/134}, \href
  {https://ui.adsabs.harvard.edu/abs/2011AJ....141..134S} {141, 134}

\bibitem[\protect\citeauthoryear{{Saito} et~al.,}{{Saito}
  et~al.}{2012}]{2012A&A...537A.107S}
{Saito} R.~K.,  et~al., 2012, \mn@doi [\aap] {10.1051/0004-6361/201118407},
  \href {https://ui.adsabs.harvard.edu/abs/2012A&A...537A.107S} {537, A107}

\bibitem[\protect\citeauthoryear{Santander-Garc{\'\i}a, Bujarrabal, Alcolea,
  Castro-Carrizo, Contreras, Quintana-Lacaci, Corradi  \&
  Neri}{Santander-Garc{\'\i}a et~al.}{2016}]{Santander-Garcia.2016}
Santander-Garc{\'\i}a M.,  Bujarrabal V.,  Alcolea J.,  Castro-Carrizo A.,
  Contreras C.~S.,  Quintana-Lacaci G.,  Corradi R. L.~M.,   Neri R.,  2016,
  \mn@doi [A\&A] {10.1051/0004-6361/201629288}, 597, A27

\bibitem[\protect\citeauthoryear{{Sargent} et~al.,}{{Sargent}
  et~al.}{2009}]{2009ApJS..182..477S}
{Sargent} B.~A.,  et~al., 2009, \mn@doi [\apjs] {10.1088/0067-0049/182/2/477},
  \href {https://ui.adsabs.harvard.edu/abs/2009ApJS..182..477S} {182, 477}

\bibitem[\protect\citeauthoryear{{Shull} \& {McKee}}{{Shull} \&
  {McKee}}{1979}]{1979ApJ...227..131S}
{Shull} J.~M.,  {McKee} C.~F.,  1979, \mn@doi [\apj] {10.1086/156712}, \href
  {https://ui.adsabs.harvard.edu/abs/1979ApJ...227..131S} {227, 131}

\bibitem[\protect\citeauthoryear{{Soker} \& {Kashi}}{{Soker} \&
  {Kashi}}{2012}]{2012ApJ...746..100S}
{Soker} N.,  {Kashi} A.,  2012, \mn@doi [\apj] {10.1088/0004-637X/746/1/100},
  \href {https://ui.adsabs.harvard.edu/abs/2012ApJ...746..100S} {746, 100}

\bibitem[\protect\citeauthoryear{Suto et~al.,}{Suto et~al.}{2006}]{Suto.2006}
Suto H.,  et~al., 2006, \mn@doi [\mnras] {10.1111/j.1365-2966.2006.10594.x},
  370, 1599

\bibitem[\protect\citeauthoryear{{Szyszka}, {Walsh}, {Zijlstra}  \&
  {Tsamis}}{{Szyszka} et~al.}{2009}]{2009ApJ...707L..32S}
{Szyszka} C.,  {Walsh} J.~R.,  {Zijlstra} A.~A.,   {Tsamis} Y.~G.,  2009,
  \mn@doi [\apjl] {10.1088/0004-637X/707/1/L32}, \href
  {https://ui.adsabs.harvard.edu/abs/2009ApJ...707L..32S} {707, L32}

\bibitem[\protect\citeauthoryear{{Szyszka}, {Zijlstra}  \& {Walsh}}{{Szyszka}
  et~al.}{2011}]{2011MNRAS.416..715S}
{Szyszka} C.,  {Zijlstra} A.~A.,   {Walsh} J.~R.,  2011, \mn@doi [\mnras]
  {10.1111/j.1365-2966.2011.19087.x}, \href
  {https://ui.adsabs.harvard.edu/abs/2011MNRAS.416..715S} {416, 715}

\bibitem[\protect\citeauthoryear{{Temim} et~al.,}{{Temim}
  et~al.}{2024}]{2024ApJ...968L..18T}
{Temim} T.,  et~al., 2024, \mn@doi [\apjl] {10.3847/2041-8213/ad50d1}, \href
  {https://ui.adsabs.harvard.edu/abs/2024ApJ...968L..18T} {968, L18}

\bibitem[\protect\citeauthoryear{{Tielens} \& {Hollenbach}}{{Tielens} \&
  {Hollenbach}}{1985}]{1985ApJ...291..722T}
{Tielens} A.~G.~G.~M.,  {Hollenbach} D.,  1985, \mn@doi [\apj]
  {10.1086/163111}, \href
  {https://ui.adsabs.harvard.edu/abs/1985ApJ...291..722T} {291, 722}

\bibitem[\protect\citeauthoryear{Toal\'a \& Arthur}{Toal\'a \&
  Arthur}{2014}]{Toala.2014zoom}
Toal\'a J.~A.,  Arthur S.~J.,  2014, \mn@doi [\mnras] {10.1093/mnras/stu1360},
  443, 3486

\bibitem[\protect\citeauthoryear{Toal\'a \& Arthur}{Toal\'a \&
  Arthur}{2016}]{Toala.2016}
Toal\'a J.~A.,  Arthur S.~J.,  2016, \mn@doi [\mnras] {10.1093/mnras/stw2307},
  463, 4438

\bibitem[\protect\citeauthoryear{{Uscanga}, {Vel{\'a}zquez}, {Esquivel},
  {Raga}, {Boumis}  \& {Cant{\'o}}}{{Uscanga}
  et~al.}{2014}]{2014MNRAS.442.3162U}
{Uscanga} L.,  {Vel{\'a}zquez} P.~F.,  {Esquivel} A.,  {Raga} A.~C.,  {Boumis}
  P.,   {Cant{\'o}} J.,  2014, \mn@doi [\mnras] {10.1093/mnras/stu1064}, \href
  {https://ui.adsabs.harvard.edu/abs/2014MNRAS.442.3162U} {442, 3162}

\bibitem[\protect\citeauthoryear{Varga et~al.,}{Varga
  et~al.}{2024}]{Varga.2024}
Varga J.,  et~al., 2024, \mn@doi [Astronomy \& Astrophysics]
  {10.1051/0004-6361/202347535}, 681, A47

\bibitem[\protect\citeauthoryear{Waters et~al.,}{Waters
  et~al.}{1998}]{1998Natur.391..868W}
Waters L. B. F.~M.,  et~al., 1998, \mn@doi [Nature] {10.1038/36052}, 391, 868

\bibitem[\protect\citeauthoryear{{Weingartner} \& {Draine}}{{Weingartner} \&
  {Draine}}{2001}]{2001ApJ...548..296W}
{Weingartner} J.~C.,  {Draine} B.~T.,  2001, \mn@doi [\apj] {10.1086/318651},
  \href {https://ui.adsabs.harvard.edu/abs/2001ApJ...548..296W} {548, 296}

\bibitem[\protect\citeauthoryear{Wells et~al.,}{Wells
  et~al.}{2015}]{Wells.2015won}
Wells M.,  et~al., 2015, \mn@doi [PASP] {10.1086/682281}, 127, 646

\bibitem[\protect\citeauthoryear{Wesson}{Wesson}{2016}]{Wesson.2016}
Wesson R.,  2016, \mn@doi [\mnras] {10.1093/mnras/stv2946}, 456, 3774

\bibitem[\protect\citeauthoryear{Wesson et~al.,}{Wesson
  et~al.}{2024}]{Wesson.2023xe}
Wesson R.,  et~al., 2024, \mn@doi [\mnras] {10.1093/mnras/stad3670}, 528, 3392

\bibitem[\protect\citeauthoryear{{Wolfire}, {McKee}, {Hollenbach}  \&
  {Tielens}}{{Wolfire} et~al.}{2003}]{2003ApJ...587..278W}
{Wolfire} M.~G.,  {McKee} C.~F.,  {Hollenbach} D.,   {Tielens} A.~G.~G.~M.,
  2003, \mn@doi [\apj] {10.1086/368016}, \href
  {https://ui.adsabs.harvard.edu/abs/2003ApJ...587..278W} {587, 278}

\bibitem[\protect\citeauthoryear{{Wolfire}, {Hollenbach}  \& {McKee}}{{Wolfire}
  et~al.}{2010}]{2010ApJ...716.1191W}
{Wolfire} M.~G.,  {Hollenbach} D.,   {McKee} C.~F.,  2010, \mn@doi [\apj]
  {10.1088/0004-637X/716/2/1191}, \href
  {https://ui.adsabs.harvard.edu/abs/2010ApJ...716.1191W} {716, 1191}

\bibitem[\protect\citeauthoryear{{Wolfire}, {Vallini}  \& {Chevance}}{{Wolfire}
  et~al.}{2022}]{2022ARA&A..60..247W}
{Wolfire} M.~G.,  {Vallini} L.,   {Chevance} M.,  2022, \mn@doi [\araa]
  {10.1146/annurev-astro-052920-010254}, \href
  {https://ui.adsabs.harvard.edu/abs/2022ARA&A..60..247W} {60, 247}

\bibitem[\protect\citeauthoryear{{Woods}, {Millar}, {Zijlstra}  \&
  {Herbst}}{{Woods} et~al.}{2002}]{2002ApJ...574L.167W}
{Woods} P.~M.,  {Millar} T.~J.,  {Zijlstra} A.~A.,   {Herbst} E.,  2002,
  \mn@doi [\apjl] {10.1086/342503}, \href
  {https://ui.adsabs.harvard.edu/abs/2002ApJ...574L.167W} {574, L167}

\bibitem[\protect\citeauthoryear{{Woods}, {Millar}, {Herbst}  \&
  {Zijlstra}}{{Woods} et~al.}{2003}]{2003A&A...402..189W}
{Woods} P.~M.,  {Millar} T.~J.,  {Herbst} E.,   {Zijlstra} A.~A.,  2003,
  \mn@doi [\aap] {10.1051/0004-6361:20030215}, \href
  {https://ui.adsabs.harvard.edu/abs/2003A&A...402..189W} {402, 189}

\bibitem[\protect\citeauthoryear{Wright, Barlow, Ercolano  \& Rauch}{Wright
  et~al.}{2011}]{Wright.2011}
Wright N.~J.,  Barlow M.~J.,  Ercolano B.,   Rauch T.,  2011, \mn@doi [\mnras]
  {10.1111/j.1365-2966.2011.19490.x}, 418, 370

\bibitem[\protect\citeauthoryear{Wright, Rieke, Glasse  \& et al.}{Wright
  et~al.}{2023}]{Wright.2023}
Wright G.~S.,  Rieke G.~H.,  Glasse A.,   et al. 2023, \mn@doi [PASP]
  {10.1088/1538-3873/acbe66}, 135, 048003

\bibitem[\protect\citeauthoryear{Zeidler, Posch, Mutschke, Richter  \&
  Wehrhan}{Zeidler et~al.}{2011}]{Zeidler.2011}
Zeidler S.,  Posch T.,  Mutschke H.,  Richter H.,   Wehrhan O.,  2011, \mn@doi
  [\aap] {10.1051/0004-6361/201015219}, 526, A68

\bibitem[\protect\citeauthoryear{Zeidler, Posch  \& Mutschke}{Zeidler
  et~al.}{2013}]{Zeidler.2013e78}
Zeidler S.,  Posch T.,   Mutschke H.,  2013, \mn@doi [\aap]
  {10.1051/0004-6361/201220459}, 553, A81

\bibitem[\protect\citeauthoryear{Zeidler, Mutschke  \& Posch}{Zeidler
  et~al.}{2015}]{Zeidler.2015}
Zeidler S.,  Mutschke H.,   Posch T.,  2015, \mn@doi [ApJ]
  {10.1088/0004-637x/798/2/125}, 798, 125

\bibitem[\protect\citeauthoryear{{van Hoof}}{{van Hoof}}{2018}]{Hoof.2018}
{van Hoof} P.~A.,  2018, \mn@doi [Galaxies] {10.3390/galaxies6020063}, 6, 63

\makeatother
\end{thebibliography}
